\documentclass[published]{nst}

\usepackage{subfigure,dcolumn}
\usepackage{epstopdf}
\usepackage{mhchem}
\usepackage{upgreek}
\usepackage{comment}
\usepackage{xspace}
\usepackage{placeins}
\usepackage{relsize}
\usepackage{multirow}

\newcommand{\pp}           {pp\xspace}

\newcommand{\PbPb}         {\mbox{Pb--Pb}\xspace}

\newcommand{\s}            {\ensuremath{\sqrt{s}}\xspace}
\newcommand{\snn}          {\ensuremath{\sqrt{s_{\mathrm{NN}}}}\xspace}
\newcommand{\pt}           {\ensuremath{p_{\rm T}}\xspace}

\newcommand{\Npart}        {\ensuremath{N_\mathrm{part}}\xspace}

\newcommand{\RAA}         {\ensuremath{R_{\rm AA}}\xspace}

\newcommand{\mee}         {\ensuremath{m_{\rm e^+e^-}}\xspace}

\newcommand{\nineH}        {$\sqrt{s}~=~0.9$~Te\kern-.1emV\xspace}
\newcommand{\seven}        {$\sqrt{s}~=~7$~Te\kern-.1emV\xspace}
\newcommand{\twoH}         {$\sqrt{s}~=~0.2$~Te\kern-.1emV\xspace}
\newcommand{\twosevensix}  {$\sqrt{s}~=~2.76$~Te\kern-.1emV\xspace}
\newcommand{\five}         {$\sqrt{s}~=~5.02$~Te\kern-.1emV\xspace}
\newcommand{\twosevensixnn}{$\sqrt{s_{\mathrm{NN}}}~=~2.76$~Te\kern-.1emV\xspace}
\newcommand{\fivenn}       {$\sqrt{s_{\mathrm{NN}}}~=~5.02$~Te\kern-.1emV\xspace}

\newcommand{\GeVc}         {Ge\kern-.1emV/$c$\xspace}
\newcommand{\MeVc}         {Me\kern-.1emV/$c$\xspace}
\newcommand{\TeV}          {Te\kern-.1emV\xspace}
\newcommand{\GeV}          {Ge\kern-.1emV\xspace}
\newcommand{\MeV}          {Me\kern-.1emV\xspace}
\newcommand{\GeVmass}      {Ge\kern-.2emV/$c^2$\xspace}
\newcommand{\MeVmass}      {Me\kern-.2emV/$c^2$\xspace}

\newcommand{\gev} {\ensuremath{\mathrm{GeV/}c}}

\newcommand{\ee}           {\ensuremath{\rm{e}^{+}\rm{e}^{-}}\xspace}

\newcommand{\dzero}        {\ensuremath{{\rm D}^{0}}\xspace}
\newcommand{\ds}{\ensuremath{{\rm D}_{\rm s}^{+}}}
\newcommand{\bs}{\ensuremath{{\rm B}_{\rm s}^{0}}}
\newcommand{\lc}{\ensuremath{\lmb_{\rm c}^{+}}}
\newcommand{\lmb}          {\ensuremath{\Lambda}\xspace}

\newcommand{\jpsi}         {\ensuremath{{\rm J}/\psi}\xspace}
      
\newcommand{\psitwos}         {\ensuremath{\psi(\rm 2S)}\xspace}

\newcommand{\abs}[1]{\ensuremath{\left|#1\right|}}

\newcommand{\cent}   [2] {$#1$--$#2\%$}

\newcommand{\pTjet}{\ensuremath{p_\mathrm{T, jet}}}
\newcommand{\pTjetch}{\ensuremath{p_\mathrm{T, ch~jet}}}
\newcommand{\pTtrig}{\ensuremath{p_\mathrm{T}^\mathrm{trig}}}
\newcommand{\rr}{\ensuremath{R}}

\newcommand {\dphi}{\ensuremath{\Delta (\mathrm{phi})}\xspace}

\newcommand{\IAApT}{\ensuremath{{I}_{\rm AA}(\pTjetch)}}
\newcommand{\IAAphi}{\ensuremath{{I}_{\rm AA}(\dphi)}}
\newcommand{\Drecoil}{\ensuremath{\Delta_\mathrm{recoil}}}

\newcommand{\DrecoilpT}{\ensuremath{\Delta_\mathrm{recoil}(\pTjet)}}

\newcommand{\TTSig}{\ensuremath{\mathrm{TT}_\mathrm{sig}}}

\begin{document}

\title{Properties of the QCD Matter: \\ A Review of Selected Results from the ALICE Experiment}
\thanks{This review is dedicated to Professor Wenqing Shen in honor of his leadership and significant impact on the Chinese heavy-ion physics community. All authors contributed equally to this work.}

\author{Qi-Ye Shou}
\affiliation{Key Laboratory of Nuclear Physics and Ion-Beam Application (MOE), Institute of Modern Physics, Fudan University, Shanghai 200433, China}
\affiliation{Shanghai Research Center for Theoretical Nuclear Physics, NSFC and Fudan University, Shanghai 200438, China}

\author{Yu-Gang Ma}
\affiliation{Key Laboratory of Nuclear Physics and Ion-Beam Application (MOE), Institute of Modern Physics, Fudan University, Shanghai 200433, China}
\affiliation{Shanghai Research Center for Theoretical Nuclear Physics, NSFC and Fudan University, Shanghai 200438, China}

\author{Song Zhang}
\affiliation{Key Laboratory of Nuclear Physics and Ion-Beam Application (MOE), Institute of Modern Physics, Fudan University, Shanghai 200433, China}
\affiliation{Shanghai Research Center for Theoretical Nuclear Physics, NSFC and Fudan University, Shanghai 200438, China}

\author{Jian-Hui Zhu}
\affiliation{Key Laboratory of Nuclear Physics and Ion-Beam Application (MOE), Institute of Modern Physics, Fudan University, Shanghai 200433, China}
\affiliation{Shanghai Research Center for Theoretical Nuclear Physics, NSFC and Fudan University, Shanghai 200438, China}

\author{Ya-Xian Mao}
\affiliation{Key Laboratory of Quark \& Lepton Physics (MOE) and Institute of Particle Physics, Central China Normal University, Wuhan 430079, China}

\author{Hua Pei}
\affiliation{Key Laboratory of Quark \& Lepton Physics (MOE) and Institute of Particle Physics, Central China Normal University, Wuhan 430079, China}

\author{Zhong-Bao Yin}
\affiliation{Key Laboratory of Quark \& Lepton Physics (MOE) and Institute of Particle Physics, Central China Normal University, Wuhan 430079, China}

\author{Xiao-Ming Zhang}
\affiliation{Key Laboratory of Quark \& Lepton Physics (MOE) and Institute of Particle Physics, Central China Normal University, Wuhan 430079, China}

\author{Dai-Cui Zhou}
\affiliation{Key Laboratory of Quark \& Lepton Physics (MOE) and Institute of Particle Physics, Central China Normal University, Wuhan 430079, China}

\author{Xin-Ye Peng}
\affiliation{School of Mathematics and Physics, China University of Geosciences (Wuhan), Wuhan 430074, China}

\author{Xiao-Zhi Bai}
\affiliation{Department of Modern Physics, University of Science and Technology of China, Hefei, China}

\author{Ze-Bo Tang}
\affiliation{Department of Modern Physics, University of Science and Technology of China, Hefei, China}

\author{Yi-Fei Zhang}
\affiliation{Department of Modern Physics, University of Science and Technology of China, Hefei, China}

\author{Xiao-Mei Li}
\affiliation{China Institute of Atomic Energy, Beijing 102413, China}

\begin{abstract}
The Large Hadron Collider (LHC), the world's largest and most powerful particle accelerator, has been a pivotal tool in advancing our understanding of fundamental physics. By colliding heavy ions (such as lead ions), the LHC recreates conditions similar to those just after the Big Bang. This allows scientists to study the Quark-Gluon Plasma (QGP), a state of matter where quarks and gluons are not confined within protons and neutrons. These studies provide insights into the strong force and the early universe's behavior.
In this paper, we provide a comprehensive overview of recent significant findings from A Large Ion Collider Experiment (ALICE) at LHC. The topics encompass measurements regarding to properties of the QGP, particle production, flow and correlations, dileptons, quarkonia and electromagnetic probes, heavy flavor, and jets. Additionally, we introduce future plans for detector upgrades of the ALICE experiment.

\end{abstract}


\maketitle

\tableofcontents

\section{Introduction} \label{sec.I}

The strong force dictates the interactions between quarks and gluons, which are the elementary particles that account for most of the visible mass in the universe. Quantum Chromodynamics (QCD), a non-Abelian gauge theory, is the mathematical framework that describes the strong force and is essential for comprehending the fundamental nature of matter under extreme conditions~\cite{wilczek_quantum_1999, gross_ultraviolet_1973, politzer_reliable_1973}.

QCD is characterized by two notable features: asymptotic freedom and color confinement. Asymptotic freedom describes how the interaction between quarks and gluons weakens as their momentum exchange increases. Color confinement dictates that quarks and gluons cannot exist in isolation; they are always bound together within composite particles called hadrons, which can not carry a net color charge and are the only observable entities.
The field of ``QCD condensed matter" investigates the behavior of quarks and gluons in a dense many-body system under conditions of high energy density. By heating such a system, often with zero net baryon density, to temperatures above 150–160 MeV, scientists can observe the creation of quark–gluon plasma (QGP).

Unlike normal nuclear matter, the QGP is a state where quarks and gluons are not confined within hadrons. It is believed that the early Universe existed in such a primordial state for the first few millionths of a second after the Big Bang, with the strong force playing a crucial role in the formation of the vast majority of visible mass in the Universe. Recreating this primordial state of matter in laboratory experiments and studying its evolution can provide insights into the organization of matter and the mechanisms governing the confinement of quarks and gluons.

\begin{figure*}[!htb]
\includegraphics[width=\linewidth]{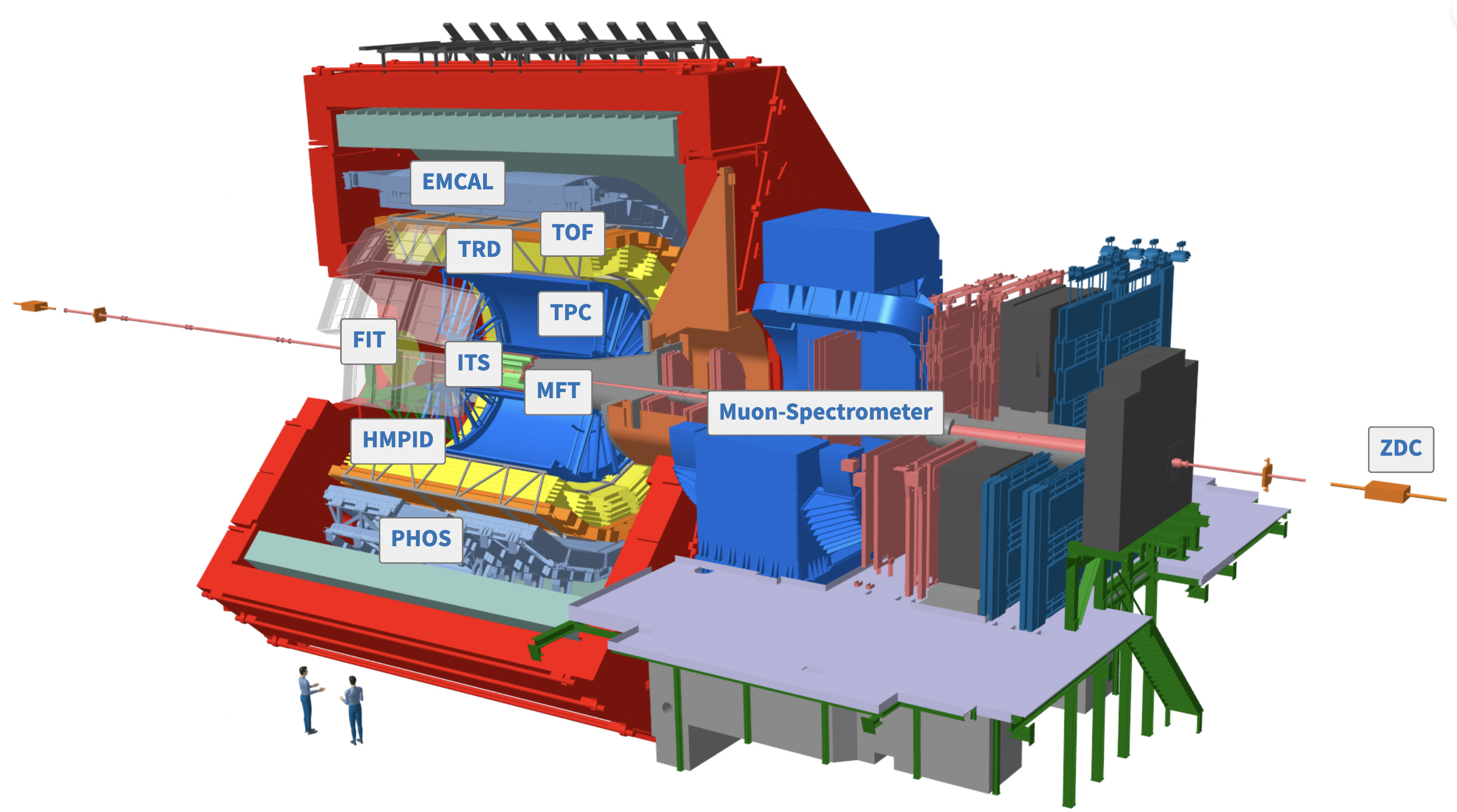}
\caption{The ALICE detector in Run 3 era. See the texts for details.}
\label{fig:alice}
\end{figure*}

Relativistic heavy-ion collisions at the BNL-STAR generate QGP, which is in the extreme states with hottest,  densest, vortical and polarized, but with the smallest viscose  fluid  ever studied in the laboratory \cite{Liang:2004ph,STAR:2017ckg,STAR:2022fan,Chen:2023hnb,Luo:2017faz,Deng2024,He-NST,Ma-CPL,SunKJ_NC,Chen-NST}, and the LHC-ALICE  heavy-ion collisions can generate even more extreme matter. The ALICE (A Large Ion Collider Experiment) detector~\cite{collaboration_alice_2008, ALICEIntro} at the LHC was specifically designed to investigate the properties of QGP produced at these high energies. In the laboratory, conditions similar to those of the early universe can be recreated by colliding heavy ions at energies in the multi-TeV range.
ALICE began physics data collection in 2009 with the first LHC $pp$ collisions at \(\sqrt{s} = 0.9\) TeV, and has since gathered data from all available collision systems and energies during Run 1 (2009–2013) and Run 2 (2015–2018). Notably, Pb-Pb collisions were studied in 2010 and 2011 at \(\sqrt{s_{\rm NN}} = 2.76\) TeV, and in 2015 and 2018 at \(\sqrt{s_{\rm NN}} = 5.02\) TeV. A brief run with Xe-Xe collisions at \(\sqrt{s_{\rm NN}} = 5.44\) TeV took place in 2017. Proton-proton collisions at the same energies as Pb-Pb collisions were also conducted over the years, serving as a reference for nucleus–nucleus reactions and for specific QCD studies. For investigating cold nuclear matter effects, $p$-Pb collisions were studied in 2013 and 2016 at \(\sqrt{s_{\rm NN}} = 5.02\) and 8.16 TeV, following a pilot run in 2012. Additionally, ALICE collected $pp$ collisions data at various energies up to \(\sqrt{s} = 13\) TeV over the years. In 2021, ALICE completed a significant upgrade of its detectors to enhance its capabilities for scientific exploration during LHC Runs 3 and 4, planned until the end of 2032. Concurrently, plans are underway for ALICE 3, the next-generation experiment for LHC Runs 5 and 6.

The ALICE detector (Fig.~\ref{fig:alice}) is positioned at the interaction point IP2 of the LHC. It comprises a central barrel that covers the full azimuthal angle and the pseudorapidity region \( |\eta| < 0.9 \). The detector is equipped with robust particle identification capabilities up to \( p_T \) of 20 GeV/c, along with excellent capabilities for reconstructing primary and secondary vertices. ALICE's main charged-particle tracking detectors include the Inner Tracking System (ITS) and a large Time Projection Chamber (TPC). Complementing the TPC, external tracking is provided by a Transition Radiation Detector (TRD) and a Time Of Flight system (TOF). Beyond the TOF, the azimuthal region houses two electromagnetic calorimeters: the high-resolution PHOton Spectrometer (PHOS) and the EMCal, as well as a High Momentum Particle Identification Detector (HMPID). The central barrel detectors are enclosed within the L3 solenoid magnet, which generates a magnetic field of up to \( B = 0.5 \) T. In forward region, ALICE has a muon spectrometer and various sets of smaller detectors, including Forward Multiplicity Detector (FMD), Photon Multiplicity Detector (PMD), V0, T0 and Zero-Degree Calorimeters (ZDC). Each year, the data collected at ALICE can easily reach sizes exceeding tens of Petabytes. Processing such vast amounts of data for reconstructing physics objects from raw data is made possible by the Worldwide LHC Computing Grid (WLCG) infrastructure, which relies on approximately 200 computing clusters distributed worldwide. In China, a newly reactivated cluster located at the Institute of High Energy Physics, CAS, Beijing, has been operational since 2024.

Below, we provide a brief introduction to selected physics topics investigated by ALICE, organized as follows. 
In Sec.~\ref{sec.II}, highlight results in macroscopic properties~(\ref{sec.II.1}), flow and correlations~(\ref{sec.II.2}), dileptons, quarkonia and electromagnetic probes~(\ref{sec.II.3}), heavy flavour~(\ref{sec.II.4}) and jets~(\ref{sec.II.5}) will be presented in turn. Future plans for detector upgrades will be discussed in Sec.~\ref{sec.III}.

\section{The study of quark–gluon plasma} \label{sec.II}

\subsection{Macroscopic properties}  \label{sec.II.1}
\begin{figure}[!htb]
\includegraphics[width=\linewidth]{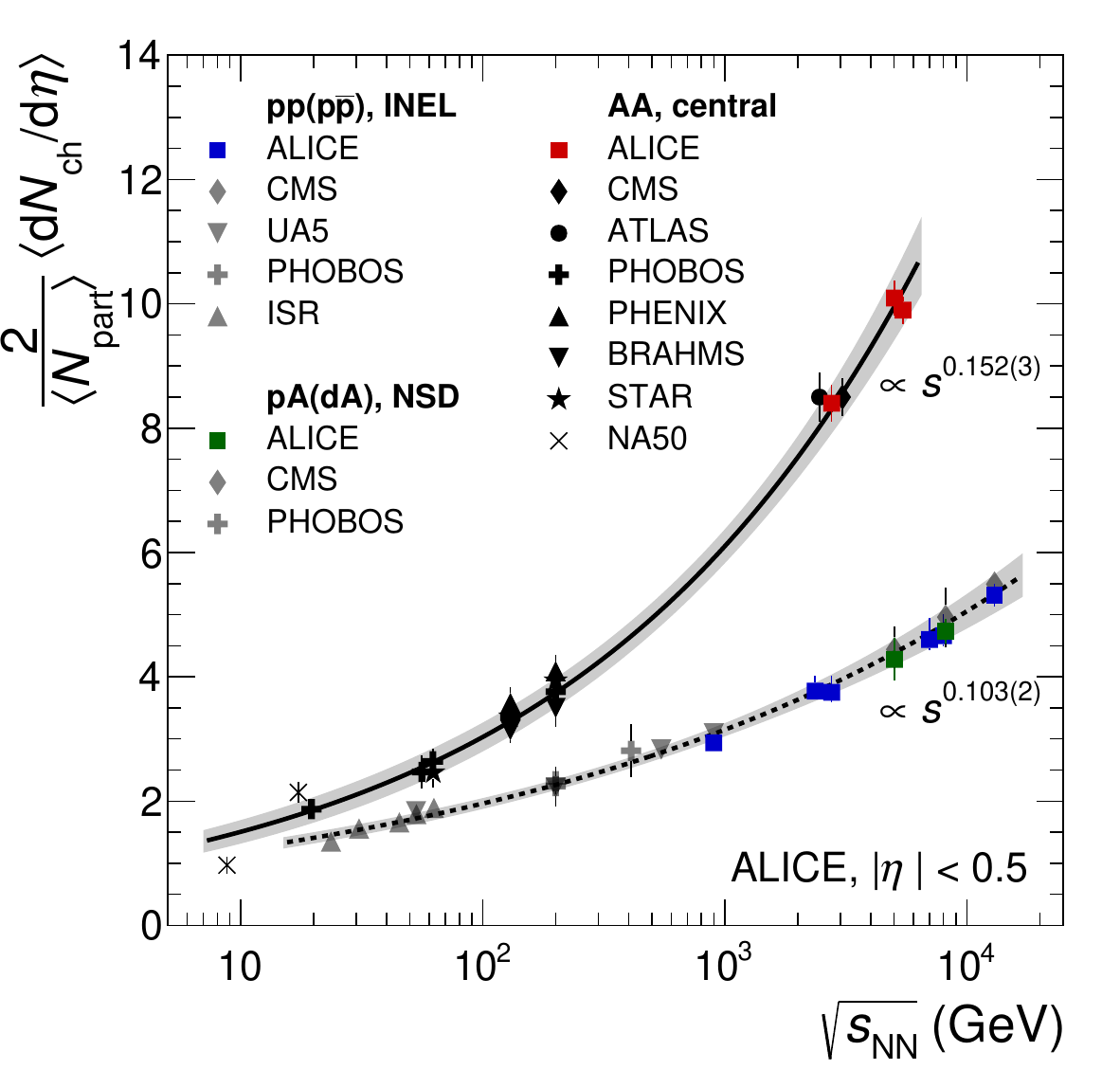}
\caption{Collision energy dependence of the charged-particle pseudorapidity density at midrapidity ($|y|<$0.5) normalised to the average number of participants, $\frac{2}{\langle N_{part}\rangle}\langle dN_{ch}/d\eta \rangle$.}
\label{fig:dNch-deta-sqrtsFig}
\end{figure}

ALICE's objective is to investigate the QGP created at center-of-mass energies ranging from a few TeV. A crucial aspect of this research is estimating the initial energy density and temperature necessary for QGP formation in collisions. The initial energy density can be inferred from the observed hadron production in the final state of the fireball across different collision centrality classes and at varying center-of-mass energies.
Figure~\ref{fig:dNch-deta-sqrtsFig} from~\cite{ALICEIntro} illustrates the scaled charged particle multiplicity measured at midrapidity ($|y| < 0.5$), normalized by $\langle N_{part} \rangle/2$, in various collision systems including $pp$, $p$${\bar p}$, $p$($d$)A, and central heavy-ion collisions, plotted against the center-of-mass energy per nucleon pair, $\sqrt{s}$. ALICE has contributed data from Pb-Pb collisions at $\sqrt{s_{\rm NN}} = 2.76$ TeV and 5.02 TeV, Xe-Xe collisions at $\sqrt{s_{\rm NN}} = 5.44$ TeV, and $pp$(${\bar p}$, Pb) collisions spanning a broad range of $\sqrt{s}$ from TeV to above ten TeV.
The dependence of $\frac{2}{\langle N_{part} \rangle}\langle \frac{dN_{ch}}{d\eta} \rangle$ on \snn was fitted with a function of $\alpha \times s^\beta$. The fitting results yielded $\beta = 0.152 \pm 0.003$ for central A-A collisions and $\beta = 0.103 \pm 0.002$ for p-p and p(d)-A collisions. This indicates that heavy-ion collisions are significantly more efficient in converting initial beam energy into particle production at midrapidity compared to $pp$ or $p$-Pb collisions~\cite{ALICEIntro}.

\begin{figure}[!htb]
\includegraphics[width=\linewidth]{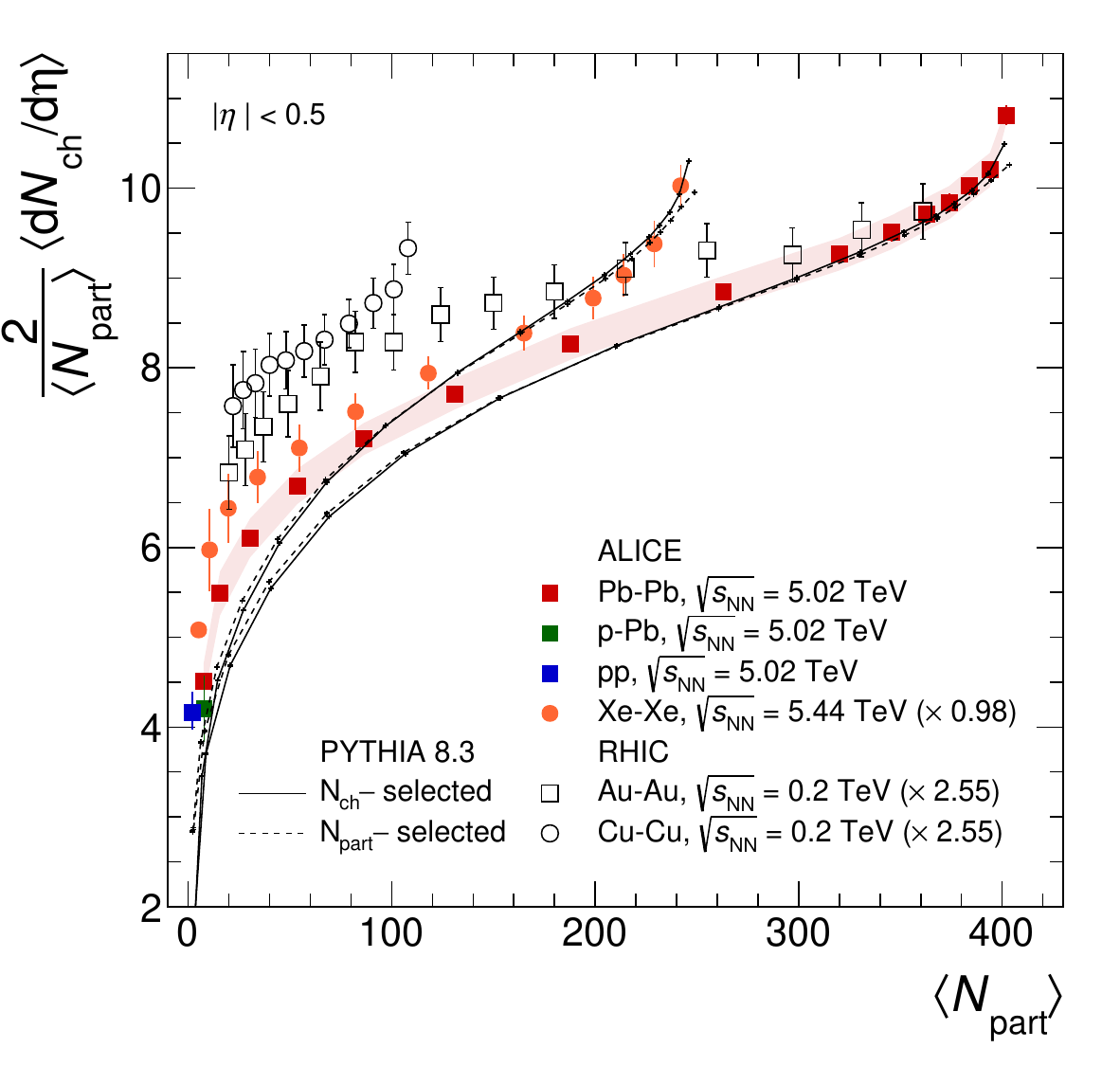}
\caption{Values of $\frac{2}{\langle N_{part}\rangle}\langle dN_{ch}/d\eta \rangle$ are compared in various collisions.}
\label{fig:dNch-npartFig}
\end{figure}

The size dependence of particle production in collision systems ranging from $p$-p(Pb) to Xe-Xe and Pb-Pb has been measured with unprecedented precision. Figure~\ref{fig:dNch-npartFig} from~\cite{ALICEIntro} illustrates the centrality dependence of $\langle N_{part}\rangle$ on $\frac{2}{\langle N_{part}\rangle}\langle \frac{dN_{ch}}{d\eta} \rangle$. Data were collected from Pb-Pb and Xe-Xe collisions at $\sqrt{s_{\rm NN}} = 5.02$ TeV and 5.44 TeV, respectively, as well as from Au-Au and Cu-Cu collisions at RHIC energy setups.
Uncertainties range from approximately 3\% for central A-A collisions at midrapidity to about 10\% for peripheral results in the forward region. Results at $\sqrt{s_{\rm NN}} = 5.02$ TeV were scaled using factors calculated from the fit function in Fig.~\ref{fig:dNch-deta-sqrtsFig} for the top 5\% most central Au-Au, Cu-Cu, and Xe-Xe collisions.
At the same $\langle N_{part}\rangle$, the shape of $\frac{2}{\langle N_{part}\rangle}\langle \frac{dN_{ch}}{d\eta} \rangle$ as a function of $\langle N_{part}\rangle$ showed slightly more variability in Xe-Xe compared to Pb-Pb, and a similar pattern was observed between Au-Au and Cu-Cu collision systems. Reference~\cite{ALICEIntro} noted that these deviations, while present, are not significant given the large uncertainties and could be attributed to different Glauber model simulations used to estimate $\langle N_{part}\rangle$ in ALICE~\cite{PhysRevLett.116.222302} and RHIC~\cite{PhysRevC.83.024913}.
Comparison to PYTHIA 8.3~\cite{Bierlich2016} calculations in Fig.~\ref{fig:dNch-npartFig}, with different centrality selection methods (e.g., $N_{ch}$-selected and $N_{part}$-selected), showed variations in the dependence of $\frac{2}{\langle N_{part}\rangle}\langle \frac{dN_{ch}}{d\eta} \rangle$, indicating fluctuations in charged-particle multiplicity at a fixed number of particle sources.

\begin{figure}[htbp!]
  \centering
  \includegraphics[width=\linewidth]{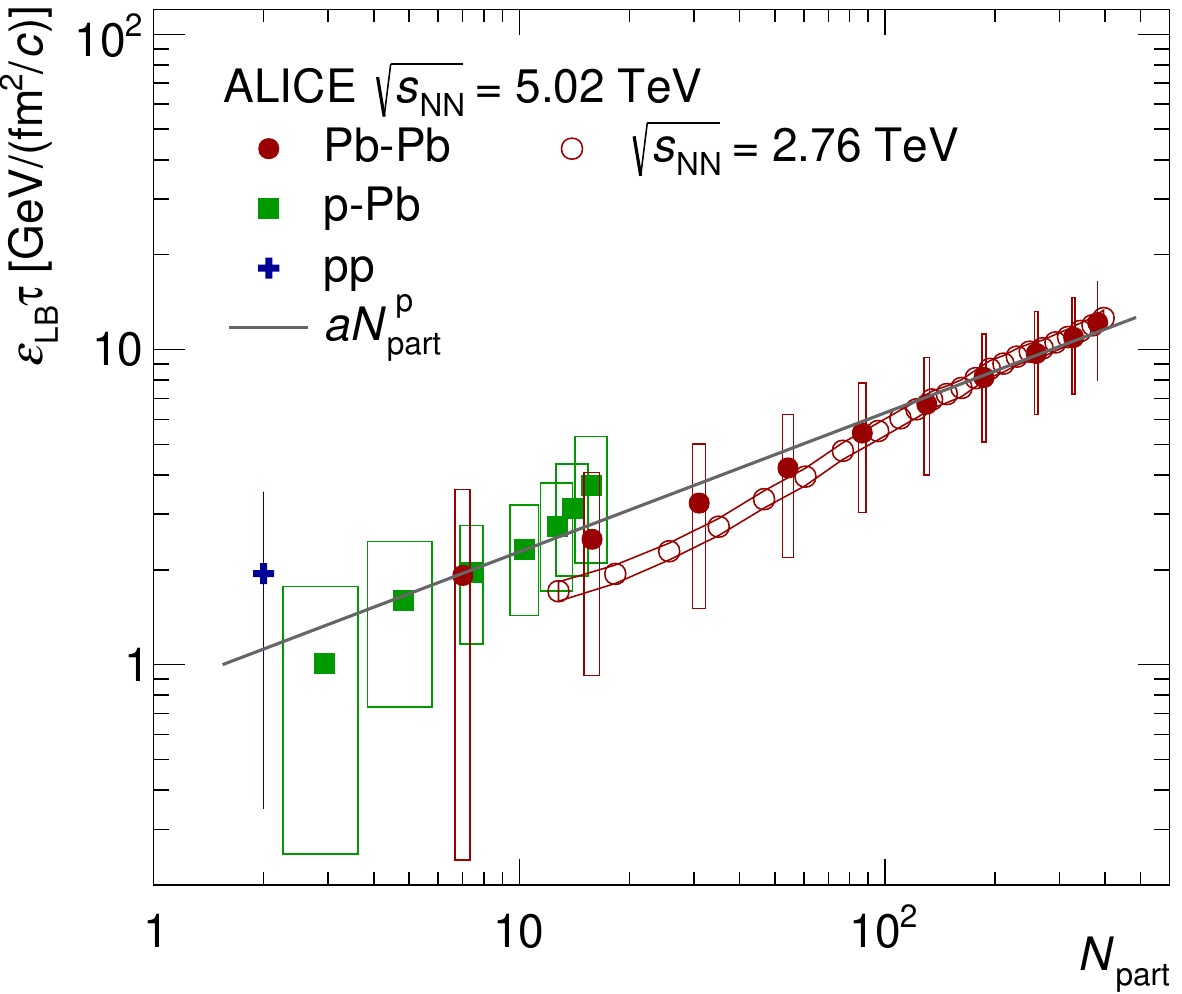}
  \caption[Lower bound estimate of the energy density in $pp$, $p$-Pb, and Pb-Pb at \snn = 5.02 TeV]{Lower-bound estimate of the energy density times the formation time $\tau$ in $pp$, $p$-Pb, and Pb-Pb collisions at \snn = 5.02 TeV as a function of the number of participants~\cite{ALICE:2022imr}.}
  \label{fig:all_ebj_2760}
\end{figure}

Particle production measurements are pivotal in estimating the initial energy density and temperature, crucial for determining whether the conditions for the QCD phase transition are met during collisions~\cite{NTQCD2023_01, NTQCD2023_04, NTQCD2023_05, NTQCD2023_06, NTQCD2023_07, NTQCD2023_12, NTQCD2023_13, ko_searching_2023}. The energy density in the collision can be estimated using the ``Bjorken estimate"~\cite{PhysRevD.27.140}: $\epsilon(\tau)$, derived from the total produced transverse momentum.
This estimation method applies to a system that undergoes free-streaming with boost-invariant longitudinal expansion and no transverse expansion, as described in references \cite{PhysRevD.27.140,ALICEIntro}. Using the measured charged-particle pseudorapidity density and assuming a normal distribution of charged particles in rapidity \cite{Adam:2016ddh,Abbas:2013bpa}, it becomes feasible to derive a lower-bound estimate of the energy density $\varepsilon_{\mathrm{LB}}$ multiplied by the formation time $\tau$ in the collisions \cite{ALICE:2022imr}.
The transverse area \( S_\mathrm{T} \) is determined using the Glauber model \cite{Loizides:2014vua}, which accounts for all participating nucleons. Figure \ref{fig:all_ebj_2760} illustrates the resulting product \( \varepsilon_\mathrm{LB} \tau \) for pp, p-Pb, and Pb-Pb collisions at \( \sqrt{s_{\mathrm{NN}}} = 5.02 \) TeV, as well as Pb-Pb collisions at \( \sqrt{s_{\mathrm{NN}}} = 2.76 \) TeV. 
A power-law fit \( aN_\mathrm{part}^p \) applied to the data indicates a notable increase in the energy density with the increasing transverse area of the initial overlap between the colliding nuclei \cite{ALICEIntro}.

\begin{figure}[!tbh]
    \begin{center}
    \includegraphics[width = 0.49\textwidth]{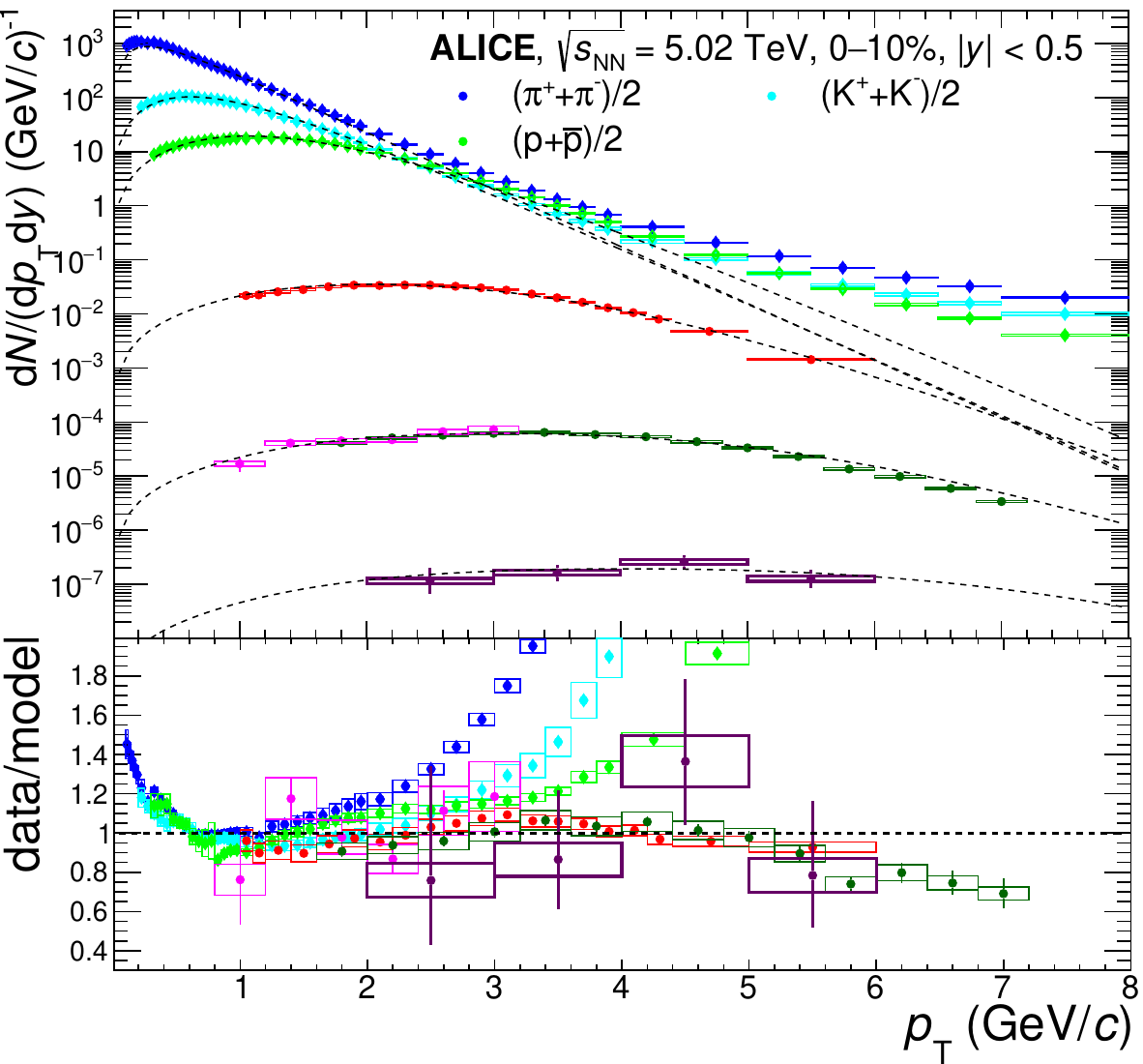}
    \includegraphics[width = 0.49\textwidth]{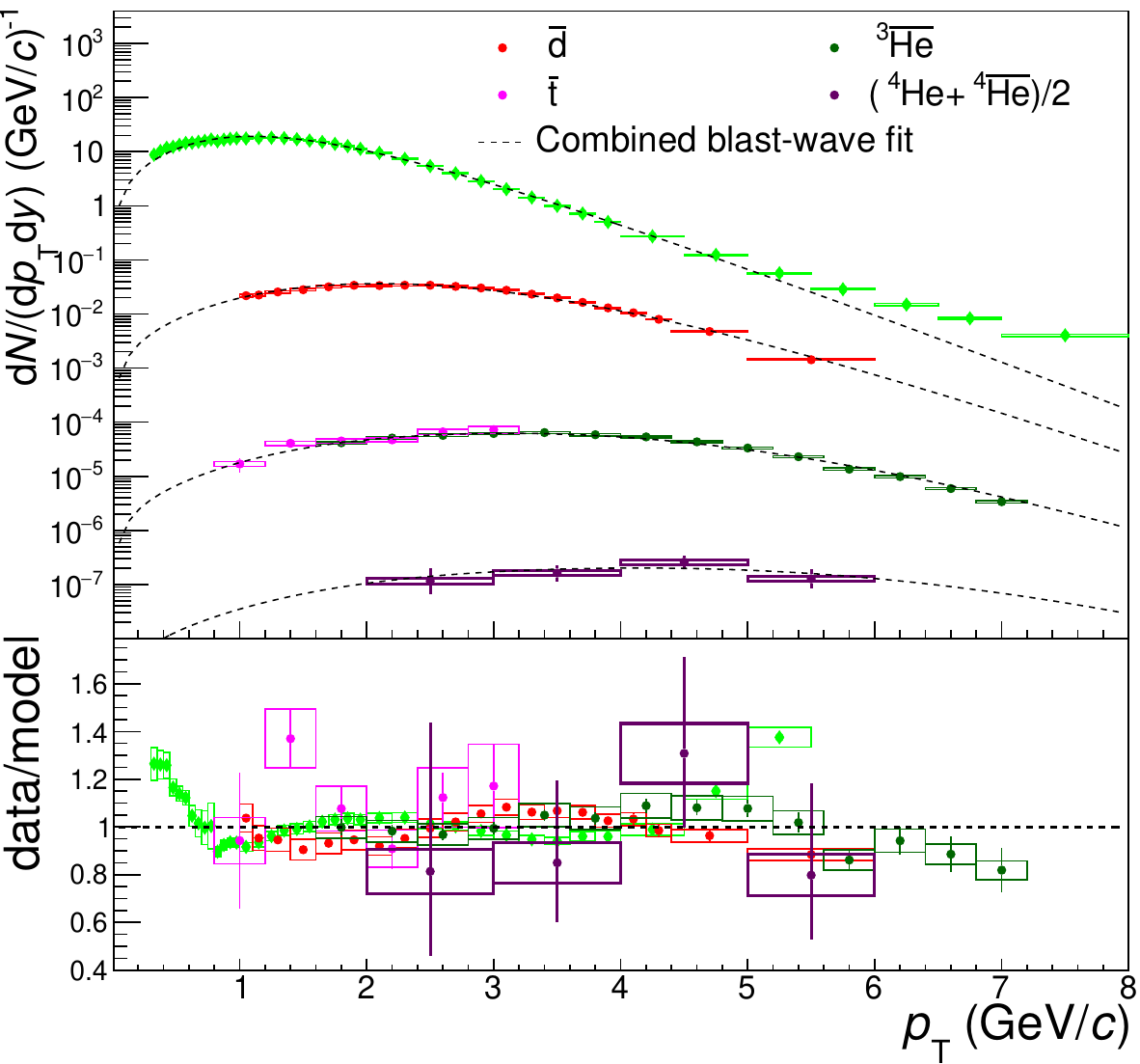}
    \end{center}
    \caption{(Color online) Combined blast-wave fit of all available light flavored hadron $p_{\rm T}$ spectra including nuclei~\cite{ALICE:2019hno,Nuclei_5TeV} (left) and only p, $\overline{\mathrm{d}}$, $\overline{\mathrm{t}}$, ${}^{3}\overline{\mathrm{He}}$ and ${}^{4}$He $p_{\rm T}$ spectra (right) in Pb--Pb collisions at $\sqrt{s_{\rm NN}} = 5.02$ TeV for 0--10\% central events (upper panels). The lower panels show the ratio between each data point and the blast-wave model fit for each species.}
    \label{fig:bw_nucl}
\end{figure}

Another critical parameter in determining the formation of QGP is the temperature of the system created in collisions. It's important to note two distinct concepts of temperature: the chemical freeze-out temperature \cite{Braun_Munzinger_2004} and the kinetic freeze-out temperature \cite{BLWPhysRevC.48.2462}. These temperatures characterize the thermodynamic properties of the fireball at the stages of chemical equilibrium and ceasing of hadron rescattering, respectively.
The measurement of the kinetic freeze-out temperature has incorporated light nuclei to explore the system's thermalization~\cite{alicealpha2024}. Due to their large mass, (anti-)$\alpha$ production yields and transverse-momentum spectra are particularly significant as they rigorously test particle production models. The combined anti-$\alpha$ and $\alpha$ spectrum, when included in a common blast-wave fit with lighter particles, indicates that the (anti-)$\alpha$ also partakes in the collective expansion of the collision medium. A blast-wave fit using only protons, (anti-)$\alpha$, and other light nuclei yields a flow velocity comparable to the fit that includes all particles. However, fitting only protons and light nuclei results in a similar flow velocity but a notably higher kinetic freeze-out temperature, as shown in Fig.~\ref{fig:bw_nucl}. Interestingly, hypernuclei exhibit similar flow velocity and kinetic freeze-out temperature as light nuclei~\cite{aliceH3Lproductionpbpb, chen_measurements_2023}, although more data from ALICE Run 3 is needed to solidify these findings~\cite{aliceH3Lproductionpbpb}.

The temperature of the early partonic phase can be experimentally accessed through sensitive probes produced in the early stages of collisions, such as heavy-flavour $\mathrm{q\bar{q}}$ states (quarkonia) and electromagnetic radiation. Since the seminal work by Matsui and Satz \cite{Matsui:1986dk}, quarkonium has been proposed as a thermometer for the QGP. The strong binding potential between quark and antiquark pairs is screened by color charges in the dense and hot medium, leading to the ``melting" of $\mathrm{q\bar{q}}$ states. This phenomenon offers the opportunity to correlate the production or suppression of quarkonia with the temperature of the QGP.
A detailed discussion on quarkonia and dileptons will be presented in Section \ref{sec.II.3}.

\begin{figure}[htb]
    \begin{center}
     \includegraphics[width = \linewidth]{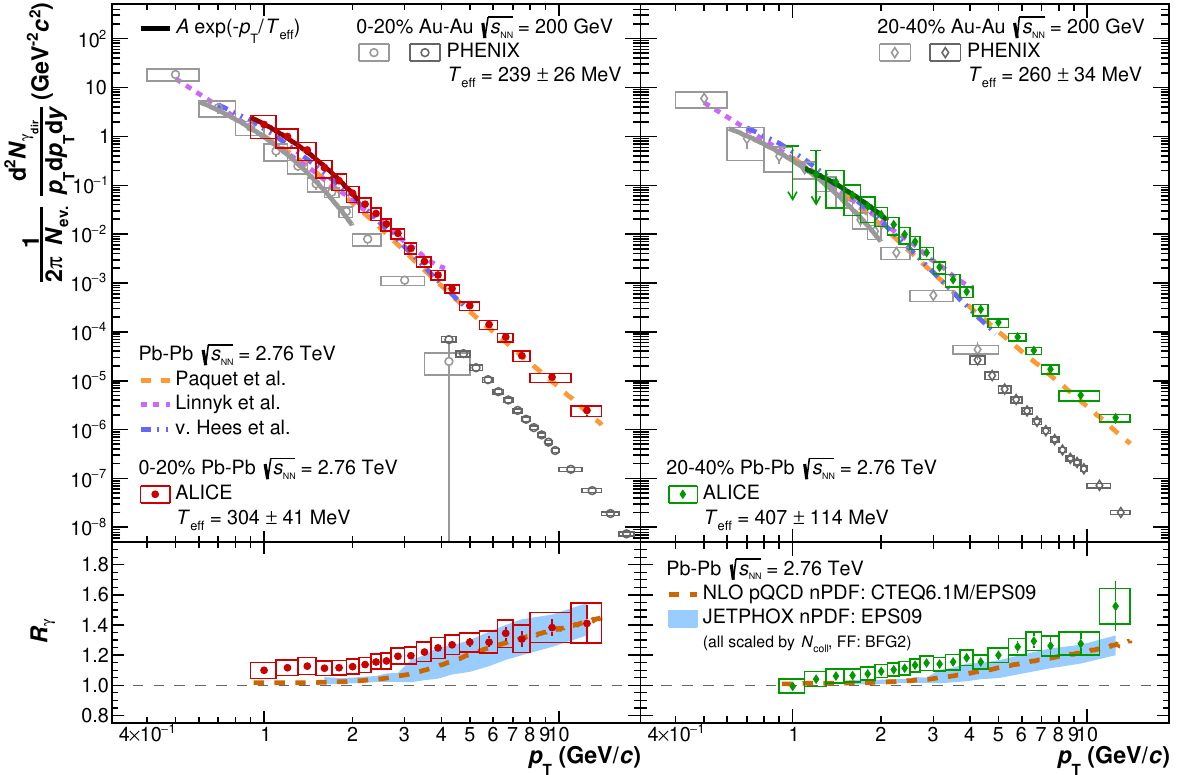}
    \end{center}
    \caption{Direct photon spectra (top) and direct photon excess $R_\gamma$ (bottom) measured in Pb-Pb collisions at 2.76~TeV~\cite{Adam:2015lda} and in Au-Au collisions at 0.2~TeV~\cite{Adler:2005,Adare:2008ab} in the 0--20\% (left) and 20--40$\%$ (right) centrality classes.}
    \label{fig:direct-photon-spectra}
\end{figure}

\begin{figure}[htb]
    \begin{center}
    \includegraphics[width = \linewidth]{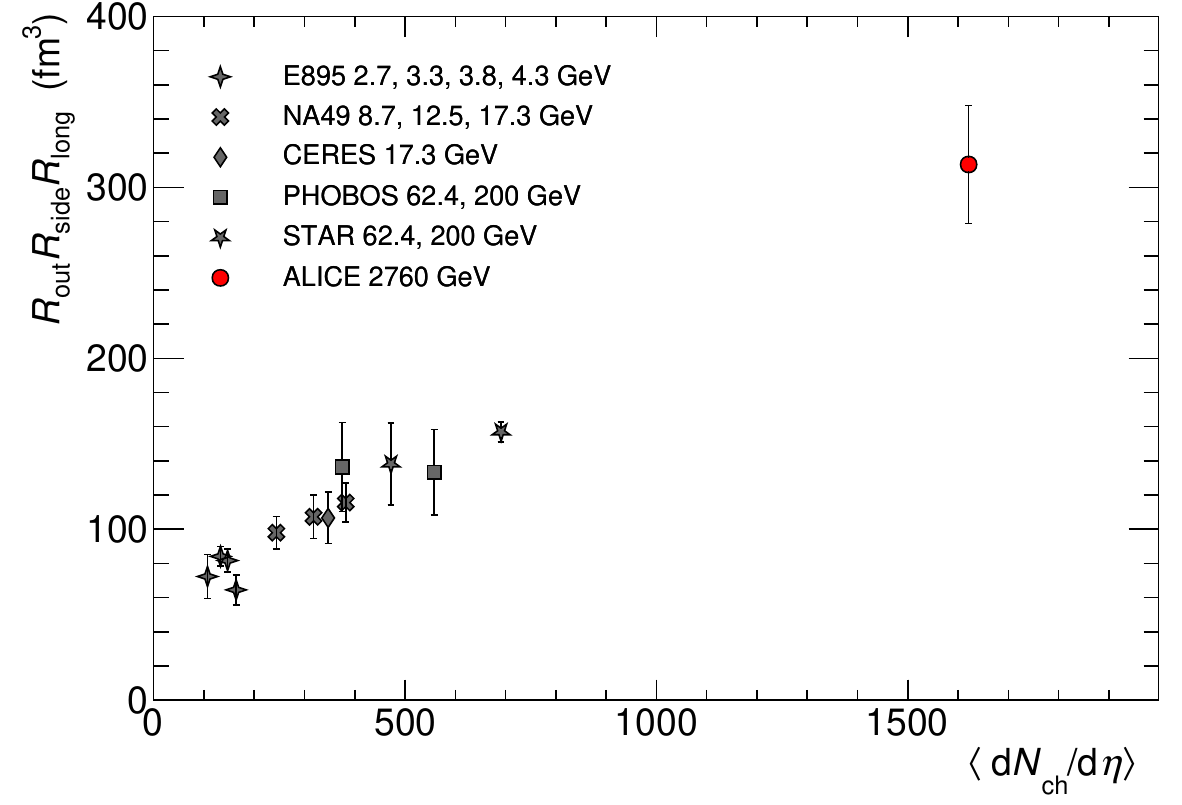}
    \includegraphics[width = \linewidth]{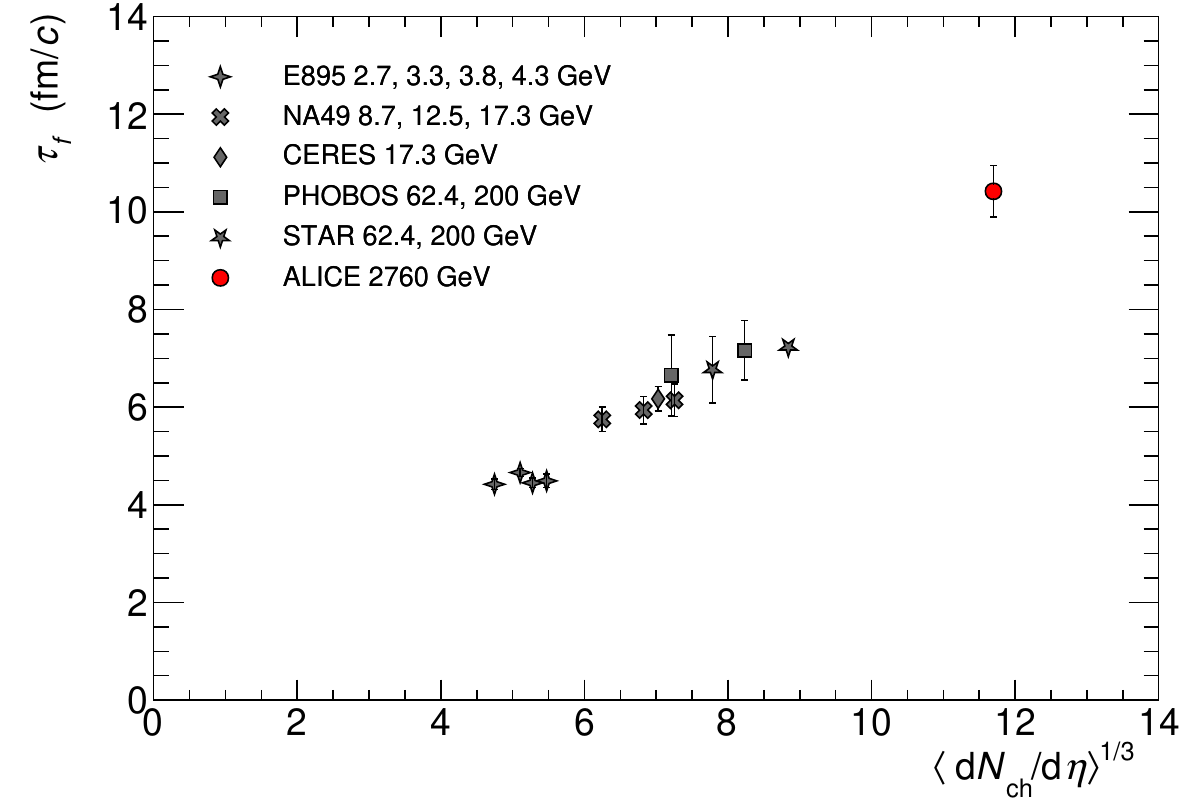}
    \end{center}
    \caption{Homogeneity volume (top) and decoupling time $\tau_f$ (bottom) measured at 2.76~TeV~\cite{Aamodt:2011mr,Adam:2015vna} compared to those obtained at experiments at lower energies. The homogeneity region is determined as the product of the three pion femtoscopic radii at $\langle k_T \rangle~=~0.3$~GeV$c$ for 0--5\% central events.}
    \label{fig:volume_and_time}
\end{figure}

Temperature can be experimentally accessed through the measurement of thermal photons emitted by the hot plasma, as their production rate provides insights into the early conditions of the QGP. Direct photons are those not originating from parton fragmentation or hadronic decays, but rather from electromagnetic interactions during various collision stages.
Figure~\ref{fig:direct-photon-spectra} presents the direct photon spectra measured in Pb-Pb collisions at 2.76 TeV by the ALICE collaboration \cite{Adam:2015lda} and in Au-Au collisions at 0.2 TeV by the PHENIX collaboration \cite{Adare:2008ab, Adler:2005} in 0-20\% and 20-40\% centrality classes. Thermal photons dominate the low transverse momentum region ($p_T \lesssim 3$ GeV/$c$), following an approximately exponential behavior characterized by $d^{2}N_{\gamma_{\mathrm{dir}}}/(p_T\mathrm{d}p_T\mathrm{d}y) \propto e^{-p_T/T_{eff}}$. The inverse logarithmic slope $T_{eff}$ accounts for the radial expansion of the system, causing a blue-shift of emitted photons \cite{vanHees:2011vb}.
At higher momenta ($p_T \gtrsim 5$ GeV/$c$), the direct photon spectrum includes contributions from ``prompt" photons, which originate from initial hard scatterings and the subsequent interaction of hard scattered partons with the medium (referred to as ``jet-photon conversion"). The invariant yield of direct photons is determined by first measuring the excess of direct photons over decay photons, denoted by $R_\gamma$. 
Figure~\ref{fig:direct-photon-spectra} presents $R_\gamma$ for central and semi-central Pb-Pb collisions at  = 2.76 TeV. At high momenta ($p_T \gtrsim 5$ GeV/$c$), the ratio $R_\gamma$ indicates consistency with prompt photon production predicted by perturbative Quantum Chromodynamics (pQCD) and JETPHOX calculations \cite{Paquet:2015lta,Aurenche_2006}. The excess of direct photons observed at low transverse momentum ($0.9 < p_T < 2.1$ GeV/$c$) suggests an abundance of thermal photons originating from the Quark-Gluon Plasma (QGP). This region, dominated by thermal direct photons, is fitted with an exponential function to extract the effective temperature ($T_{eff}$), resulting in values of $T_{eff} = (304 \pm 41)$ MeV and $T_{eff} = (407 \pm 114)$ MeV for central and semi-central Pb–Pb collisions, respectively. Comparing these temperatures to the slopes of the spectra measured by PHENIX, an increase in the effective temperature from RHIC to the LHC is evident. However, directly determining the initial temperature of the fireball is challenging and has not been achieved, as it requires model calculations that consider the evolution of the QGP medium and radial flow effects, as discussed in \cite{ALICEIntro}.

The system size and lifetime are crucial indicators of the fireball properties created in collisions, assessed through particle momentum correlations known as femtoscopy at the final kinetic freeze-out state. Historically referred to as “HBT interferometry” after its originators Hanbury-Brown and Twiss in the 1950s and 1960s in astronomy \cite{HanburyBrown:1954amm, HanburyBrown:1956bqd}, femtoscopy involves constructing relative momentum correlations and fitting them to extract particle interaction and source parameters (see Sec.~\ref{sec.II.2}).
The volume measured by femtoscopy corresponds to the size of the emitting source (the homogeneity volume), which generally differs from the total volume of the system at freeze-out \cite{Akkelin:1995gh}.

The upper portion of Fig.~\ref{fig:volume_and_time} illustrates how the size of the homogeneity region, inferred from femtoscopic pion radii, scales with the charged-particle pseudorapidity density. These measurements were conducted in central $p$-Pb and Au-Au collisions across various energy regimes. Notably, the size of the homogeneity region increases approximately threefold from AGS energies to the LHC~\cite{ALICEIntro}. The decoupling time of the system is commonly approximated using the decoupling time of pions, \( \tau_f \), due to their predominant abundance (approximately 80\%) within the system. The lower panel of Fig.~\ref{fig:volume_and_time} illustrates \( \tau_f \) alongside global data. It shows a linear increase with the cube root of the charged-particle pseudorapidity density, starting from 4-5 fm/\( c \) at AGS energies, rising to 7-8 fm/\( c \) at the highest RHIC energies, and reaching 10-11 fm/\( c \) in central Pb-Pb collisions at \( \sqrt{s_{\mathrm{NN}}} = 2.76 \) TeV.

\subsection{Flow and correlations}  \label{sec.II.2}
The dynamic behavior of the QGP provides crucial insights into the strongly-interacting matter. These properties are primarily defined by measurements of collective motion of final state particles generated in heavy-ion collisions. Anisotropic flow, quantified by Fourier coefficients of the particle azimuthal distributions with respect to symmetry plane angles, $\Psi$, stands as a key observable. A cornerstone in exploring the strongly-coupled QGP paradigm involves extensive measurements of elliptic flow ($v_2$) and triangular flow ($v_3$), the second- and third-order component of anisotropic flow respectively. Characteristic features regarding $v_2$ measurements include mass ordering, particle species (i.e. mesons and baryons) grouping and number of constituent quarks (NCQ) scaling, which can be interpreted as the interplay of the dominant ellipsoidal geometry in the initial state, the collective expansion of the system and the hadronization through quark coalescence. Overall, these measurements suggest the creation of an ``ideal fluid" in relativistic heavy-ion collisions.

\begin{figure}[!htb]
\includegraphics[width=\linewidth]{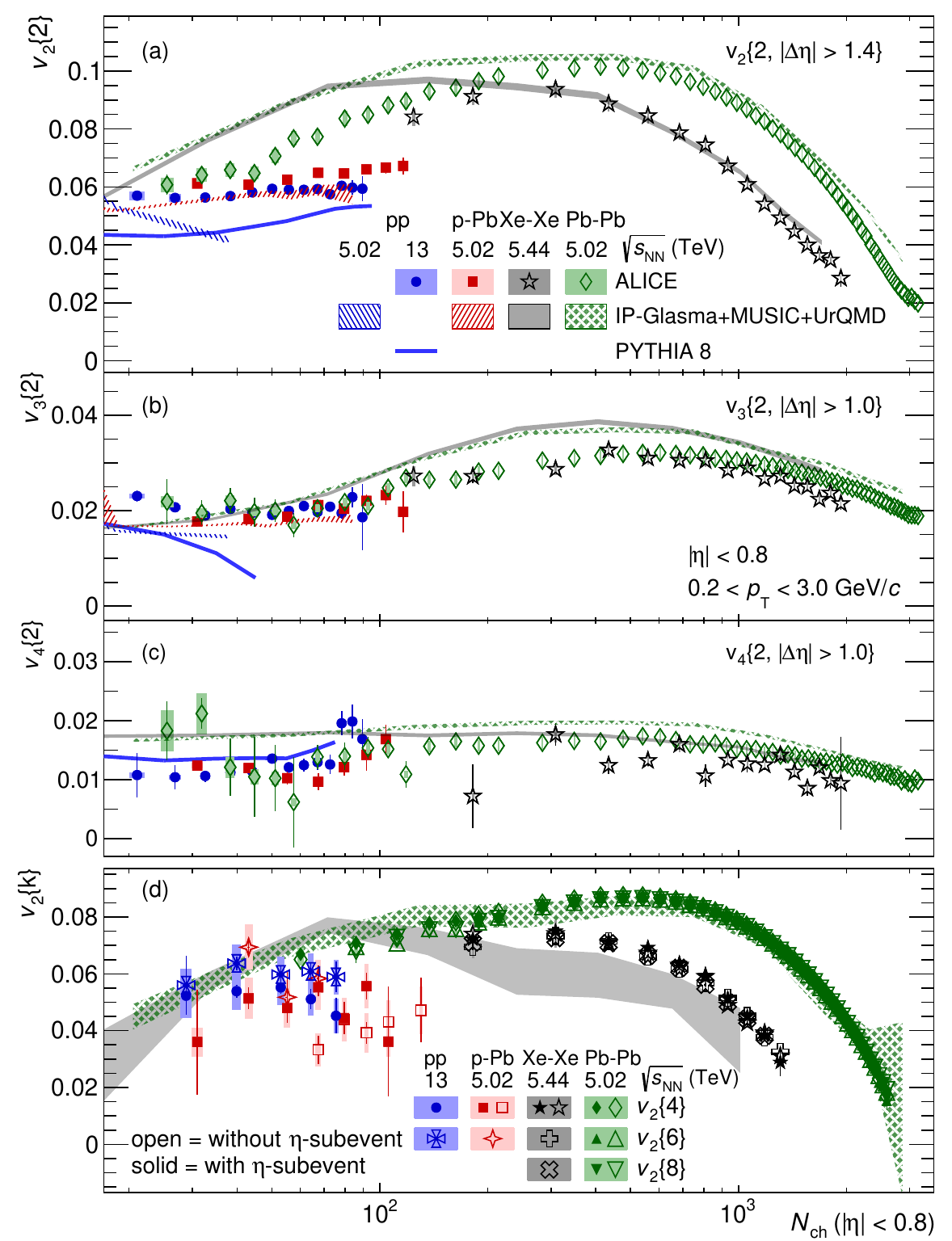}
\caption{Multiplicity dependence of $v_n$ for $pp$, $p$-Pb, Xe-Xe and Pb-Pb collisions.~\cite{bibcf4}}
\label{fig:cf1}
\end{figure}

Beyond measuring $v_2$, studying $v_2$ fluctuations can uncover deeper insights into the collective behavior of the QGP. In~\cite{bibcf1}, elliptic flow of identified hadrons in Pb–Pb collisions at \snn = 5.02 TeV are measured using two- ($v_2\{2\}$) and four- ($v_2\{4\}$) particle cumulants. By combining data for both $v_2\{2\}$ and $v_2\{4\}$, ALICE presents the first measurements of the mean elliptic flow, elliptic flow fluctuations, and relative elliptic flow fluctuations for various hadron species, which probe the event-by-event eccentricity fluctuations in the initial state and the contributions from the dynamic evolution of the expanding QGP. When compared with hydrodynamic calculations incorporating quark coalescence, differences in the relative flow fluctuations for different particle species are observed, suggesting that final state hadronic interactions further modify these fluctuations.

The correlations between event-by-event fluctuations of two different flow amplitudes are commonly quantified using “symmetric cumulant” (SC) observables. Building on previous measurements~\cite{bibcf2}, ALICE has extended event-by-event correlations between three flow amplitudes to higher-order SC observables~\cite{bibcf3}. Different three-harmonic correlations emerge during the collective evolution of the medium and differ from those present in the initial state, which cannot be explained by previous lower-order flow measurements. They provide the first constraints on the nonlinear response contribution in $v_5$ from $v_2$ and $v_3$, enhancing our understanding of the event-by-event flow fluctuation patterns in the QGP.

\begin{figure}[!htb]
\includegraphics[width=\linewidth]{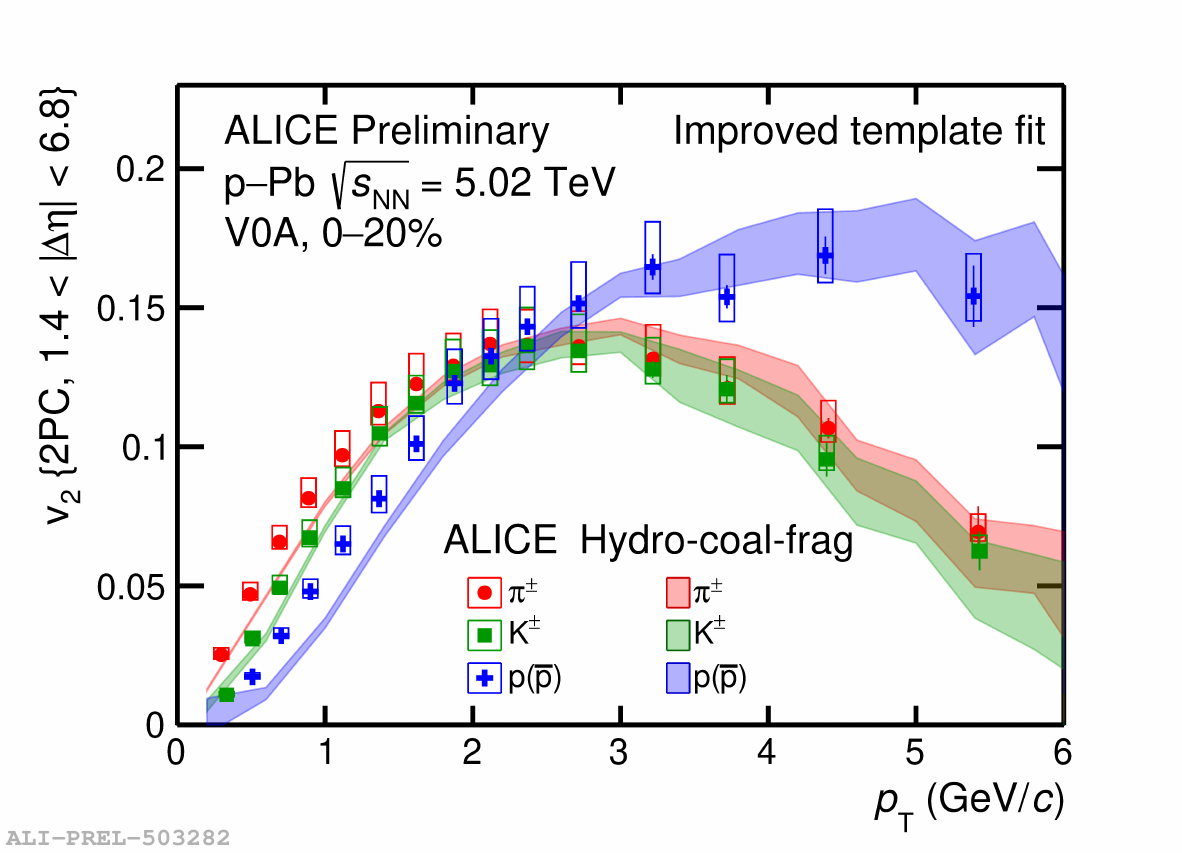}
\caption{$p_{\rm T}$-differential $v_2$ measured with two-particle correlation for various hadrons in p–Pb collisions at \snn = 5.02 TeV is compared with calculations from the Hydro-Coal-Frag model.~\cite{bibcf5}}
\label{fig:cf2}
\end{figure}

\begin{figure}[!htb]
\includegraphics[width=\linewidth]{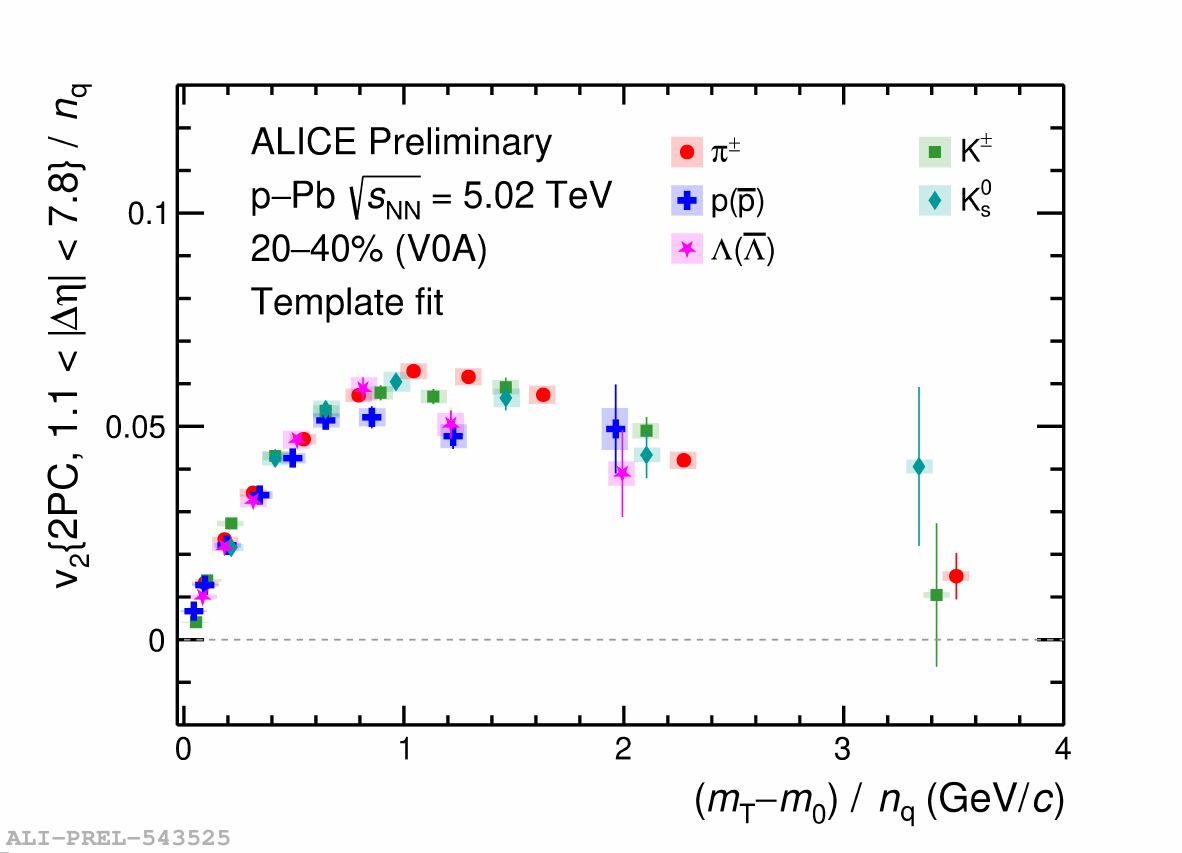}
\caption{The NCQ scaling for various baryons and mesons measured in p–Pb collisions at \snn = 5.02 TeV, indicating the existence of partonic collectivity.~\cite{bibcf6}}
\label{fig:cf3}
\end{figure}

In recent years, striking similarities between numerous observables, including the ``ridge” structure have been observed in A-A and high multiplicity $p$-A and $pp$ collisions at both RHIC and LHC energies, revealing the presence of collectivity in small system collisions. To investigate the “smallest” ($p$-A, $pp$, $ee$...) and “dilutest” (lower multiplicity) limit of collectivity onset, various measurements are carried out in ALICE. In~\cite{bibcf4}, flow coefficients and their cross-correlations using two- and multiparticle cumulants are studied and compared in collisions of $pp$ at \s =13 TeV, $p$-Pb at \snn = 5.02 TeV, Xe-Xe at \snn = 5.44 TeV, and Pb-Pb at \snn = 5.02 TeV, as shown in Fig.~\ref{fig:cf1}. The multiplicity dependence of $v_n$ is measured over a wide range. An ordering of the coefficients $v_2$ $>$ $v_3$ $>$ $v_4$ is observed in $pp$ and $p$-Pb collisions, similar to that seen in large collision systems. However, a weak $v_2$ multiplicity dependence is found compared to A-A collisions within the same range.
In Fig.~\ref{fig:cf2}, the $v_2$ measurements for charged $\pi$, $K$ and $p$ are presented and compared with state-of-the-art Hydro-Coal-Frag calculations incorporating quark-coalescence~\cite{bibcf5}. A characteristic grouping and splitting of $v_2$ for baryons and mesons is observed. The NCQ scaling is further examined and found to approximately hold~\cite{bibcf6}, as shown in Fig.~\ref{fig:cf3}. These results together confirm the existence of partonic collectivity.

In Ref~\cite{bibcf7}, a novel subevent method is employed, revealing that $v_2$ measured with four-particle cumulants aligns well with that from six-particle cumulants in $pp$ and $p$-Pb collisions. The correlation magnitude between $v^2_n$ and $v^2_m$, assessed with the aforementioned SC, is consistently positive across all multiplicities for $v_2$ and $v_4$. Conversely, for $v_2$ and $v_3$, the correlation is negative and changes sign at low multiplicity, indicating a different fluctuation pattern for $v_n$.

\begin{figure}[!htb]
\includegraphics[width=\linewidth]{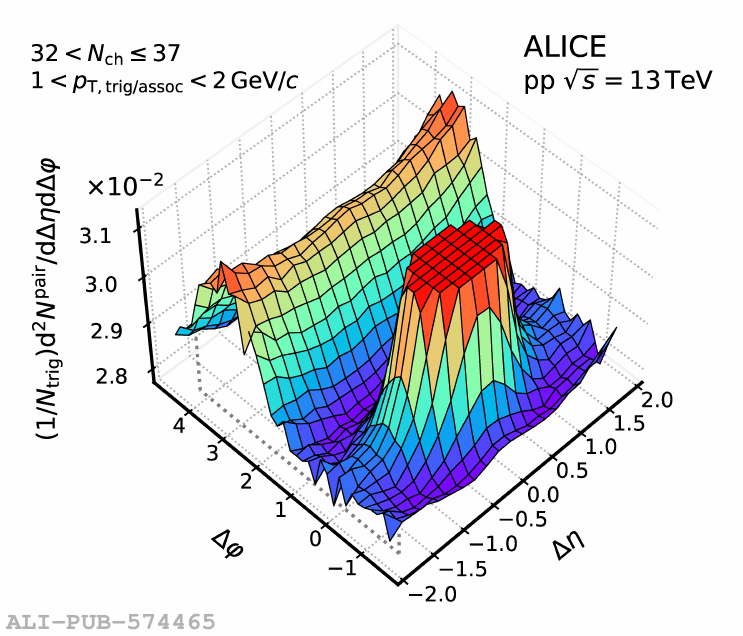}
\caption{Two-particle per-trigger yield measured for charged track pairs within the multiplicity range 32 $<$ $N_{\rm ch}$ $<$ 37. The jet fragmentation peak has been truncated for a better visibility.~\cite{bibcf8}}
\label{fig:cf4}
\end{figure}

\begin{figure}[!htb]
\includegraphics[width=\linewidth]{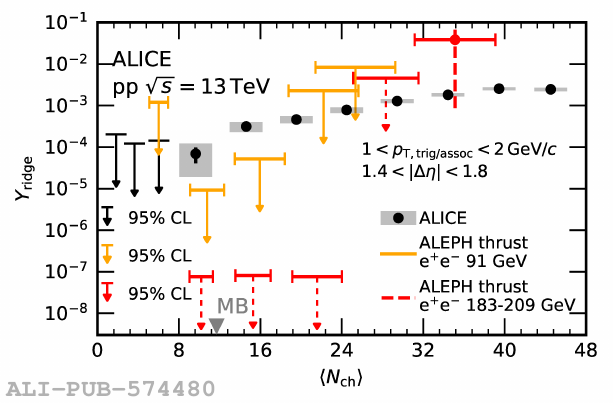}
\caption{Ridge yield as a function of multiplicity, compared to the upper limits on the ridge yield in e$^+$e$^-$ collisions.~\cite{bibcf8}}
\label{fig:cf5}
\end{figure}

The measurement of near-side associated per-trigger yields (ridge yields) from the analysis of angular correlations of charged hadrons is performed in $pp$ collisions at \s =13 TeV~\cite{bibcf8}.
A prominent jet-fragmentation peak, resulting from the correlations of particles originating from the fragmentation of the same parton, is clearly observed. A broad away-side structure emerges from the correlations of tracks from back-to-back jet fragments, spreading across the entire \(\Delta\eta\) region. As illustrated in Fig.~\ref{fig:cf4}, an enhancement of the correlation, known as the "ridge" structure, is visible at \(|\Delta\eta| > 1.4\) and \(\Delta\phi = 0\). This ridge, observed in previous measurements, is interpreted in heavy-ion collisions as indicative of the collective expansion of the QGP medium.
Long-range ridge yields are extracted, extending into the low multiplicity region where a strongly interacting medium is typically considered unlikely to form. The precision of these new low multiplicity results enables the first direct quantitative comparison with results obtained in $e^+e^-$ collisions, as shown in Fig.~\ref{fig:cf5}. In $e^+e^-$ collisions, initial-state effects such as preequilibrium dynamics and collision geometry are not expected to play a role. In the multiplicity range where the $e^+e^-$ results are precise, ridge yields in $pp$ collisions are substantially larger than those observed in $e^+e^-$ annihilations, indicating that the processes involved in $e^+e^-$ annihilations do not significantly contribute to the emergence of long-range correlations in $pp$ collisions.

The study of nuclear structure in heavy-ion collisions has draw extensive attentions in recent years. Pearson correlations between mean transverse momentum $[p_{\rm T}]$ and $v_2$ ($v_3$) are measured as a function of centrality in Pb–Pb (spherical) and Xe–Xe (deformed) collisions at 5.02 and 5.44 TeV respectively, in ALICE~\cite{bibcf90, bibcf9}. 
The positive correlations of \(\rho(v_2^2, [p_{\rm T}])\) and \(\rho(v_3^2, [p_{\rm T}])\), along with the negative higher-order correlation, are well-explained by hydrodynamic models with IP-Glasma initial conditions. This suggests that geometric effects in the initial state play a significant role. The data obtained from Xe-Xe collisions offer a novel avenue for investigating nuclear structure through relativistic heavy-ion collisions at the LHC. Further studies can be seen in~\cite{bibcf91}. 

Heavy-ion collisions generate extremely strong electromagnetic fields, predominantly caused by spectator protons. These fields are estimated to reach magnitudes of \(10^{18}\) to \(10^{19}\) Gauss within the first 0.5 fm of the collision at LHC energy levels.
In Ref.~\cite{bibcf10, zhao_electromagnetic_2024}, such strong electromagnetic fields are probed with charge-dependent directed flow ($v_1$)~\cite{APS2023_072504} at \snn = 5.02 TeV Pb-Pb collisions. The difference in \(v_1\) between positively and negatively charged hadrons exhibits a positive slope with respect to \(\eta\). Measurements for \(D^0\) mesons and their anti-particles reveal a value three orders of magnitude greater than that for charged hadrons. This significant disparity provides new insights into the effects of the strong electromagnetic field, the initial tilt of matter, and the differing sensitivities of charm and light quarks to the early dynamics of heavy-ion collisions.
The interplay between the chiral anomaly and intense magnetic (vortical) fields is suggested to be able to generate several chiral anomalous effects~\cite{NTQCD2023_08, NTQCD2023_11, APS2023_112502, shou_progress_2023}, such as the Chiral Magnetic Effect (CME), the Chiral Magnetic Wave (CMW) and the Chiral Vortical Effect (CVE). Utilizing various azimuthal- and charge-dependent correlators and methods, the fractions and upper limits (at 95\% C.L.) of the CME and the CMW have been experimentally extracted with unprecedented high precision, as listed in Tab.~\ref{tab:cf1}. Notably, the CMW fraction is a pioneering achievement among all experiments and a unified background of the Local Charge Conservation (LCC) entwined with the collective flow is further proposed to simultaneously interpret the CME and CMW measurements~\cite{bibcf15}. The Chiral Vortical Effect has also been experimentally measured using $\Lambda$-$p$ pairs~\cite{bibcf16}.

\begin{table}[h]
\caption{Fractions and upper limits of CME and CMW signals measured at ALICE. \newline}
\begin{tabular}{c || c | c | c}
\hline
 & Dataset & Fraction & Upper limit  \\
\hline
\multirow{3}{*}{CME} & Pb-Pb 2.76 TeV~\cite{bibcf11} & -2.1\%$\pm$4.5\% & 15\%  \\
& Pb-Pb 5.02 TeV~\cite{bibcf11} & 0.3\%$\pm$2.9\% & 18\%  \\
& Pb-Pb 5.02 TeV~\cite{bibcf12} & 15\%$\pm$6\% & 25\% \\
& Xe-Xe 5.44 TeV~\cite{bibcf12} & -0.1\%$\pm$1\% & 2\% \\
\hline
CMW & Pb-Pb 5.02 TeV~\cite{bibcf14} & 8\%$\pm$5\% & 26\% \\
\hline
\end{tabular}
\label{tab:cf1}
\end{table}

\begin{figure}[!htb]
\includegraphics[width=\linewidth]{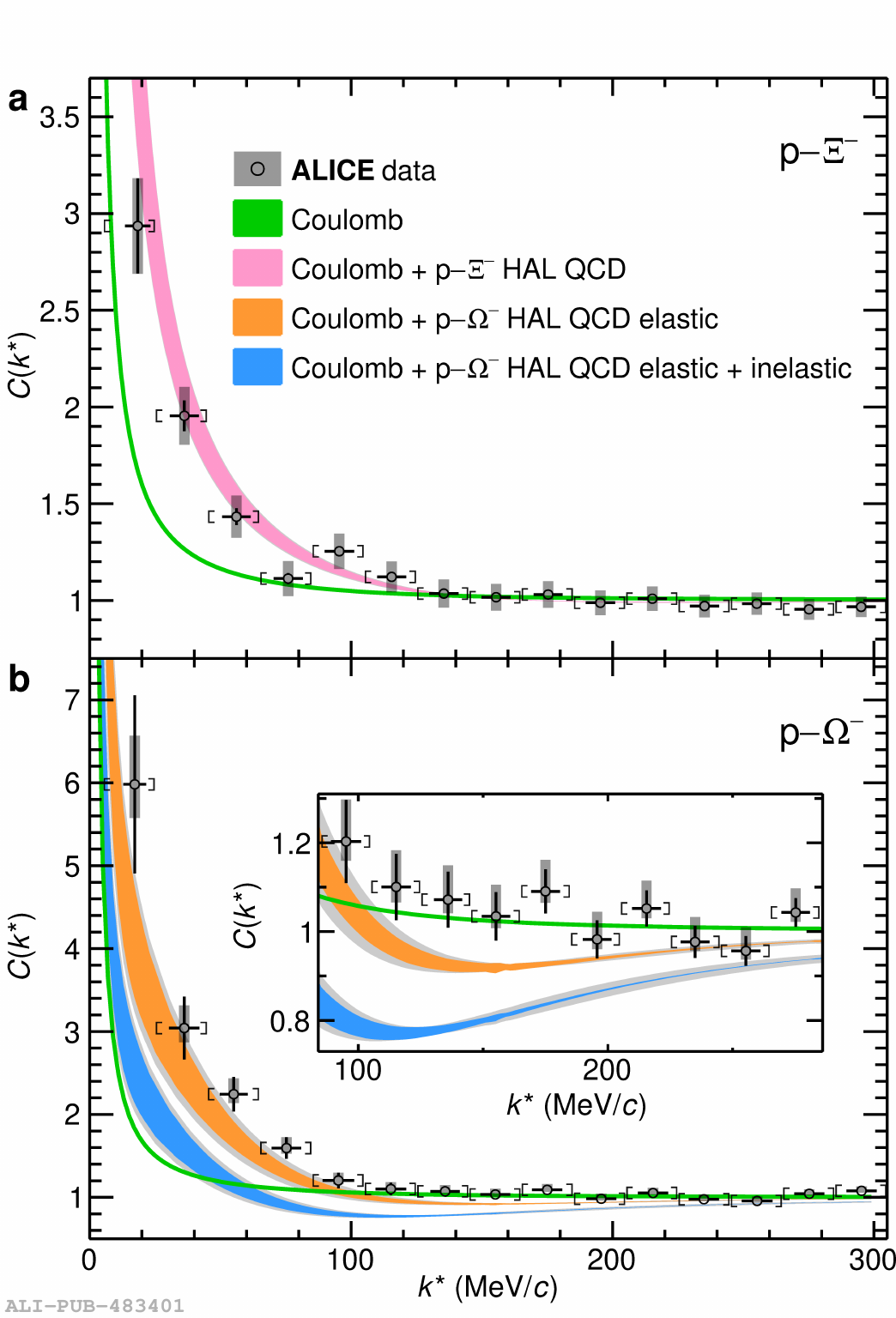}
\caption{Experimental \(p-\Xi^-\) and \(p-\Omega^-\) correlation functions, indicating the presence of an additional strong attractive interaction.~\cite{bibcf18}}
\label{fig:cf6}
\end{figure}

In addition to azimuthal correlations, ALICE serves as a novel laboratory to determine space-time characteristics of relativistic heavy ion collisions and hadron-hadron interactions through momentum correlations~\cite{bibcf17, ma_hypernuclei_2023}. 
In Ref.~\cite{bibcf18}, momentum correlations between hyperon–proton pairs are measured in \(\sqrt{s} = 13\) TeV \(pp\) and \(\sqrt{s_{\rm NN}} = 5.02\) TeV \(p\)-Pb collisions with high precision. The influence of strong interactions is examined by comparing experimental data with lattice calculations. In Fig.~\ref{fig:cf6}, it is observed that the signal for \(p-\Omega^-\) pairs is up to twice as large as that for \(p-\Xi^-\) pairs. The measured correlations exhibit significant enhancement compared to the Coulomb prediction, indicating the presence of an additional strong attractive interaction.
In Ref.~\cite{bibcf20}, $p-\Lambda$ correlations are measured in \(\sqrt{s} = 13\) TeV \(pp\) collisions. The significance of the coupling between $p-\Lambda$ and $N-\Sigma$ is evident from a cusp-like enhancement observed at the corresponding threshold energy. This marks the first direct experimental observation, offering an opportunity to refine theoretical calculations for the coupled $N-\Sigma \leftrightarrow N-\Lambda$ system.
In another study, Ref.~\cite{bibcf21} investigates the $p$-$\Sigma^0$ interaction directly, reconstructing $\Sigma^0$ through the $\Lambda\gamma$ channel. The measured results suggest a shallow strong interaction.
In Ref.~\cite{bibcf22}, $p$-$\phi$ correlations are obtained from \(\sqrt{s} = 13\) TeV \(pp\) collisions. By fitting the data with theoretical calculations, the scattering length and effective range are extracted. The measured results conclusively exclude inelastic processes in the $p$-$\phi$ interaction, contributing valuable experimental data for achieving a self-consistent description of the $N$-$\phi$ interaction.

\begin{figure}[!htb]
\includegraphics[width=\linewidth]{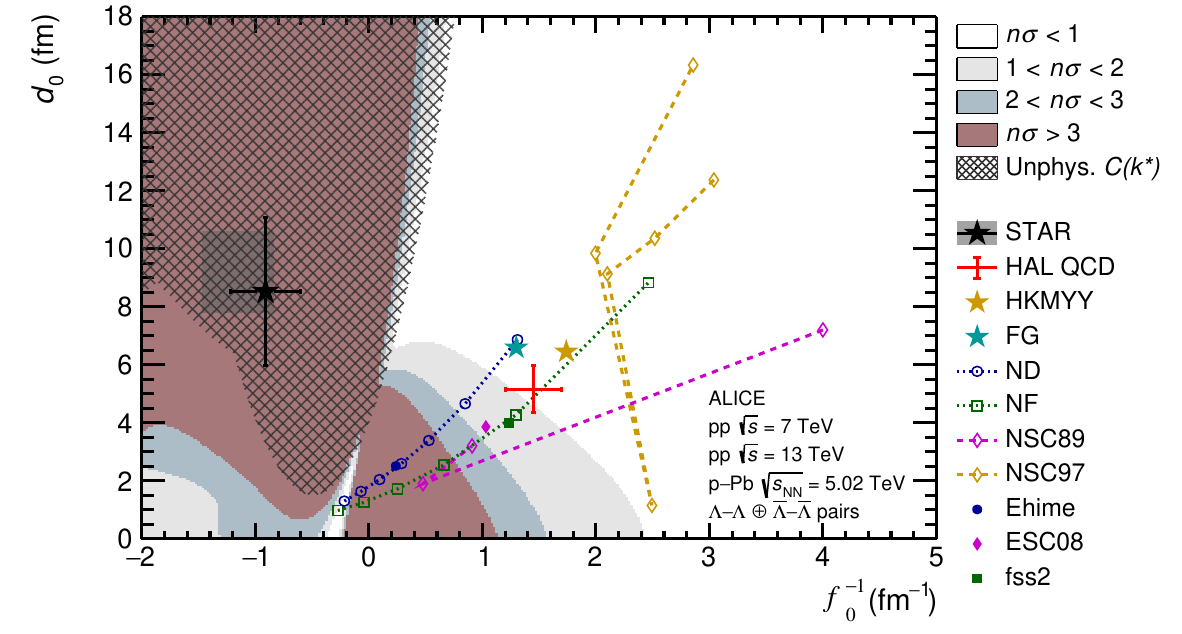}
\caption{Exclusion plot for the $\Lambda$-$\Lambda$ scattering parameters. Different colors represent the confidence level of excluding a set of parameters.~\cite{bibcf24}}
\label{fig:cf7}
\end{figure}

In Ref.~\cite{bibcf23}, Kaon-Proton correlations are measured in \(\sqrt{s} = 13\) TeV \(pp\) collisions, both near and above the kinematic threshold. A significant structure is observed around a relative momentum of 58 MeV/$c$ in the correlation function of opposite-charge \(Kp\) pairs. This observation, with high statistical significance, represents the first experimental evidence for the opening of the neutral \(Kn\) isospin breaking channel, providing the most precise experimental information to date on the \(KN\) interaction.

In the search for the possible $\Lambda$-$\Lambda$ bound state, known as the H-dibaryon, femtoscopic correlations of $\Lambda$-$\Lambda$ pairs are studied in $pp$ and $p$-Pb collisions~\cite{bibcf24}. By comparing measured data with model calculations, the scattering parameter space, characterized by the inverse scattering length and effective range, is constrained in Fig.~\ref{fig:cf7}. The data reveal a shallow attractive interaction, consistent with findings from hypernuclei studies and lattice computations.

Besides strange hadrons, similar measurements have also been extended to charm hadrons. In Ref.~\cite{bibcf25}, momentum correlations of $p$-$D^-$ and their anti-particle pairs are measured in \(\sqrt{s} = 13\) TeV collisions. The data are consistent with either a Coulomb-only interaction hypothesis or a shallow $N$-$D$ strong interaction, contrary to predictions of a repulsive interaction.
In Ref.~\cite{bibcf26}, various $\pi$-$D$ and $K$-$D$ pairs are also studied. For all particle pairs, the data can be adequately described by Coulomb interaction alone, and the extracted scattering lengths are consistent with zero. This suggests a shallow interaction between charm and light-flavor mesons.

Three-body nuclear forces are crucial in understanding the structure of nuclei, hypernuclei, and the dynamics of dense baryonic matter, such as in neutron stars. In Ref.~\cite{bibcf27}, ALICE presents the first direct measurement of three-particle correlations involving $p$-$p$-$p$ and $p$-$p$-$\Lambda$ systems in $pp$ collisions at \(\sqrt{s} = 13\) TeV.
Using the Kubo formalism, three-particle cumulants are extracted from the correlation functions, where the contribution of the three-particle interaction can be isolated by subtracting known two-body interaction terms. A negative cumulant is observed for the $p$-$p$-$p$ system, suggesting the presence of a residual three-body effect. Conversely, the cumulant for the $p$-$p$-$\Lambda$ system is consistent with zero, indicating no significant three-particle correlations in this case.
Studying correlations between a given particle species and deuterons ($d$) is an alternative way to explore the three-body interactions since $d$ is composed of a neutron and a proton. $K^+$-$d$ and $p$-$d$ femtoscopic correlations are measured in Ref.~\cite{bibcf28}. The relative distances at which deuterons and $p$/$K^+$ are produced are approximately 2 fm. Importantly, only a full three-body calculation that considers the internal structure of the deuteron can adequately explain the observed data. These measurements highlight the feasibility of probing three-body correlations at the LHC, providing insights into the complex dynamics involving nuclei and their constituents.

\subsection{Dileptons, quarkonia and electromagnetic probes}  \label{sec.II.3}
Heavy quarkonia, which are bound states of charm-anticharm (charmonium) or bottom-antibottom (bottomonium) pairs, have been extensively studied in various experiments. Quarkonia are essential probes for studying the QGP and its properties. The binding of heavy-quark pairs is affected due to the screening of the QCD force by the high density of free color charges in the QGP, leading to quarkonium dissolution. This concept was proposed in 1986~\cite{Matsui:1986dk}, with initial studies aiming to link quarkonium suppression directly to the temperature of the deconfined phase. The binding energies of quarkonia, ranging from a few \MeV to over 1 \GeV, suggest a ``sequential suppression" with increasing temperature, where strongly bound states persist up to higher dissociation temperatures. In nuclear collisions, varying the QGP temperature by adjusting collision centrality or energy could potentially make quarkonia an ideal thermometer for the medium.

\begin{figure}[!htb]
\begin{center}
  \includegraphics[width = \linewidth]{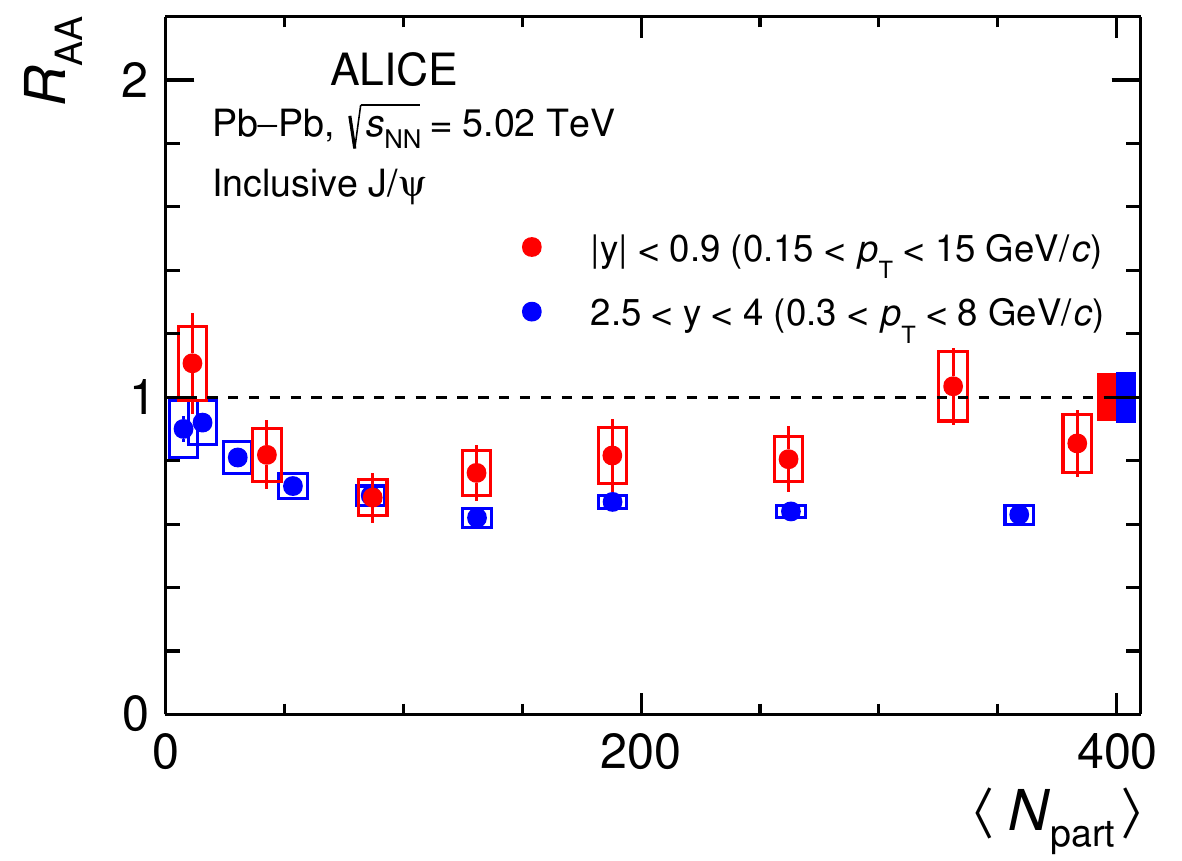}
  \includegraphics[width = \linewidth]{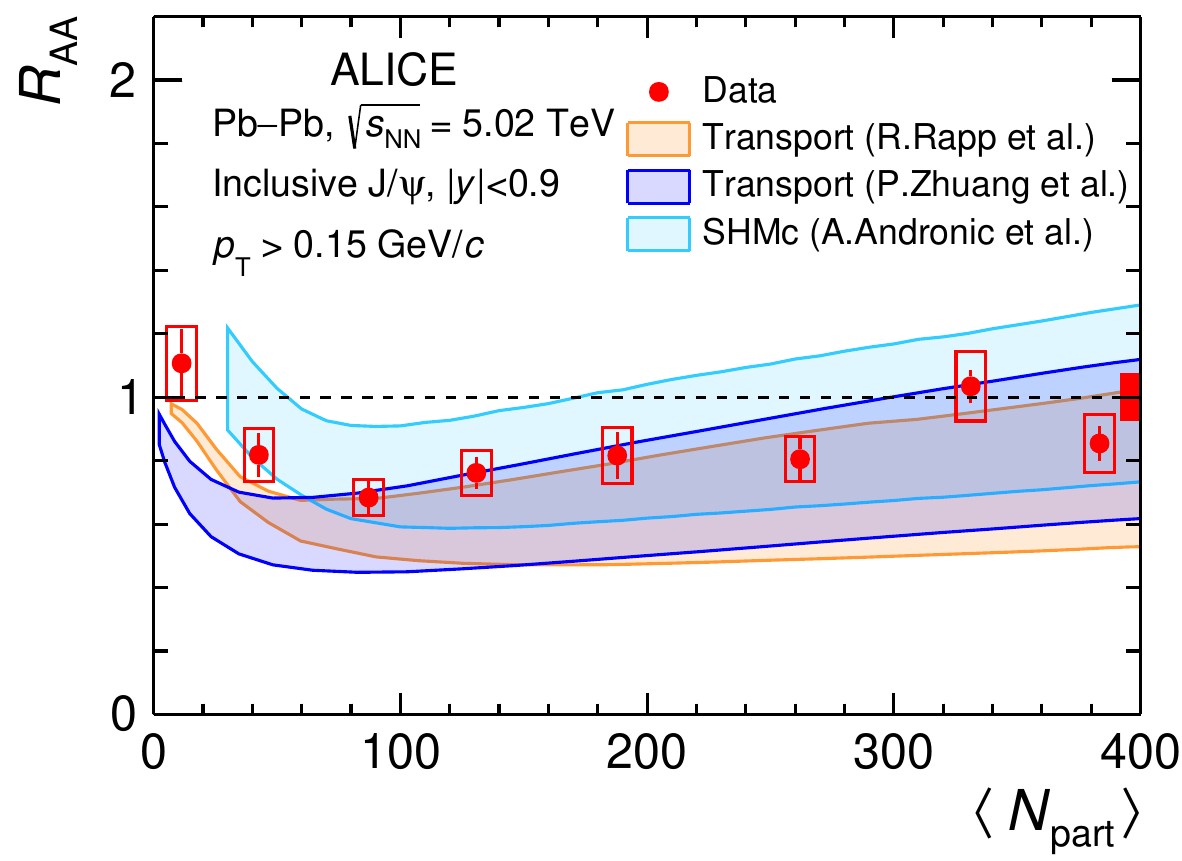}
\caption{\pt-integrated inclusive \jpsi \RAA at midrapidity and forward rapidity, as a function of the number of participants in \PbPb collisions at \snn = 5.02 TeV (upper panel)~\cite{ALICE:2023gco}, the \RAA at midrapidity are compared with the calculations from two microscopic transport models~\cite{Zhou:2014kka,Zhao:2010nk} and the statistical hadronisation model~\cite{Andronic:2019wva}(lower panel).}	  
\label{fig:Raa_vs_cent}
\end{center}
\end{figure}

The nuclear modification factor (\RAA) is calculated using the measured yields from \PbPb and $pp$ collisions at the same center-of-mass energies. Figure~\ref{fig:Raa_vs_cent} presents the \pt-integrated \jpsi \RAA as a function of \Npart in \PbPb collisions at \snn = 5.02 TeV~\cite{ALICE:2016flj,ALICE:2023gco}. \jpsi produced via photo-production processes~\cite{ALICE:2019tqa,ALICE:2021gpt}, particularly in peripheral collisions~\cite{ALICE:2015mzu,ALICE:2022zso}, are excluded by applying \pt thresholds greater than 0.15~\GeVc at midrapidity and greater than 0.3~\GeVc at forward rapidity.

A strong suppression of the \jpsi \RAA is observed in semi-central and central \PbPb collisions, especially at forward rapidity. At midrapidity, \RAA exhibits a slightly increasing trend from semi-central to central collisions, with \RAA values at midrapidity slightly larger than those at forward rapidity, showing a significant difference of 2.2$\sigma$ in the 0--10\% centrality interval. The larger \RAA in central collisions at midrapidity provides strong evidence for (re)generation effects of \jpsi production at LHC energies~\cite{Andronic:2019wva,Zhou:2014kka,Wu:2020zbx}.

\begin{figure}[!htb]
	\begin{center}
		\includegraphics[width = \linewidth]{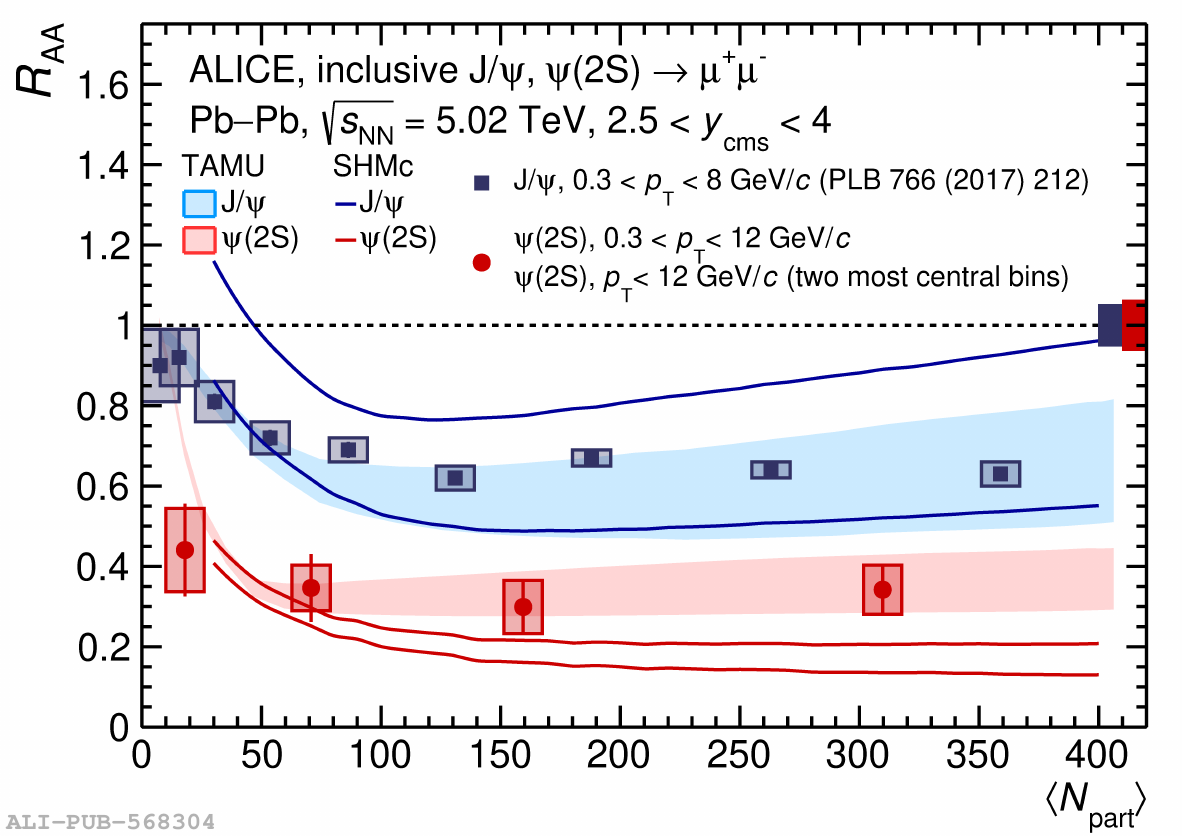}
		\includegraphics[width = \linewidth]{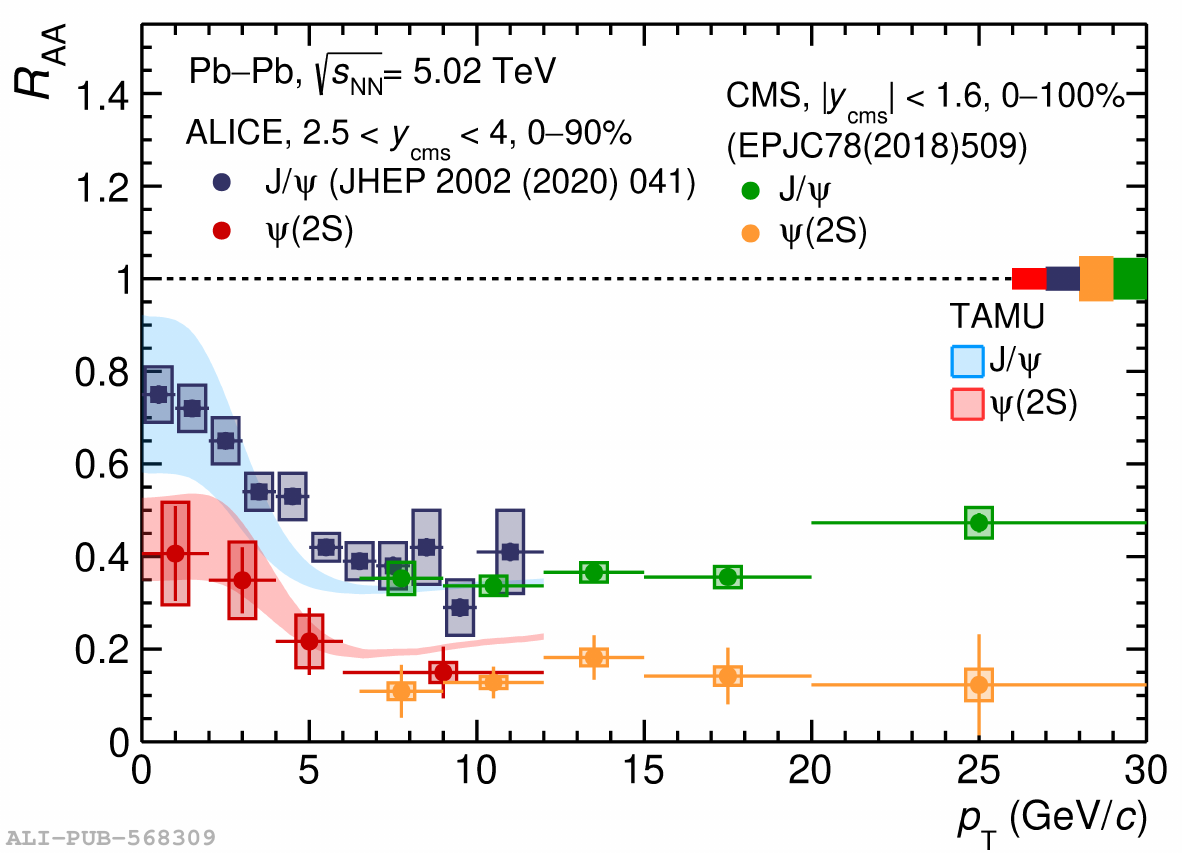}
		\caption{\pt-integrated (upper panel) and \pt-differential (lower panel) \psitwos and \jpsi \RAA at forward rapidity in \PbPb collisions~\cite{ALICE:2022jeh}, the  results are compared with the calculations from  microscopic transport model~\cite{Zhao:2010nk} and the statistical hadronisation model~\cite{Andronic:2019wva}.}	  
		\label{fig:Psi2sRaa}
	\end{center}
\end{figure}

The production of the \psitwos was measured in \PbPb collisions at \snn = 5.02 TeV, \pt down to 0 for the first time at LHC energies, utilizing the dimuon decay channel at forward rapidity (2.5 $<$ y $<$ 4) by the ALICE experiment~\cite{ALICE:2022jeh}. The nuclear modification factors \RAA of \psitwos and \jpsi are compared as a function of \Npart in the upper panel of Fig.~\ref{fig:Psi2sRaa}. The \RAA values for \psitwos are systematically lower than those for \jpsi. The results are compared with calculations from a microscopic transport model~\cite{Zhao:2010nk} and the statistical hadronization model~\cite{Andronic:2019wva}. The agreement between the data and the transport model is slightly better than with the statistical hadronization model, particularly for \psitwos.

The \pt-differential \psitwos and \jpsi \RAA are compared in the lower panel of Fig.~\ref{fig:Psi2sRaa}. The ALICE results are also compared with similar measurements from CMS, which is in the region of $|y|$ $<$ 1.6 and 6.5 $<$ \pt $<$ 30 \GeVc. The CMS data agree with ALICE measurements within uncertainties in the common \pt range. An increasing trend of the \RAA toward low \pt for \psitwos, similar to \jpsi, is observed. This is a hint of (re)generation process of charm and anticharm quarks in the \psitwos production. The strong suppression of the \psitwos \RAA at high \pt agrees within uncertainties with those of ALICE in the common \pt range.

The polarization of quarkonia in high-energy nuclear collisions can be significantly influenced by the unique conditions within the quark-gluon plasma (QGP). The fast-moving charges of the nuclei produce an intense magnetic field that is oriented perpendicular to the reaction plane and decreases rapidly~\cite{Skokov:2009qp}. Additionally, heavy quark pair production occurs early in the collision process, and the subsequent evolution into bound states allows the strong magnetic field to potentially influence charmonium polarization. Another factor affecting quarkonia polarization is the orbital angular momentum of the medium~\cite{Becattini:2007sr}. These conditions suggest that both the magnetic field and the vorticity of the QGP can play crucial roles in altering quarkonium polarization~\cite{Liang:2004ph}.

\begin{figure}[!htb]
	\begin{center}
		\includegraphics[width = \linewidth]{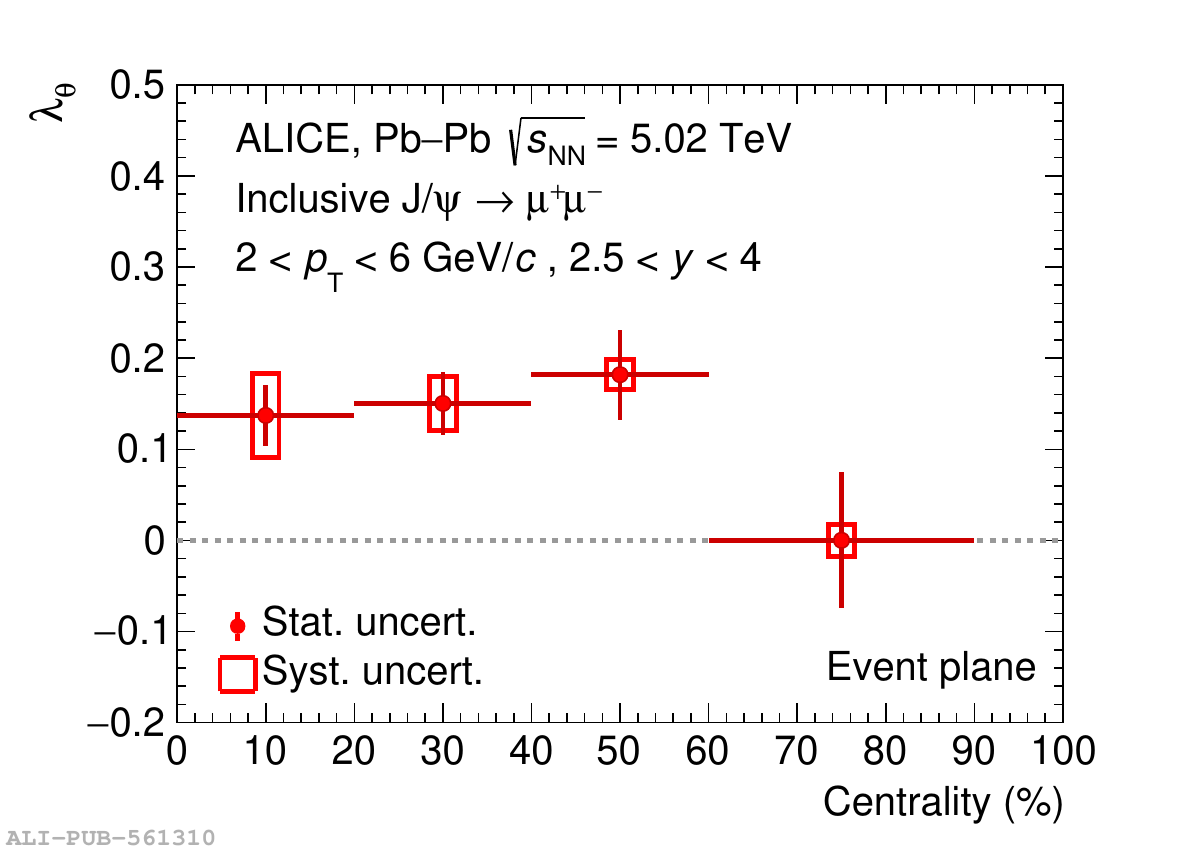}
		\includegraphics[width = \linewidth]{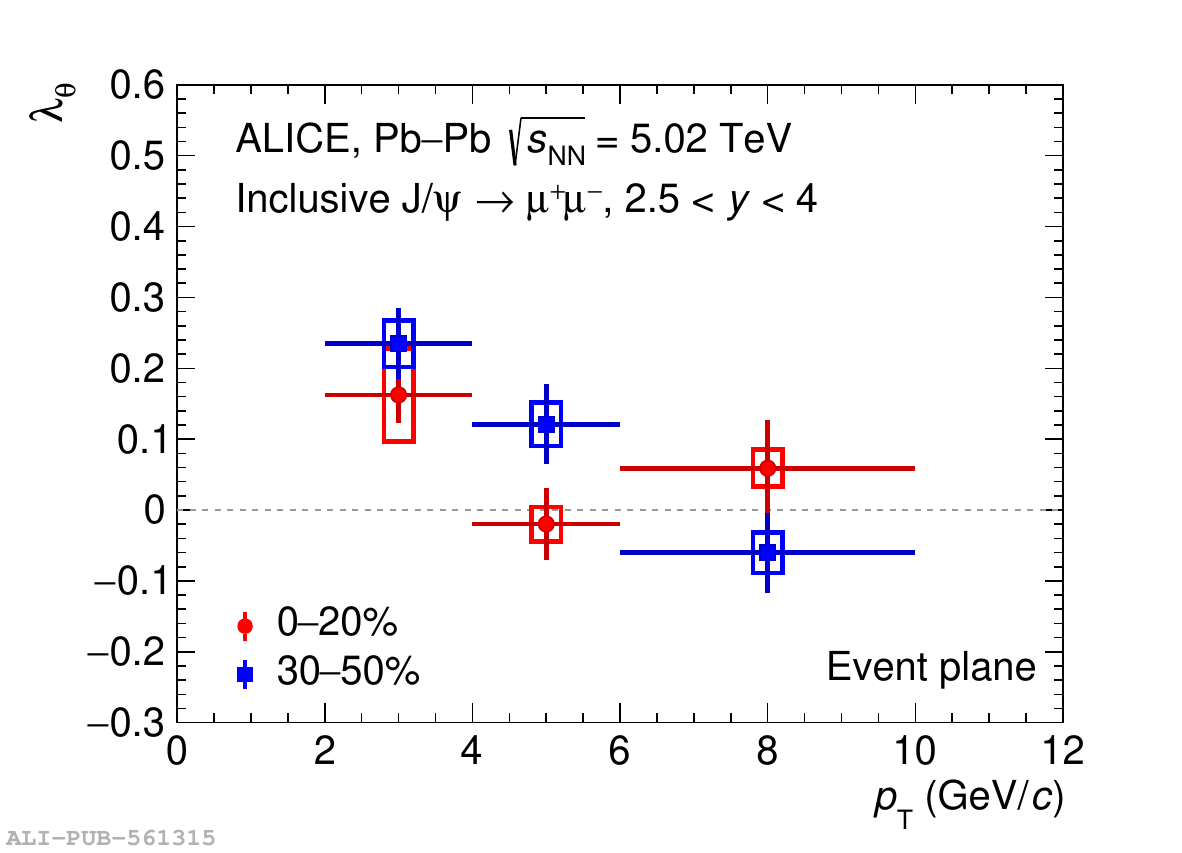}
		\caption{\pt-integrated (upper panel) and \pt-differential (lower panel)  polarization parameters of the inclusive \jpsi  $\lambda_\theta$ with respect to the reaction plane in \PbPb collisions at forward rapidity via the dimuon decay channels~\cite{ALICE:2022dyy}.}
		\label{fig:Jpsi_spin_aliment}
	\end{center}
\end{figure}

The measured polarization parameter $\lambda_\theta$ of the \pt-integrated (2 $<$ \pt $<$ 6 \GeVc) inclusive \jpsi as a function of centrality in \PbPb collisions at forward rapidity (2.5 $<$ y $<$ 4) via the dimuon decay channels is shown in the upper panel of Fig.~\ref{fig:Jpsi_spin_aliment}~\cite{ALICE:2022dyy}. Similar measurements as a function of \pt are shown in the lower panel of Fig.~\ref{fig:Jpsi_spin_aliment}. A significant polarization is found in central and semi-central collisions, with a 3.5$\sigma$ effect observed. The \pt-dependence indicates that the deviation from zero is more pronounced at lower \pt, with the maximum deviation occurring in the (2 $<$ \pt $<$ 4 \GeVc) range during semi-central (30-50\%) collisions, where a 3.9$\sigma$ effect is noted when considering the total uncertainties.

ALICE also measured the spin alignment of \(K^{*0}\) and \(\phi\) mesons in \PbPb collisions at mid-rapidity~\cite{ALICE:2019aid}. The spin matrix element \(\rho_{00}\) for the polarization parameters is found to be less than 1/3 at low \pt in semi-central \PbPb collisions. For \jpsi mesons, the maximum \(\lambda_\theta\) value is approximately 0.2, corresponding to a spin matrix element \(\rho_{00} \approx 0.25\). Thus, the measured spin matrix elements \(\rho_{00}\) for \(K^{*0}\) and \(\phi\) mesons in \PbPb collisions are all less than 1/3. Future theoretical studies on charmonium production, along with experimental improvements and increased integrated luminosity, are essential to confirm and understand the mechanisms underlying the spin alignment of vector mesons in heavy-ion collisions.

As aforementioned, the magnetic field produced by the highly Lorentz-contracted passing ultra-relativistic heavy-ion collisions can be up to $10^{15}$ Tesla, which are anticipated to give rise to various exotic phenomena. The measurements of thermal dileptons production from QGP and the hot hadron gas produced in heavy-ion collisions have long been recognized as a precise and powerful probe for studying the time evolution of the medium's properties~\cite{Turbide:2003si}. The first measurement of low mass \ee pair production at low \pt in \PbPb collisions at \snn = 5.02 TeV is presented~\cite{ALICE:2023jef,ALICE:2022hvk}. 

\begin{figure}[!htb]
	\begin{center}
         \includegraphics[width = \linewidth]{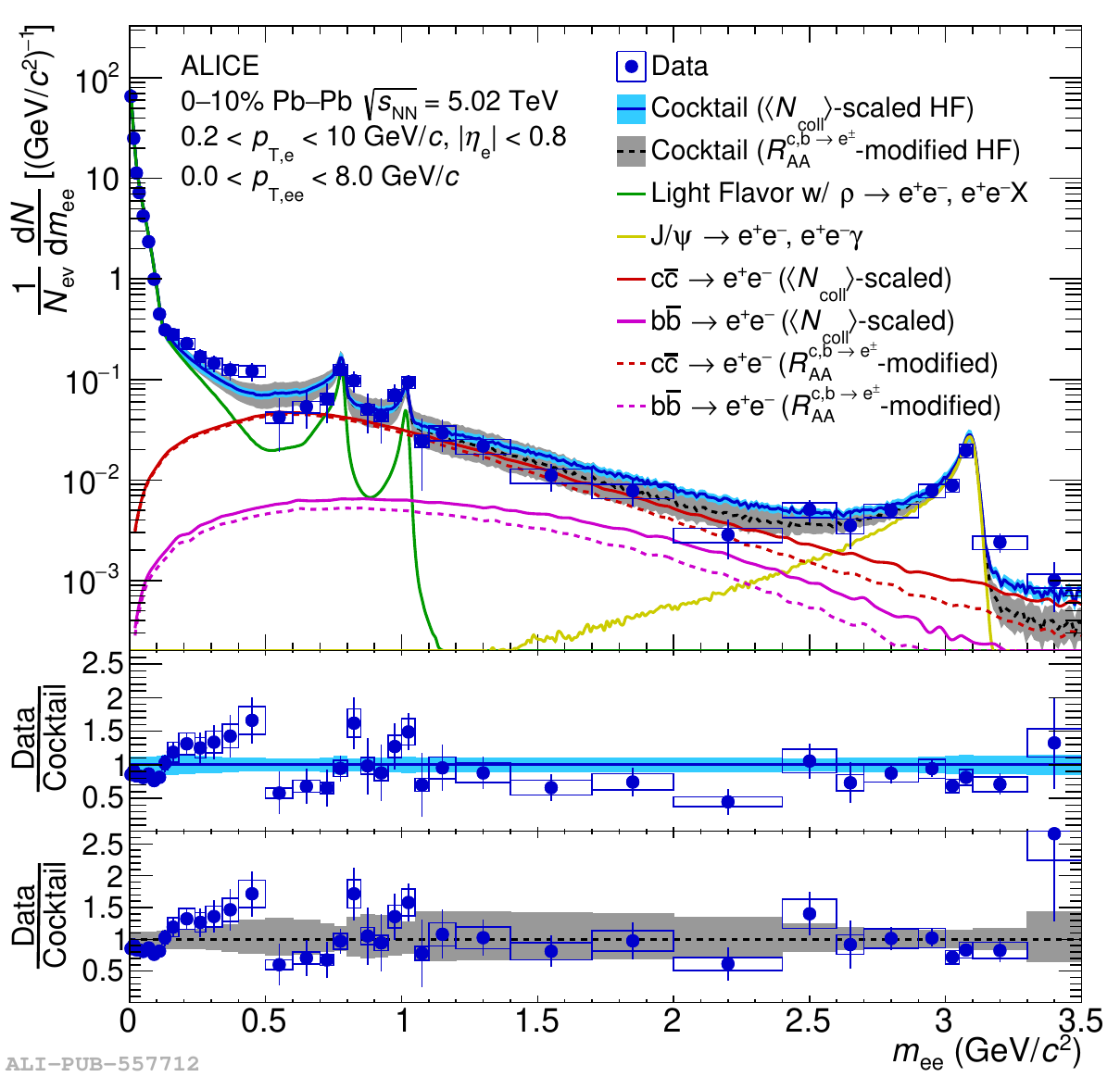}
         \includegraphics[width = \linewidth]{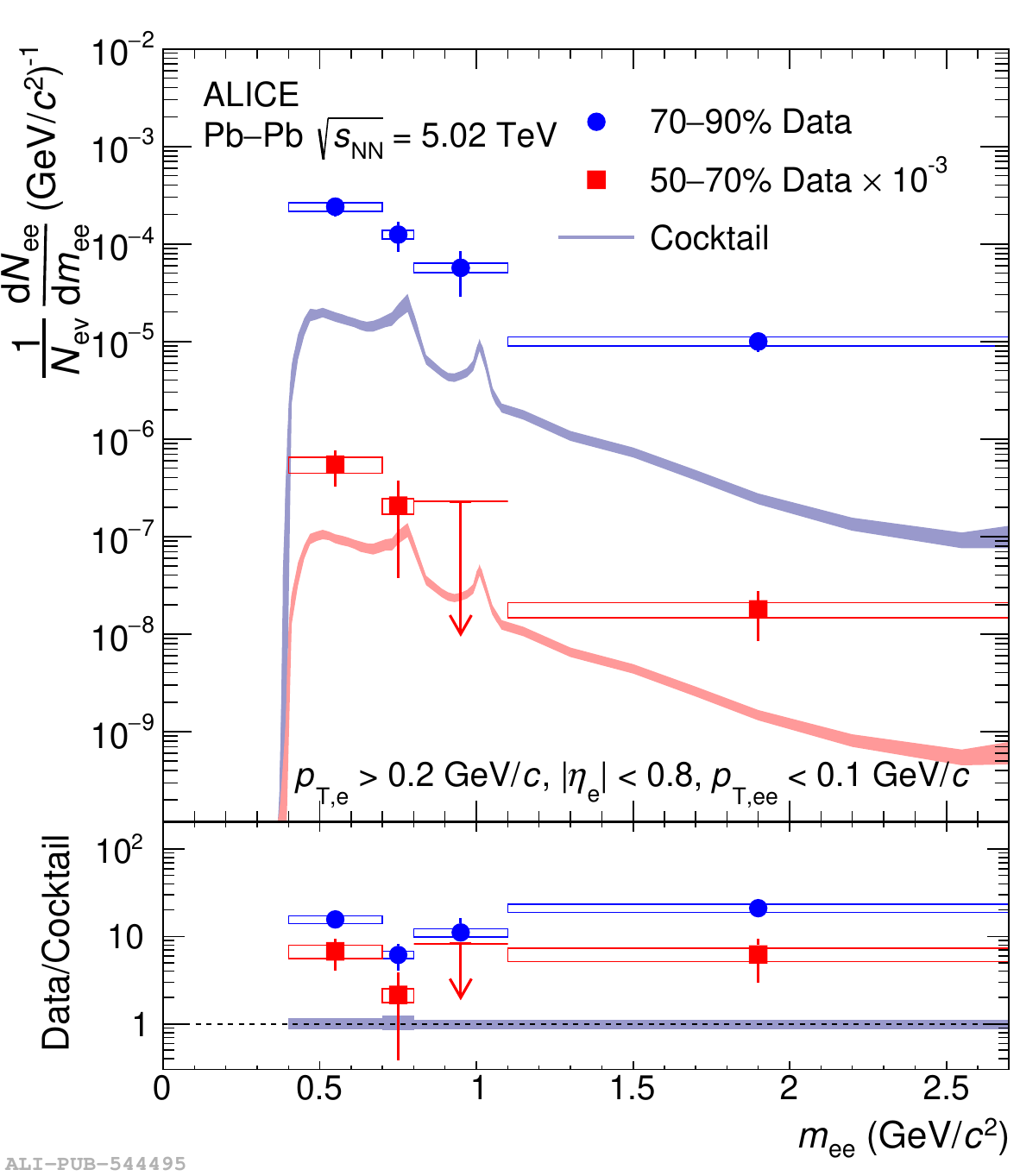}   
	\caption{ Dielectron \mee-differential yields comparison with the the expected \ee contributions from the hadronic productions decays, in \PbPb collision \snn = 5.02 TeV at 0--10\%  centrality interval (upper panels), similar comparisons in 50--70\% and 70--90\% centrality classes (lower panels)~\cite{ALICE:2023jef,ALICE:2022hvk}.}
		\label{fig:lee}
	\end{center}
\end{figure}

The invariant mass spectra of \ee pairs production are shown in Fig.~\ref{fig:lee}. 
The yield of \ee pair production in the range \( 0.2 < p_{\rm T} < 10 \, \text{GeV}/c \) is shown as a function of \( m_{\mathrm{ee}} \) for 0-10\% central \PbPb collisions at \snn = 5.02 TeV (upper panel). The measurements are compared with expected contributions from known hadronic sources, specifically the \( R_{\mathrm{AA}}^{\mathrm{c}, \mathrm{b} \rightarrow \mathrm{e}^{\pm}} \)-modified heavy-flavor (HF) cocktail and the \( N_{\text{coll}} \)-scaled HF cocktail.
At low invariant masses (\( 0.18 < m_{\mathrm{ee}} < 0.5 \, \mathrm{GeV}/c^2 \)), the ratios indicate a potential excess that is consistent with unity within certain confidence levels. Notably, this excess does not significantly depend on the method used for estimating heavy flavor. The contribution from \(\rho\) mesons, excluding medium effects, accounts for approximately 18\% of the total hadronic yield; however, thermally produced \( \mathrm{e}^{+} \mathrm{e}^{-} \) pairs from \(\rho\) mesons are expected to be significant. 
In the intermediate mass range (\( 1.2 < m_{\mathrm{ee}} < 2.6 \, \mathrm{GeV}/c^2 \)), the data are better described by the modified heavy-flavor cocktail, although this region still faces considerable uncertainties.
The lower panel presents the non-central (70-90\%) and semi-central (50-70\%) \PbPb collisions at \snn = 5.02 TeV, measured within the ALICE acceptance at mid-rapidity and for \( p_{\rm T} > 0.2 \, \text{GeV}/c \). The yields of the produced \ee pairs are compared to the expected \ee production from hadronic sources, represented as "cocktails." A clear excess is observed relative to the hadronic cocktail in both centrality classes, with enhancement factors being larger in the (70-90\%) collisions compared to the (50-70\%) collisions. The hadronic cocktail contribution is subtracted from the inclusive \ee pairs. Contributions from thermal dielectrons, arising from both the partonic and hadronic phases, are estimated using an expanding thermal fireball model that incorporates an in-medium broadened \(\rho\) spectral function \cite{Rapp:1999us,Rapp:2013nxa}. It is anticipated that thermal radiation from the medium will be at least an order of magnitude smaller than the measured \( \mathrm{e}^{+} \mathrm{e}^{-} \) pairs.

\subsection{Heavy flavour probes}  \label{sec.II.4}
In ultra-relativistic heavy-ion collisions, hard scatterings involving parton constituents of nucleons can produce a range of energetic final states, collectively referred to as ``hard probes".
Hard probes are energetic partons created in the high momentum-transferred (high-$Q^{2}$) partonic scattering processes at the early stage of heavy-ion collisions.
Since their formation time is less than the QGP medium, they suffer the entire medium evolution.
Due to the high-momentum and short-wavelength properties of the hard probes, their interactions with medium constituents are sensitive to the microscopic structure and quasi-particle nature of the medium and provide a tomography of the medium in a wide virtuality (wavelength) range.
The high-$Q^{2}$ scale also results in the production cross-sections of the hard probes being calculable with controlled and improvable accuracy using the pQCD tools.
Furthermore, the interactions between hard probes and the QGP medium can be calculated starting from the pQCD formulation of elementary collisions or transport theory, giving a firm conceptual basis to such modeling approaches.
Heavy-flavor quarks, jets and photons are all hard probes studied by ALICE during Runs 1 and 2 data taking.

For heavy quarks, i.e., charm and beauty quarks, the high-$Q^{2}$ scale is imported by their heavy masses ($m_{\rm charm}\simeq 1.5$~GeV/$c^2$ and $m_{\rm beauty}\simeq 4.5$~GeV/$c^2$)~\cite{ParticleDataGroup:2022pth}, which are much larger than both $\Lambda_{\rm QCD}$ and the medium temperature~\cite{ALICE:2015xmh}.
Their production, even at low transverse momentum \pt, is governed predominantly by early-stage hard partonic scatterings.
The contribution from the medium evolution and additional thermal production is negligible.
Thus, heavy quarks service also as self-normalized probes of the QCD medium.

When propagating through the QGP medium, heavy quarks interact with the medium constituents via both elastic (collisional) and inelastic (induced radiative) processes at low and high \pt, respectively, resulting in their in-medium energy loss~\cite{Dong:2019byy,Prino:2016cni}.
Experimentally, the energy loss is explored by measuring the nuclear modification factor \RAA.
Assuming the QGP is not formed in $pp$ collisions, $\RAA = 1$ if AA collisions are considered as a superposition of nucleon--nucleon collisions. Moreover, if a heavy quark deposits most of its energy in the QGP medium, it may participate in the medium hydrodynamic expansion and approach thermalization.

ALICE experiment measures the fully reconstructed open heavy-flavour hadrons from their hadronic decays at midrapidity ($\abs{y} < 0.5$)~\cite{ALICE:2012inj,ALICE:2012ab,ALICE:2012gkr,ALICE:2014xjz,ALICE:2017thy,ALICE:2017dja,ALICE:2017pbx,ALICE:2018lyv,ALICE:2018hbc}.
In addition, the open heavy-flavour production is measured using semi-electronic~\cite{ALICE:2012mzy,ALICE:2015zhm,ALICE:2019nuy,ALICE:2020hdw} and semi-muonic decays~\cite{ALICE:2012sxy,ALICE:2015xyt,ALICE:2017fsl,ALICE:2020sjb,ALICE:2020try} at midrapidity ($\abs{y}<$ 0.8 for low- and intermediate-$\pt$ and $\abs{y} < 0.6$ for high-$\pt$) and forward rapidity ($2.5 < y < 4$), respectively.
The semi-electronic decays are also adopted for the partial reconstruction of charmed baryons~\cite{ALICE:2017dja,ALICE:2021psx,ALICE:2021bli}.
Furthermore, beauty production is measured via non-prompt $\jpsi$ from the decay mode ${\rm B}\to\jpsi+X$~\cite{ALICE:2012vpz,ALICE:2015nvt,ALICE:2018szk} and non-prompt D mesons from the decay mode ${\rm B}\to{\rm D}+X$~\cite{ALICE:2021mgk,ALICE:2022tji}.

\begin{figure*}[!htbp]
\centering
\includegraphics[width=\linewidth]{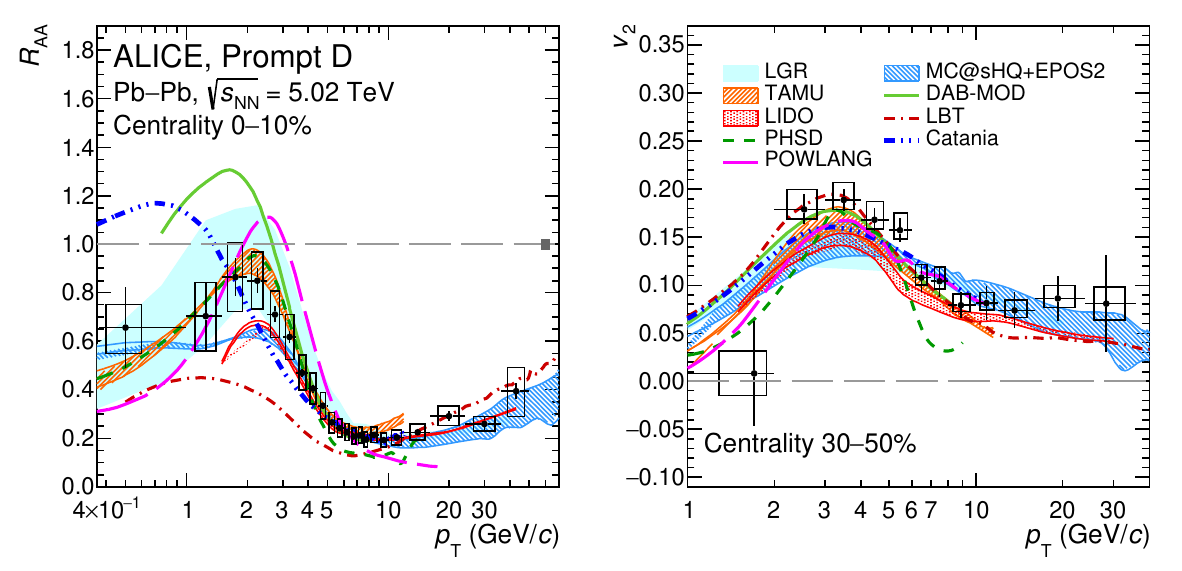}
\caption{$\pt$-differential $\RAA$ (left) and $v_{2}$ (right) of prompt D mesons measured, respectively, in \cent{0}{10} and \cent{30}{50} centrality classes at midrapidity in Pb--Pb collisions at \fivenn by ALICE~\cite{ALICE:2020iug,ALICE:2021rxa}.}
\label{fig:PromptDRAAv2}
\end{figure*}

The left and right panels of Fig.~\ref{fig:PromptDRAAv2} show, respectively, $\RAA$ and the second-order anisotropy flow (elliptic flow) coefficient $v_{2}$ as a function of $\pt$ for prompt D mesons measured at midrapidity in central (\cent{0}{10}) and semi-central (\cent{30}{50}) Pb--Pb collisions at \fivenn with ALICE~\cite{ALICE:2020iug,ALICE:2021rxa}.
A significant suppression, up to a factor of five, on the yields of D mesons is observed in the intermediate- and high-$\pt$ regions in the most $10\%$ central Pb--Pb collisions, illustrating charm quarks undergo strong in-medium energy loss in the QGP.
Consequently, a positive $v_{2}$ of D mesons is observed at intermediate $\pt$ in the \cent{30}{50} semi-central collisions, indicating charm quarks participate in the collective motion of the QGP resulting from their large energy deposition in the medium.
Thus, at low- and intermediate-$\pt$, i.e., $\pt\lesssim 10$~\GeVc, charm quarks exchange energy and momentum via multiple soft and incoherent collisions within the hydrodynamically expanding medium.
The interaction of charm quarks with the medium is treated in a diffusion approach based on Fokker-Planck or Langevin dynamics.
The behaviour can be described in terms of Brownian motion in the QGP medium.
Hence, the coupling between the medium and charm quarks can be expressed by the spatial diffusion coefficient $D_{\rm s}$, which is almost independent of the quark mass and encodes the transport properties of the medium~\cite{Prino:2016cni,Rapp:2009my,Svetitsky:1987gq}.
Furthermore, $D_{\rm s}$ is proportional to the relaxation time $\tau_{\rm Q}$ of heavy particles, i.e.,
$\tau_{\rm Q} = (m_{\rm Q}/T)D_{\rm s}$,
where $m_{\rm Q}$ and $T$ denotes the heavy quark mass and medium temperate, respectively.

In Fig.~\ref{fig:PromptDRAAv2}, the measurements are also compared with model calculations based on charm-quark transport in a hydrodynamically expanding QGP.
All models are qualitatively in agreement with the data, including the triangle flow coefficient $v_{3}$ of D mesons~\cite{ALICE:2020iug}.
Although tension is observed in particular at low-$\pt$, models that agree with data at the level $\chi^{2}/{\rm ndf} < 2$ yield a value $1.5 < 2\pi D_{\rm s}T < 4.5$ with temperature $T$ at the critical temperature of the QCD deconfinement phase transition $T_{\rm c}\simeq 155$~\MeV~\cite{HotQCD:2014kol,Borsanyi:2013bia}.
By adopting $m_{\rm charm} = 1.5$~GeV/$c^2$, one obtains the corresponding charm quark relaxation time $\tau_{\rm charm}$ in the range $3 < \tau_{\rm charm} < 9$~fm/$c$.
These values are similar in magnitude to the estimated lifetime of the QGP $\tau_{\rm QGP}\simeq 10$~fm/$c$ at LHC energies~\cite{ALICE:2011dyt}, indicating that charm quarks may thermalized completely in the QGP medium created at LHC energies.

It is worth noticing that $\RAA$~\cite{ALICE:2012sxy,ALICE:2020sjb} and $v_{2}$~\cite{ALICE:2015xyt} of open heavy-flavour decay muons measured at forward rapidity agree with that of prompt D mesons at mid-rapidity within uncertainties at low- and intermediate-$\pt$, where the muons are dominated by the charm hadron decays, indicating charm quarks are undergone strong interaction and formalized in the QGP medium in a wide rapidity region.
The complementary measurements at forward rapidity bring significant constraints on the modeling of the longitudinal dependence of the QGP transport properties~\cite{Prado:2019ste}.
In addition, the measurements of $\RAA$~\cite{ALICE:2020sjb} of open heavy flavours at forward rapidity have much higher precision than mid-rapidity at high $\pt$, providing further constraints on the medium-induced gluon radiation behaviours of beauty quarks.

\begin{figure*}[!htbp]
\centering
\includegraphics[width=.52\textwidth]{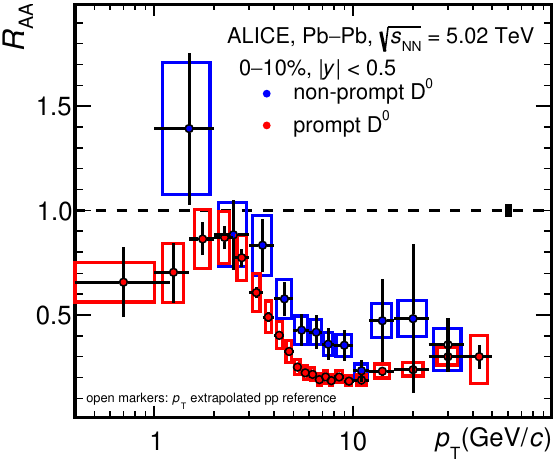}
\includegraphics[width=.47\textwidth]{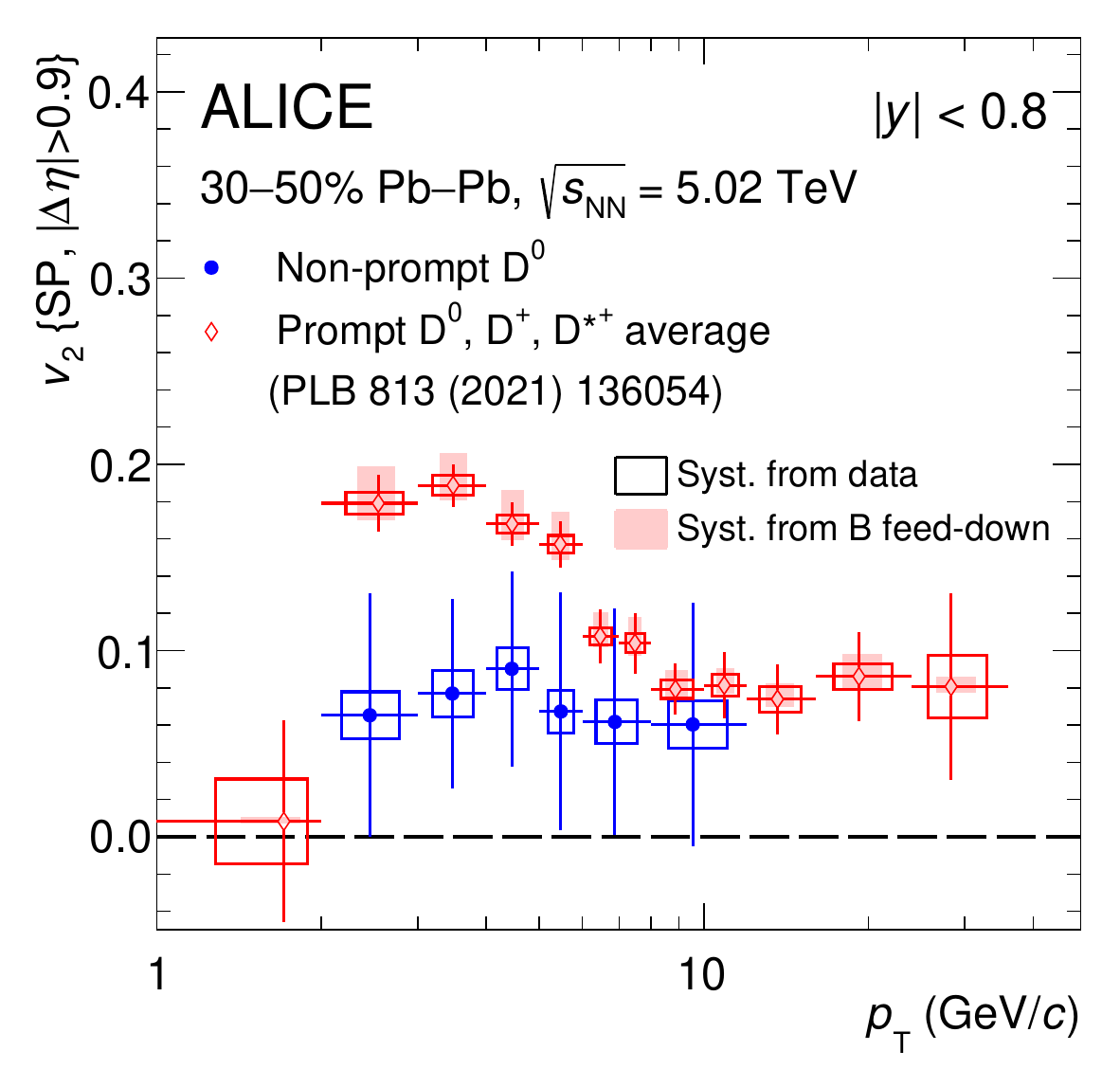}
\caption{$\pt$-differential $\RAA$ (left) and $v_{2}$ (right) of non-prompt $\dzero$ mesons measured, respectively, in \cent{0}{10} and \cent{30}{50} centrality classes at midrapidity in Pb--Pb collisions at \fivenn by ALICE~\cite{ALICE:2022tji,ALICE:2023gjj}. Results are compared with the corresponding measurement of prompt D mesons~\cite{ALICE:2020iug,ALICE:2021rxa}.}
\label{fig:npDRAAv2}
\end{figure*}

To further explore the beauty-quark transport properties in the QGP medium, the non-prompt D mesons are measured at midrapidity by ALICE.
The left and right panels of Fig.~\ref{fig:npDRAAv2} show, respectively, the measurements of $\pt$-differential $\RAA$ and  $v_{2}$ of non-prompt $\dzero$ mesons in central (\cent{0}{10}) and semi-central (\cent{30}{50}) Pb--Pb collisions at \fivenn with ALICE~\cite{ALICE:2022tji,ALICE:2023gjj}.
The results are compared with the corresponding measurement of prompt D mesons~\cite{ALICE:2020iug,ALICE:2021rxa}.
Similar to the prompt D mesons, a significant suppression is observed in the yield of non-prompt $\dzero$ mesons at intermediate and high $\pt$ in the most $10\%$ central Pb--Pb collisions, referring beauty quarks also undergo large in-medium energy loss in the QGP.
Compared to the prompt D mesons, an enhancement on the non-prompt $\dzero$-meson $\RAA$ illustrates the mass-dependent in-medium energy of heavy quarks resulting from the dead-cone effect~\cite{Djordjevic:2014tka}, which leads to less in-medium energy loss for beauty quarks with respect to charm quarks.

Being consistent with the large yield suppression, a positive $v_{2}$, with a significance of $2.7$ standard deviations ($\sigma$), is observed for non-prompt $\dzero$ mesons at intermediate $\pt$ in \cent{30}{50} semi-central collisions.
The measured $v_{2}$ of non-prompt $\dzero$ is lower than that of prompt D mesons with $3.2\sigma$ significance in $2 < \pt < 8$~\GeVc.
Considering the spatial diffusion coefficient $D_{\rm s}$ is independent of heavy-quark masses and $m_{\rm beauty}$ is around 3 times larger than $m_{\rm charm}$, the beauty-quark relaxation time $\tau_{\rm beauty}$ is close or even larger than the $\tau_{\rm QGP}$ at the LHC, indicating beauty quarks are less thermalized in the QGP medium than charm quarks and resulting in the observed smaller $v_{2}$ of non-prompt $\dzero$ mesons than prompt D mesons at intermediate $\pt$.
The measured $v_{2}$ of non-prompt $\dzero$ mesons and prompt D mesons seems to be consistent at high $\pt$, reflecting the interplay between path-length dependent heavy-quark in-medium energy loss and density evolution and density fluctuations of the QGP medium.

The degree of thermalization of charm and beauty quarks that emerges from interacting with the QGP medium not only leads them to participate in the collective motion of the medium but also suggests their hadronization via recombination with the quarks and di-quark pairs in the medium.
In particular, recombination is expected to affect the $\pt$ distributions and the abundances of different heavy-flavour hadron species in AA collisions compared to those in $pp$ collisions~\cite{He:2012df}.
If heavy quarks hadronize via recombination, the yield of charm and beauty hadrons with strange-quark content (e.g., $\ds$ and $\bs$ mesons)  relative to non-strange hadrons is expected to be larger in AA collisions compared to $pp$ collisions, because of the strange-quark production yield enhancement in the QGP medium~\cite{Kuznetsova:2006bh,Andronic:2007zu}, and the production of baryons relative to that of mesons is expected to be enhanced at intermediate-$\pt$, i.e., $2\lesssim \pt \lesssim 8$~\GeVc~\cite{Lee:2007wr,Oh:2009zj,Das:2016llg,Plumari:2017ntm,He:2019vgs,Beraudo:2022dpz}.
In addition, the collective radial expansion of the QGP medium, which determines a flow-velocity profile common to the thermalized particles, could also increase the baryon-to-meson yield ratio at intermediate $\pt$~\cite{Plumari:2017ntm,He:2019vgs,Beraudo:2022dpz}.
A precise description of the hadronisation process in the hot nuclear matter is crucial to understand the transport properties of the QGP~\cite{Rapp:2018qla}.

\begin{figure*}[!htbp]
\centering
\includegraphics[width=.54\textwidth]{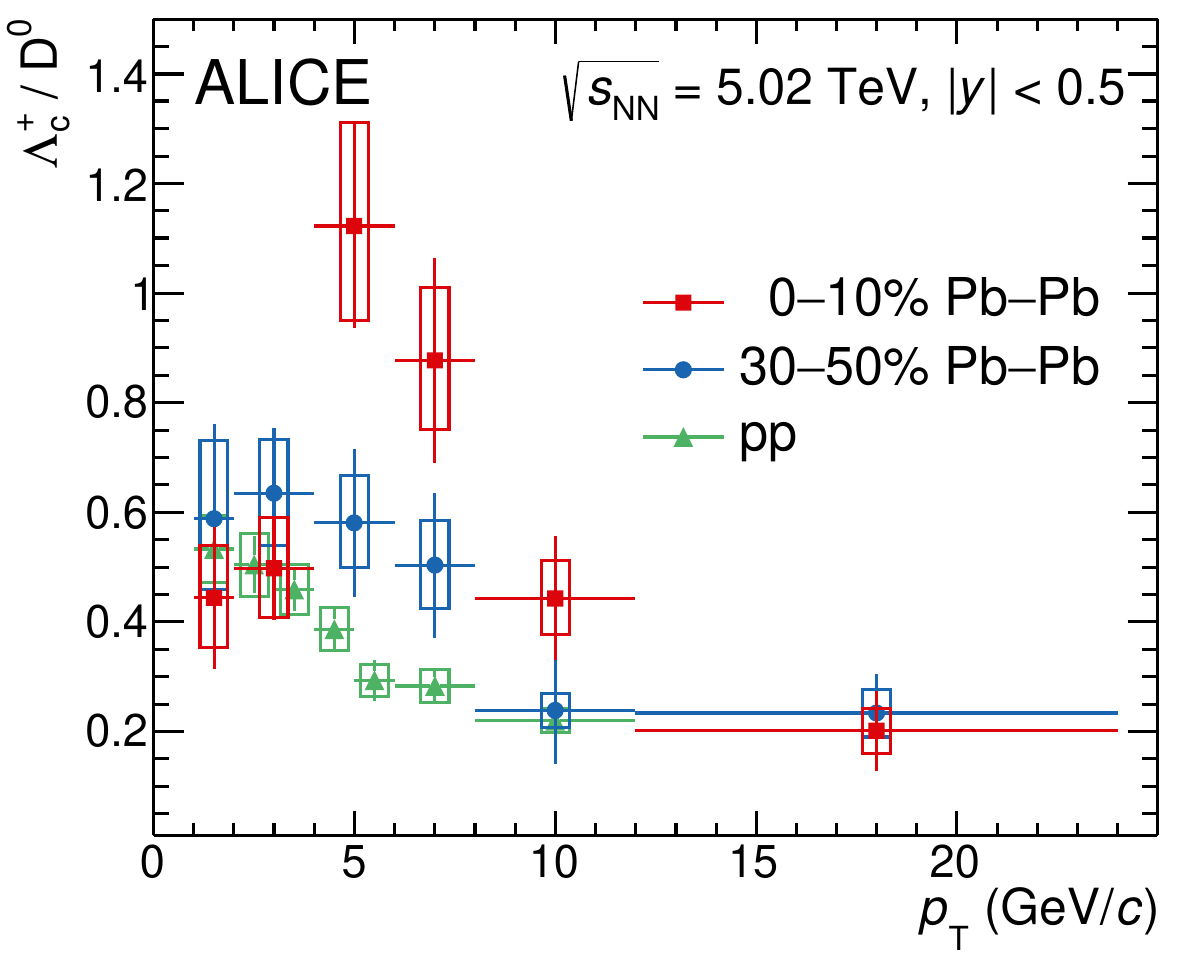}
\includegraphics[width=.45\textwidth]{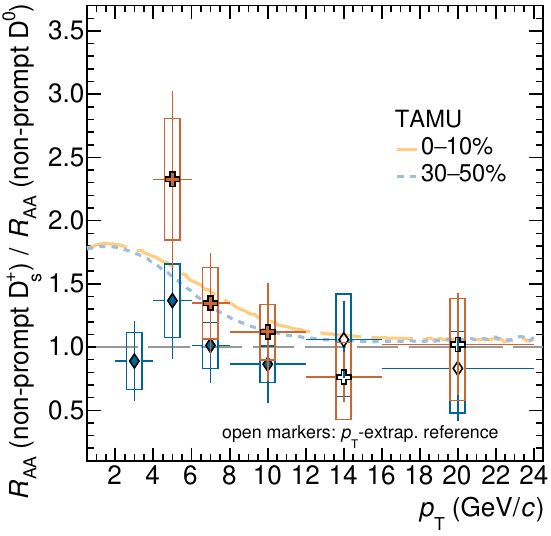}
\caption{Left: $\lc$/$\dzero$ ratio measured at midrapidity in central (\cent{0}{10}) and semi-central (\cent{30}{50}) Pb--Pb collisions at \fivenn by ALICE~\cite{ALICE:2021bib}.
The measurements are compared with that obtained from $pp$ collisions at the some binary centre-of-mass energy~\cite{ALICE:2020wla}.
Right: $\pt$-differential double ratio of $\RAA$ between non-prompt $\ds$ and non-prompt $\dzero$ measured at midrapidity for \cent{0}{10} and \cent{30}{50} centrality intervals in Pb--Pb collisions at \fivenn by ALICE~\cite{ALICE:2022xrg}.
The measurements are compared with TAMU model predictions~\cite{He:2014cla}.}
\label{fig:HFhad}
\end{figure*}

The left panel of Fig.~\ref{fig:HFhad} shows the $\pt$-differential $\lc$/$\dzero$ yield ratio measured in midrapidity in central (\cent{0}{10}) and semi-central (\cent{30}{50}) Pb--Pb collisions at \fivenn by ALICE.
The measurements are compared with that obtained from $pp$ collisions in at the some binary centre-of-mass collisions~\cite{ALICE:2020wla}.
The ratios increase from $pp$ to mid-central and central Pb--Pb collisions for $4 < \pt < 8$~\GeVc with a significance of $2.0\sigma$ and $3.7\sigma$, respectively.
This trend is qualitatively consistent with charm-quark transport in an expanding medium with the Langevin approach and hadronization primarily via coalescence in Pb--Pb collisions~\cite{He:2019vgs,Ravagli:2007xx}.
The right panel of Fig.~\ref{fig:HFhad} reports the $\pt$-differential $\RAA$ double ratio of non-prompt $\ds$ mesons to that of non-prompt $\dzero$ mesons for \cent{0}{10} and \cent{30}{50} centrality intervals in Pb--Pb collisions at \fivenn measured by ALICE~\cite{ALICE:2022xrg}.
Measuring non-prompt $\ds$ mesons together with non-prompt $\dzero$ mesons provides the potential to reveal the beauty-quark hadronization mechanisms in the QGP medium since about $50\%$ of non-prompt $\ds$ mesons are produced in strange-beauty meson $\bs$ decays in $pp$ collisions~\cite{ParticleDataGroup:2022pth,ALICE:2021mgk}.
The double ratio measured in the most $10\%$ central collisions suggests a possible enhancement with respect to unity in $4 < \pt < 12$~\GeVc.
The rise at low $\pt$ might be a consequence of enhanced production of strange-beauty mesons compared to the non-strong beauty mesons in heavy-ion collisions, as expected in a scenario of the abundance of strange quarks thermally produced in the QGP medium and the dominance of the beauty-quark hadronization via recombination with surrounding quarks in the strangeness-enriched environment.
The TAMU model~\cite{He:2014cla}, including beauty quark hadronization via recombination with light quarks from the medium, describes the data within the experimental uncertainties.

An accurate interpretation of all aforementioned measurements carried out in heavy-ion collisions, which is crucial for the characterization of the properties of the QGP medium, relies on the reference in which the QGP medium is not formed.
Traditionally, small system (e.g., $pp$ and $p$--Pb) collisions, due to the partonic environment they created being dilute, service the AA collision baseline.
In addition, measuring heavy-flavor production in $pp$ collisions also provides fundamental test pQCD calculations in the \TeV domain~\cite{Mangano:1991jk,Cacciari:1998it}.

The production cross sections of ${\rm c}\overline{\rm c}$~\cite{ALICE:2021dhb} and ${\rm b}\overline{\rm b}$~\cite{ALICE:2021mgk,ALICE:2012vpz,ALICE:2014aev,ALICE:2012acz,ALICE:2021edd} at midrapidity in $pp$ collisions are measured by ALICE.
For the ${\rm c}\overline{\rm c}$ measurements, the results are higher than the upper edge of the pQCD-based FONLL and NNLO calculations, though compatible within $\sim 1\sigma$ of the experimental uncertainties.
The measured ${\rm b}\overline{\rm b}$ production cross sections are found to be compatible with FONLL and NNLO calculations.
It is worth noting that the data uncertainties of the ${\rm c}\overline{\rm c}$ measurements are significantly smaller than the predictions and stricter constraints on the theory calculations.
Similar observations are derived from the ($\pt,y$)-double-differential comparison of FONLL predictions with the cross section of muons from semi-leptonic decays of charm and beauty hadrons measured at forward rapdity~\cite{ALICE:2012msv,ALICE:2019rmo}.
The measurements set additional constraints for pQCD calculations in a kinematic region important for probing parton distribution functions (PDFs) at low Bjorken-$x$ values, down to about $\sim 10^{-5}$.

\begin{figure*}[!htbp]
\centering
\includegraphics[width=.47\textwidth]{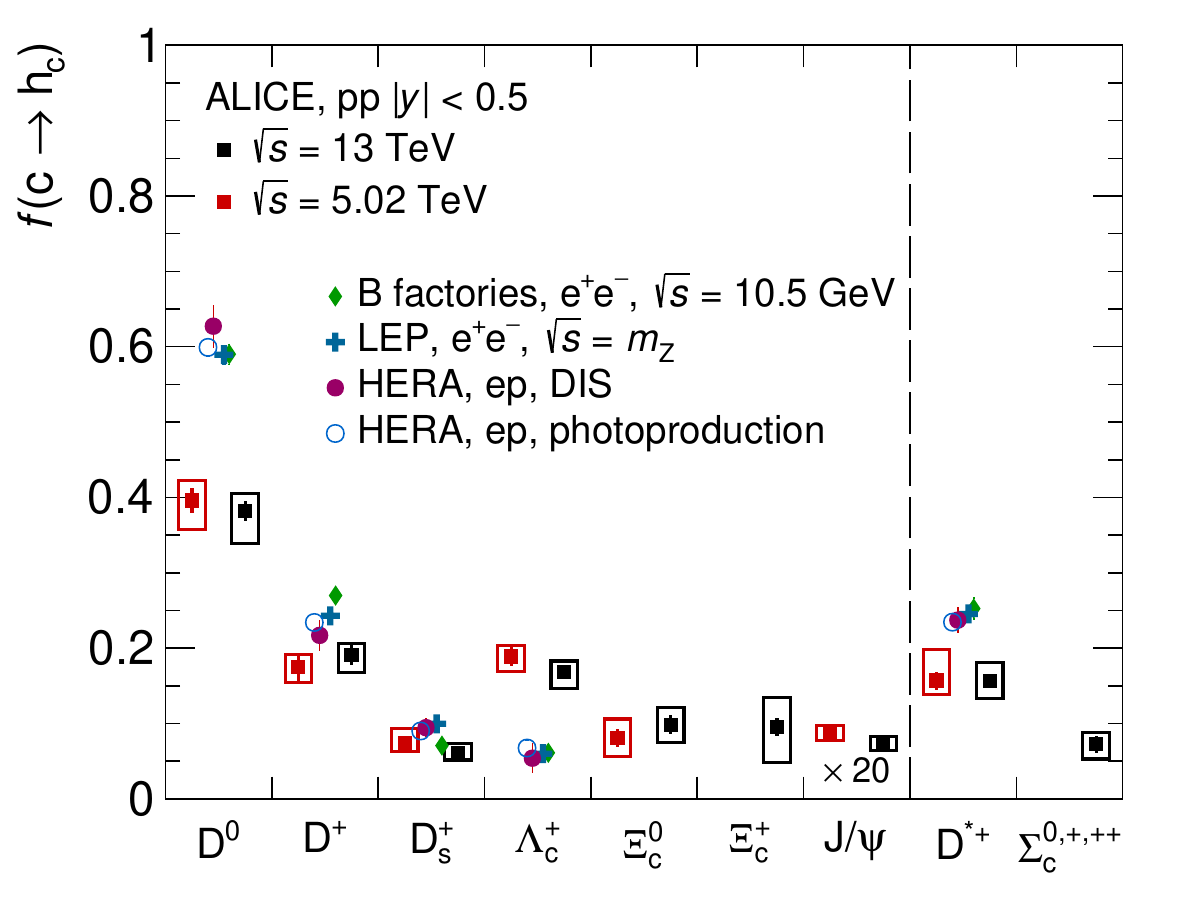}
\includegraphics[width=.52\textwidth]{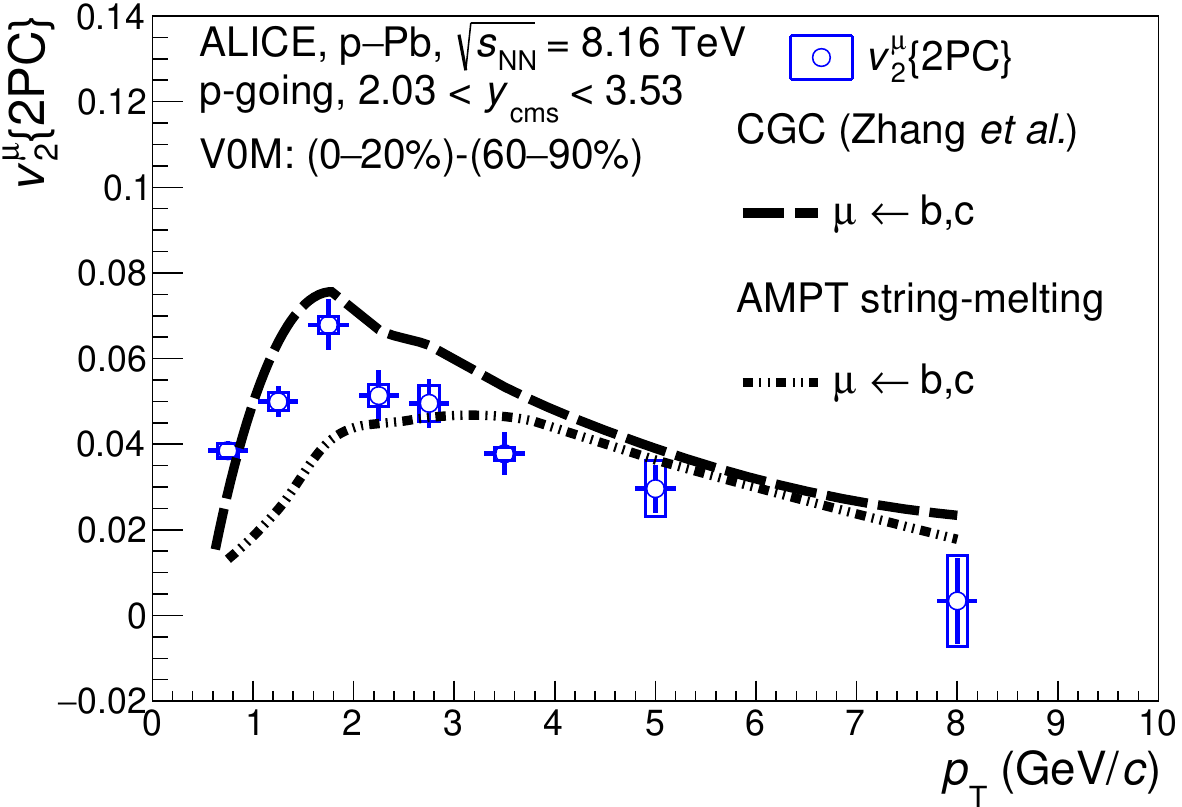}
\caption{Left: charm-quark fragmentation fractions $f({\rm c}\to{\rm h}_{\rm c})$ measured at midrapidity in $pp$ collisions at $\s = 5.02$ and $13$~\TeV by ALICE~\cite{ALICE:2023sgl}.
The results are compared with the those obtained from $e^{-}e^{+}$ and $e{\rm p}$ collisions~\cite{Lisovyi:2015uqa}.
Right: $\pt$-differential $v_{2}$ coefficient of muons measured using two-particle correlation (2PC) method at forward rapidity in high-multiplicity $p$--Pb collisions at $\snn = 8.16$~\TeV by ALICE~\cite{ALICE:2022ruh}.
The measurements are compared with AMPT~\cite{Lin:2021mdn} and CGC~\cite{Zhang:2020ayy}calculations.}
\label{fig:SmallSysHF}
\end{figure*}

The mentioned theoretical calculations are based on the pQCD factorisation approaches~\cite{Collins:1989gx,Catani:1990eg}, in which the fragmentation functions are typically parametrized from measurements performed in $e^{-}e^{+}$ or $e{\rm p}$ collisions, under the assumption that the hadronisation of heavy quarks into hadrons is a universal process across different colliding systems.
However, this assumption may be broken by higher-twist effects or other effects related to the heavy-quark kinematics and the $pp$ underlying event~\cite{Collins:1998rz,Ellis:1982cd,Alekhin:2017kpj}.
This effect is observed in the recent measurements of charmed meson-to-baryon yield ratios at midrapidity in $pp$ collisions at various collision energies for $\lc/\dzero$~\cite{ALICE:2020wla,ALICE:2020wfu,ALICE:2017thy}, $\Sigma_{\rm c}^{0,++}/\dzero$~\cite{ALICE:2021rzj}, $\Xi_{\rm c}^{+,0}/\dzero$~\cite{ALICE:2021psx,ALICE:2021bli,ALICE:2017dja}, and $\Omega_{\rm c}^{0}/\dzero$~\cite{ALICE:2022cop}.
The ALICE results evidence higher baryon-to-meson cross section ratios at low-$\pt$ related to $e^{-}e^{+}$ and $e{\rm p}$ collisions.
It is worth noting that $\Sigma_{\rm c}^{0,++}$ and $\Xi_{\rm c}^{+,0}$ cross sections are measured for the first time in hadronic collisions.
A consequence of the significant difference between the charmed baryon-to-meson yield ratios measured in $pp$ and $e^{-}e^{+}$ and $e{\rm p}$ collisions is that charm-quark fragmentation fractions, $f({\rm c}\to{\rm h}_{\rm c})$,  i.e., the probabilities of a charm quark to hadronize as a given charmed hadron species ${\rm h}_{\rm c}$, obtained in $pp$ collisions are different from that estimated from $e^{-}e^{+}$ and $e{\rm p}$ data, as presented in the left panel of figure~\ref{fig:SmallSysHF}.
In addition, $f({\rm c}\to{\rm h}_{\rm c})$ is also extracted from $p$--Pb collisions by ALICE~\cite{ALICE:2024ocs}.
The results are almost identical to that in $pp$ collisions, illustrating no significant modification of the charm-quark hadronization between the two colliding systems despite the larger system size and higher charged-particle multiplicity density in $p$--Pb collisions.
Such observations provide evidence that the assumption of universality (colliding-system independence) of parton-to-hadron fragmentation functions is not valid for charmed hadron production.

Furthermore, the measurements in $p$--Pb collisions are important to constrain the cold nuclear matter (CNM) effects~\cite{Armesto:2018ljh}, such as the nuclear modified PDFs (nPDFs), multiple scatterings in nucleons that collide with more than one other nucleon, the Cronin Effect, parton energy loss in CNM, and absorption of the produced hadrons by the nucleus.
ALICE measures the nuclear modification factors of open heavy-flavour particles in $p$--Pb collisions for D mesons~\cite{ALICE:2019fhe}, $\lc$~\cite{ALICE:2022exq} and $\Xi_{\rm c}^{0}$~\cite{ALICE:2024ozd} baryons, and open heavy-flavour hadron decay electrons~\cite{ALICE:2015zhm} at midrapidity and open heavy-flavour decay muons~\cite{ALICE:2017fsl} at forward rapdity.
The measurements of open heavy-flavour hadron decay muons at forward rapidity, together with the measurement of ${\rm W}^{\pm}$-boson production~\cite{ALICE:2022cxs} at the same rapidity region, also provide an important constraint on nPDFs at Bjorken-$x$ down to $10^{-6}$.
In addition, the nuclear modification factors of beauty-quark production in $p$--Pb collisions for beauty-hadron decay electrons~\cite{ALICE:2016uid} and b-tagged jets~\cite{ALICE:2021wct} are also measured by ALICE at midrapidity.
All those measurements are consistent with predictions adopting only the CNM effects within both experimental and theoretical uncertainties.

Up to now, all aforementioned observations in small system collisions exhibit that such systems can be considered as plain references, in which the conditions to form the QGP could not be reached.
Surprisingly, as illustrated in the right panel of figure~\ref{fig:SmallSysHF}, the non-zero $v_{2}$ coefficient of open heavy-flavour hadron decay muons, considered as emblematic signatures of QGP formation, is observed at forward rapidity in high-multiplicity $p$--Pb collisions at $\snn = 8.16$~\TeV by ALICE~\cite{ALICE:2022ruh}.
A similar observation is also found at midrapidity for open heavy-flavour hadron decay electrons~\cite{ALICE:2018gyx}.
The measurements are compared with AMPT~\cite{Lin:2021mdn} and CGC~\cite{Zhang:2020ayy}calculations.
In AMPT, the anisotropy is derived from the escape mechanism via partonic interactions.
On the other hand, the CGC calculations are based on the dilute-dense formalism, where interactions between partons from the proton projectile and dense gluons inside the target Pb nucleus at the early stage of the collision generate azimuthal anisotropies.
For muons measured in the ALICE forward muon spectrometer, the heavy-flavor contribution is the main source in $\pt > 2$~\GeVc.
Despite CGC calculations overestimating the measurements in the first few data $\pt$ intervals, the two predictions generally agree with the observed anisotropy within experimental uncertainties.
The model comparison suggests either initial- or final-state partonic interactions are possible to generate the hydrodynamic-like azimuthal anisotropy in momentum space for heavy-flavour particles.
Besides, the existence of small QGP droplets at high multiplicity in small system collisions is still under debate.

In the future, the high-multiplicity LHC project for the ALICE LHC Run 3 and Run 4 and the ALICE 3 detector upgrade project for the LHC Run 5 and beyond (see Sec.~\ref{sec.III}) will be crucial to reveal the origin of the observed non-university in charm quark hadronization and the QGP-like effects presented in small system collisions.

\subsection{Jets}  \label{sec.II.5}
\begin{figure*}[!htb]
\includegraphics[width=.45\hsize]{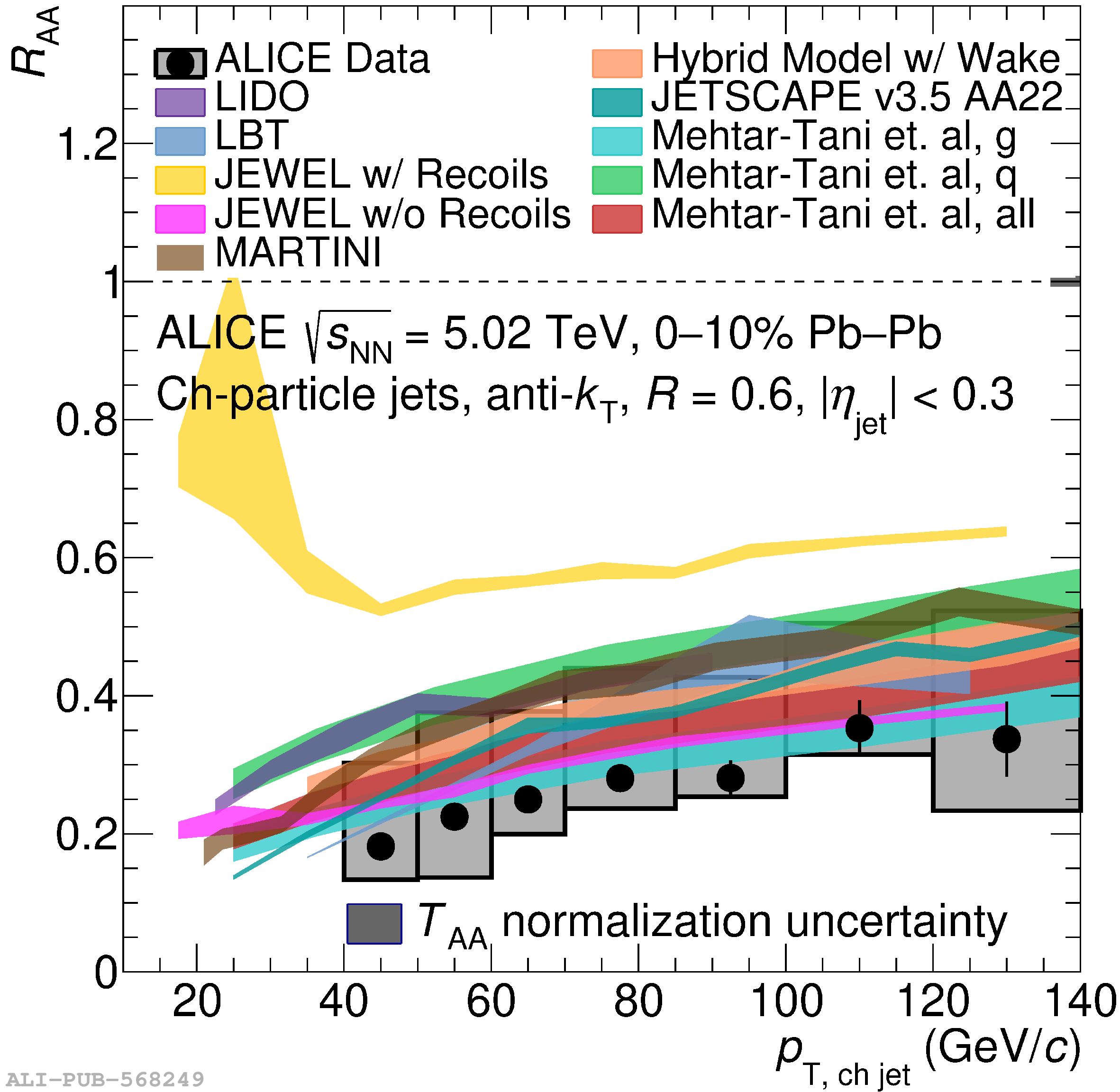}
\includegraphics[width=.45\hsize]{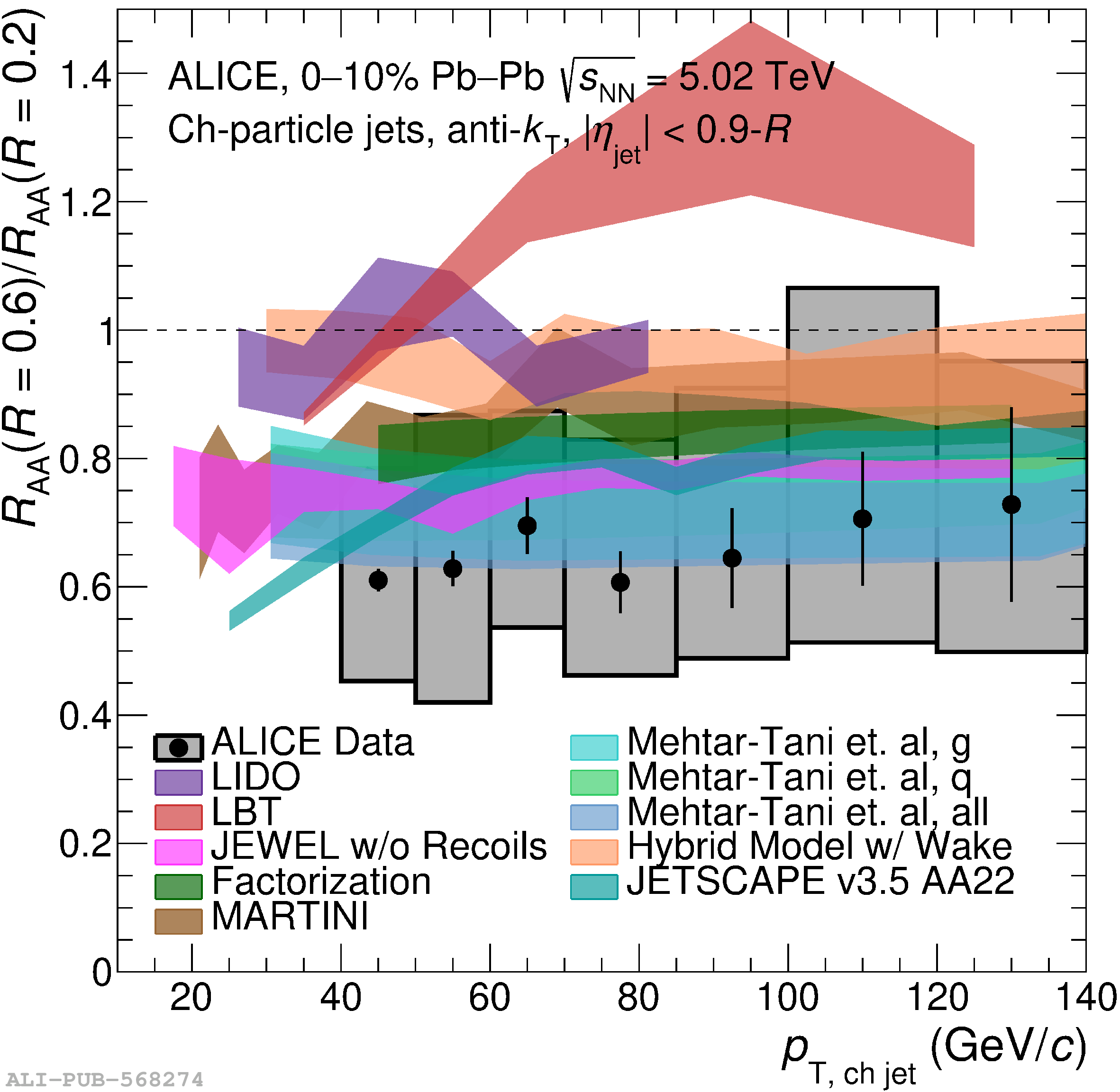}
\caption{(Color online) The nuclear modification factors for $R = 0.6 $ (left) and Double ratio of jet nuclear modification factors between $R = 0.6 $ and $R = 0.2 $ (right), shown for 0--10\%  central Pb--Pb collisions compared to theoretical calculations incorporating jet quenching~\cite{jet1}. }
\label{jetRAA}
\end{figure*}

Jets play a crucial role in ultra-relativistic heavy-ion collisions, serving multiple key functions. They help identify the initial momentum scale of hard scatterings, especially for photons and other colorless probes. Moreover, jets offer valuable insights into the interactions within the dense QCD medium, particularly those involving quarks and gluons, which carry QCD color charge. By studying jet substructure, heavy-flavor quarks, and the highest transverse momentum jets, researchers can thoroughly explore QCD medium interactions across various topologies and kinematic ranges.

\begin{figure}[!htb]
\includegraphics[width=\linewidth]{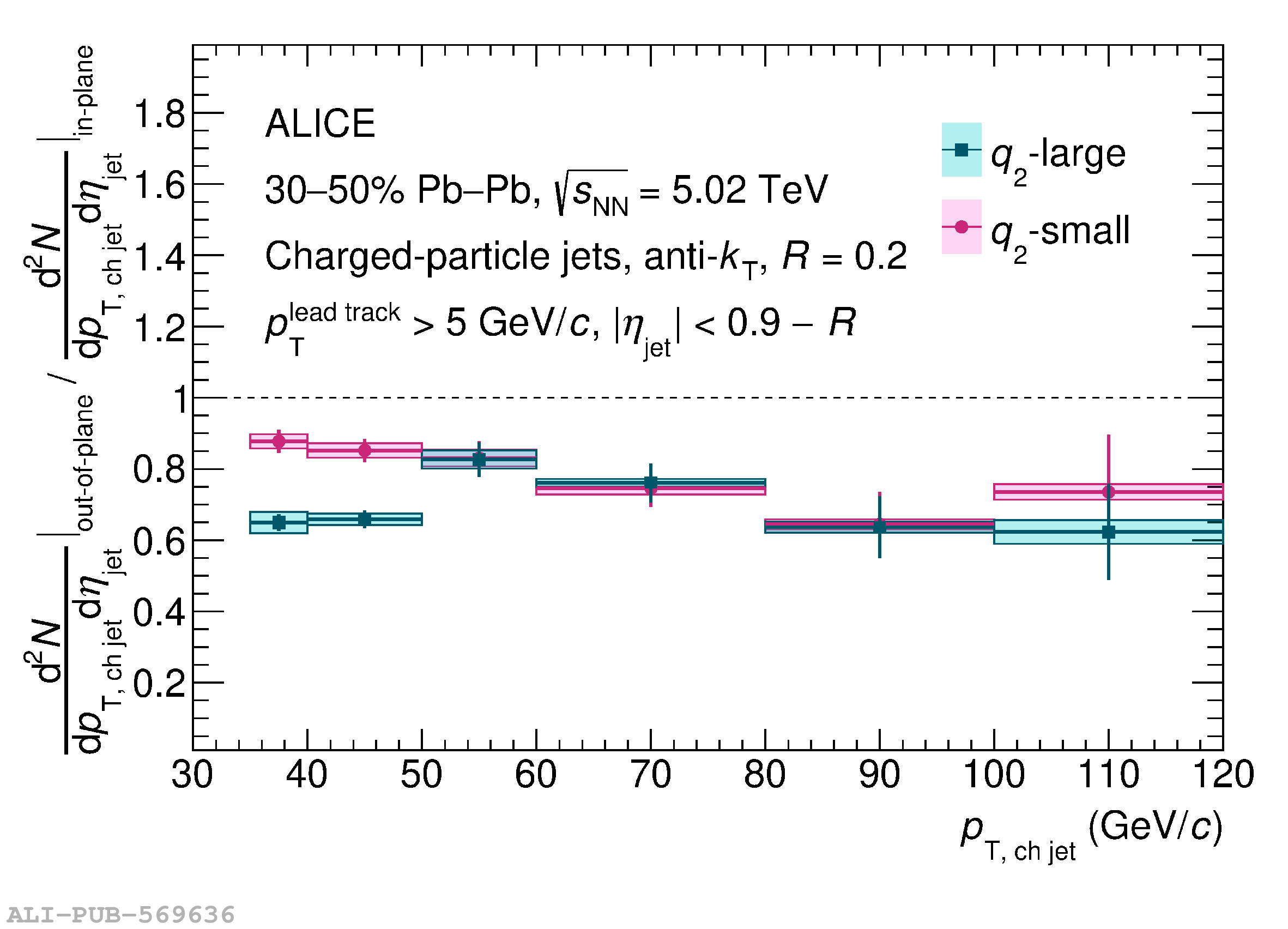}
\caption{(Color online) Ratios of out-of-plane to in-plane charged-particle jet yields for the $q_{2}$-large and $q_{2}$-small event classes in Pb-Pb collisions at $\snn = 5.02$ TeV~\cite{jet2}. }
\label{ESEjet}
\end{figure}

\begin{figure*}[!htb]
\includegraphics[width=.45\hsize]{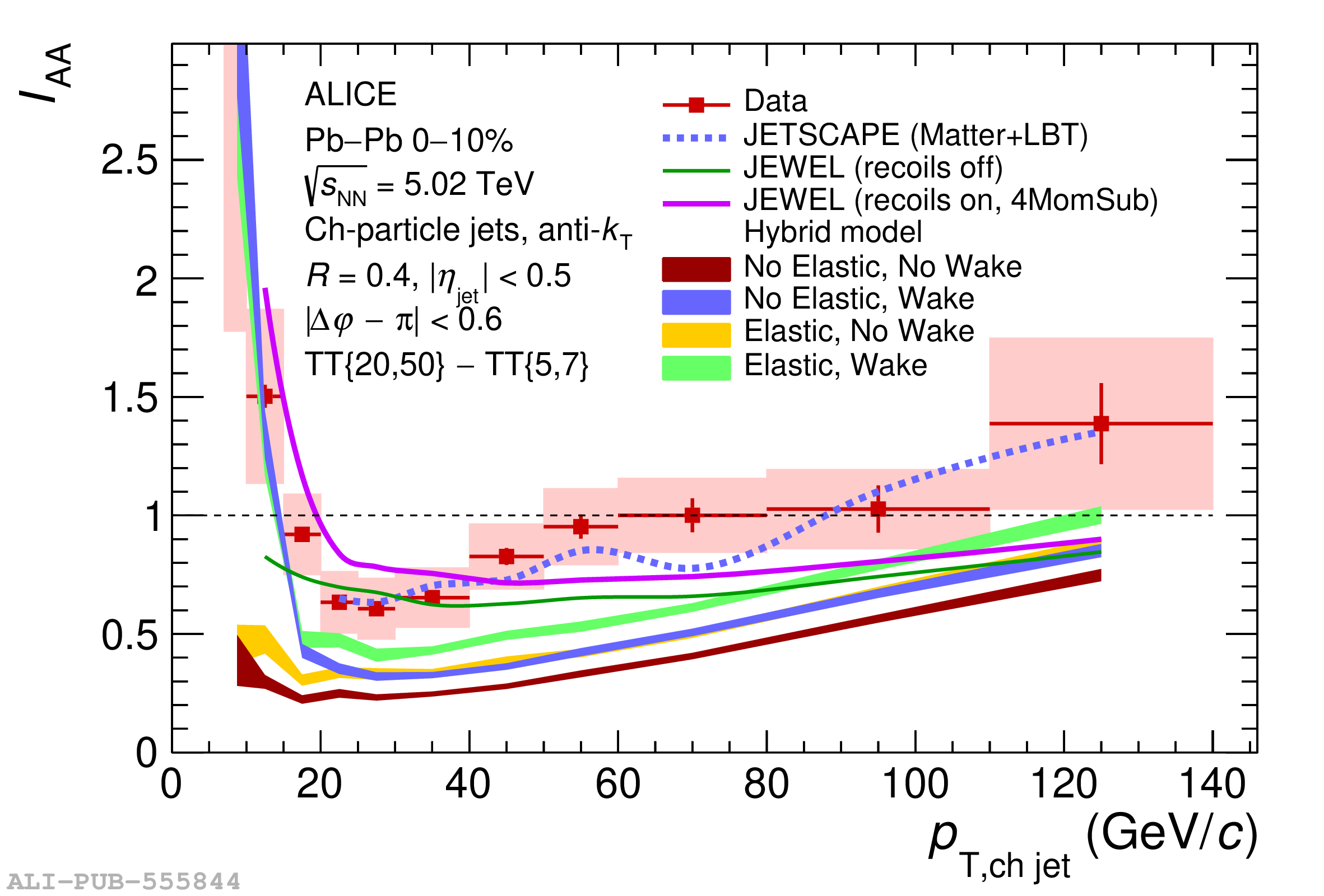}
\includegraphics[width=.45\hsize]{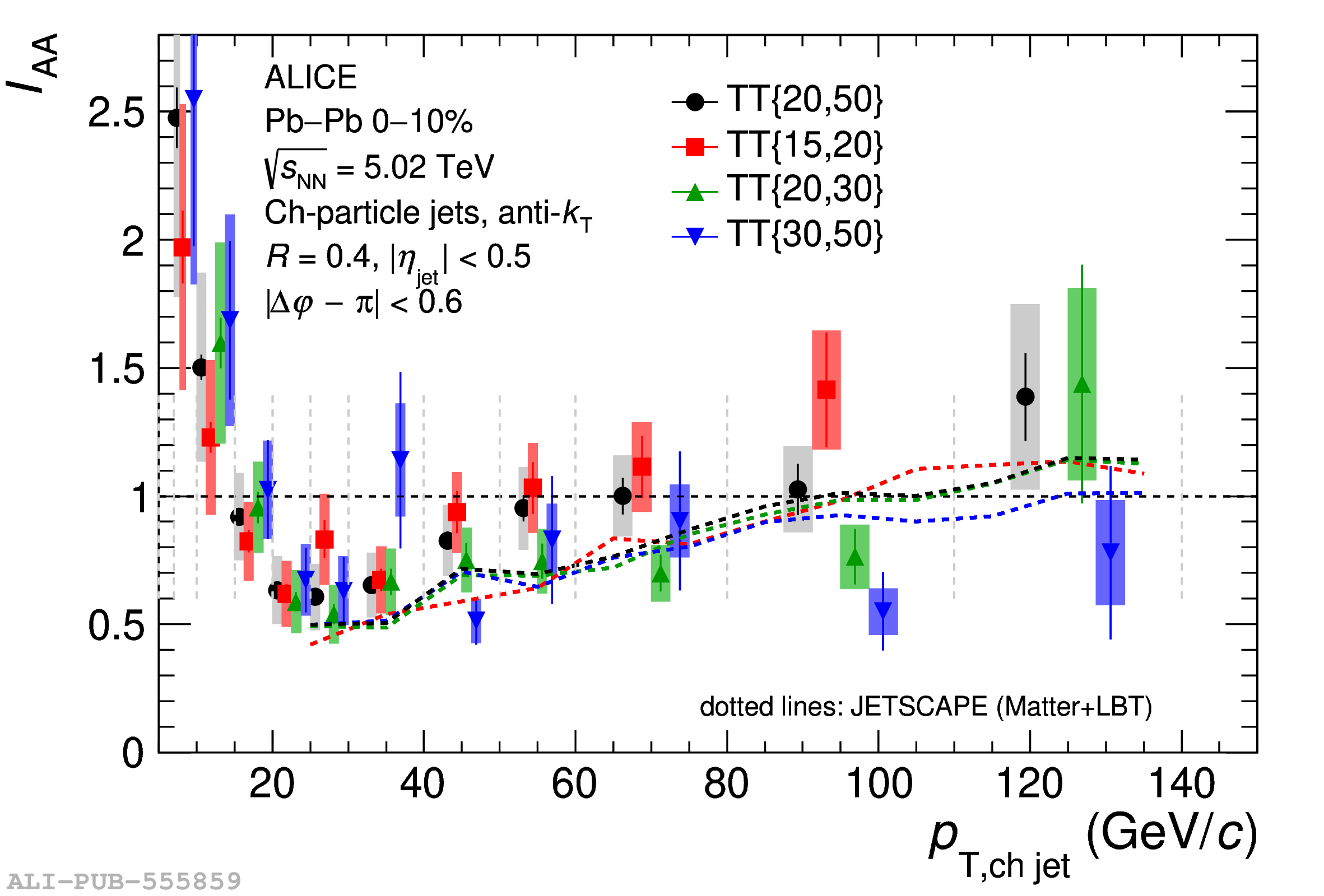}
\caption{(Color online) \IAApT\ from the \DrecoilpT\ distributions measured for $R = 0.4$ with default \TTSig\ selection (left) and with varied \TTSig\ selections (right) in central Pb-Pb collisions~\cite{ jet4}. }
\label{jetIAA}
\end{figure*}

In Fig.~\ref{jetRAA}, the nuclear modification factors are depicted for $R = 0.6$ (left) and the double ratio of jet nuclear modification factors between $R = 0.6$ and $R = 0.2$ (right), presented for 0--10\% central Pb-Pb collisions and compared to theoretical calculations incorporating jet quenching~\cite{jet1}. The double ratio \( R_{\rm AA} \) is a critical measure for evaluating the \( R \)-dependence of energy loss in jets. Values below unity indicate greater suppression for jets with larger \( R \), values at unity suggest no \( R \)-dependence or a balancing of effects, and values above unity imply less suppression for larger \( R \) jets. In 0--10\% central collisions, the \( R_{\rm AA}^{0.6/0.2} \) ratio shows suppression below unity at lower jet \( p_{\rm T} \) values, indicating a potential \( R \)-dependence within the uncertainties.

Event-shape engineering (ESE) is an experimental technique that classifies events based on their anisotropies using the magnitude of the reduced flow vector, providing a novel method to constrain the path-length dependence of jet energy loss. To investigate the event shape dependence, events were categorized based on the magnitude of the reduced flow vector \( q_{2} \), measured with the forward detector V0C. Figure~\ref{ESEjet} shows the ratio of out-of-plane to in-plane jet yields for the \( q_{2} \)-small and \( q_{2} \)-large event classes, focusing on jets with \( R = 0.2 \) in mid-central 30-50\% Pb-Pb collisions at \( \sqrt{s_{\mathrm{NN}}} = 5.02 \) TeV~\cite{jet2}. The observed ratios are significantly below unity, indicating that jets lose more energy on average when traveling out-of-plane compared to in-plane. This finding supports the notion that the extent of jet energy loss is influenced, at least partially, by the path length through the medium.

\begin{figure*}[!htb]
\includegraphics[width=.8\hsize]{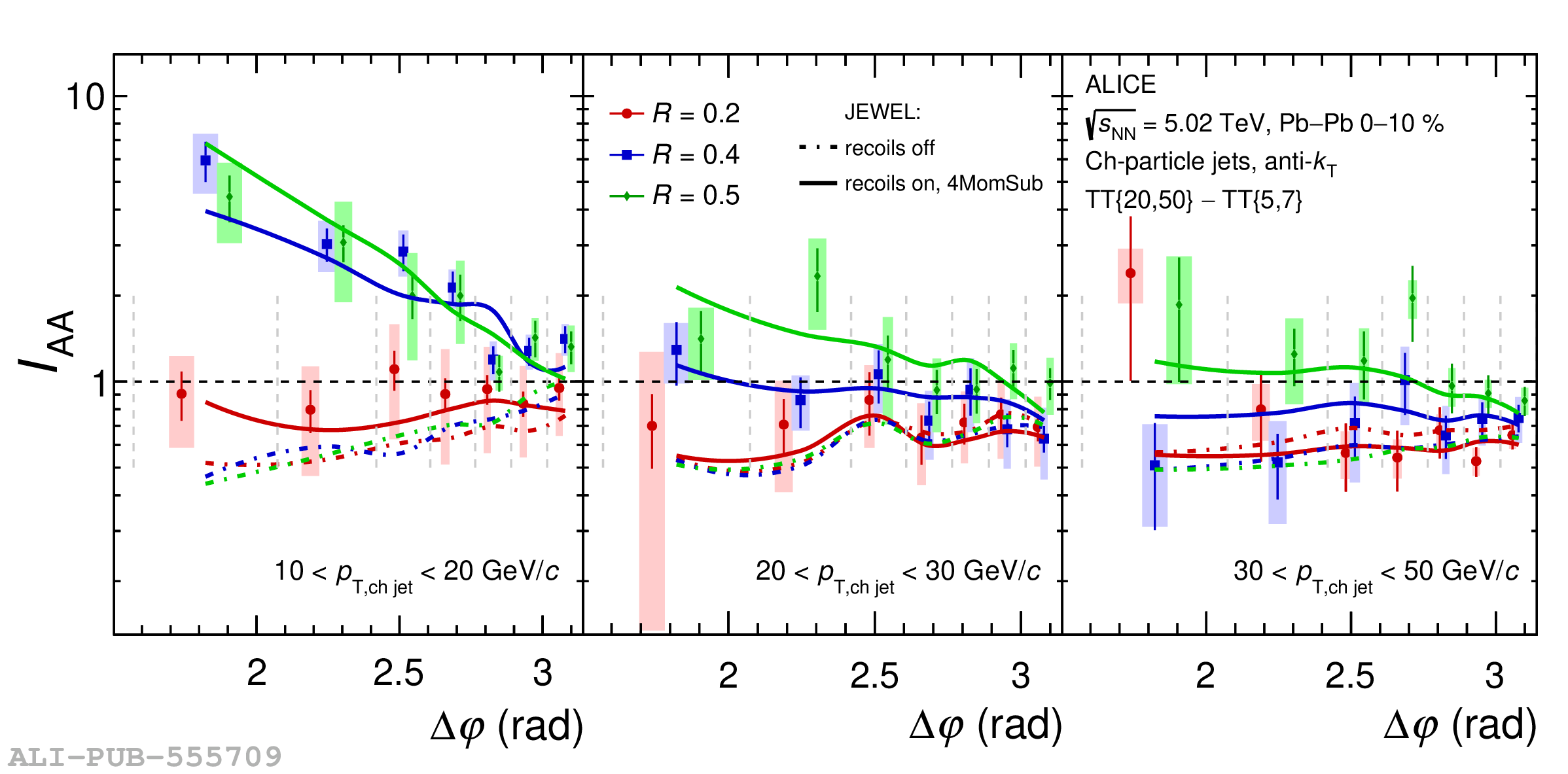}
\caption{(Color online) \IAAphi\ for $\rr=0.2$, 0.4 and 0.5, for intervals in recoil \pTjetch:
[10,20], [20,30], and [30,50] \gev. Predictions from JEWEL are also shown~\cite{jet3}. }
\label{jetIAAdPhi}
\end{figure*}

Medium-induced yield modification is measured by \( I_{AA (p_{T_{ch,jet}})} = \Drecoil (\rm Pb-Pb)/\Drecoil (pp) \), the ratio of the \(\DrecoilpT\) distributions measured in Pb-Pb and $pp$ collisions. Figure~\ref{jetIAA} presents the \IAApT\ distributions measured for \( R = 0.4 \) with the default \TTSig\ selection (left panel) and with varied \TTSig\ selections (right panel) in central Pb-Pb collisions. The \IAApT\ distributions show a notable dependence on \pTjetch. For \(\rr = 0.4\), JEWEL (recoils on) exhibits a significant increase in \IAApT\ towards low \pTjetch\ for \(\pTjetch < 20~\gev\), mirroring the trend in the data for \(\rr = 0.4\). Overall, JETSCAPE most accurately describes both the magnitude and \pTjetch\ dependence of \IAApT\ in the range \(\pTjetch > 20 \gev\). The rising trend in data towards low \pTjetch\ for \(\pTjetch < 20 \gev\) is captured by both the Hybrid Model and JEWEL, but only with the inclusion of medium-response effects~\cite{jet3, jet4}.
Figure~\ref{jetIAA} (right) displays the \IAApT distribution for \(\rr=0.4\) measured for several \pTtrig\ intervals used in the \TTSig\ event selection. A higher \pTtrig\ threshold corresponds to larger $\tilde{z}$, where the assumptions underlying the surface-bias picture may better apply. The results indicate that as the lower \pTtrig\ bound is raised, the rate of increase in \IAApT\ at large \pTjetch\ diminishes.

Figure~\ref{jetIAAdPhi} presents the first observation of medium-induced jet acoplanarity broadening in the QGP. The broadening is significant in the range of $10<\pTjetch<20$ \gev\ for $\rr=0.4$ and 0.5 but is negligible for $\rr=0.2$, and is negligible at larger \pTjetch\ for all \rr. This rapid transition in the shape of the acoplanarity distribution as a function of both \pTjetch\ and \rr\ is striking. Possible mechanisms for generating acoplanarity broadening include jet scattering from QGP quasi-particles, medium-induced wake effects, and jet splitting, where medium-induced radiation from a high-\pTjetch\ jet is reconstructed at low \pTjetch\ with a large deviation from $\dphi\sim\pi$~\cite{jet3, jet4}.

\begin{figure*}[!htb]
\includegraphics[width=.8\hsize]{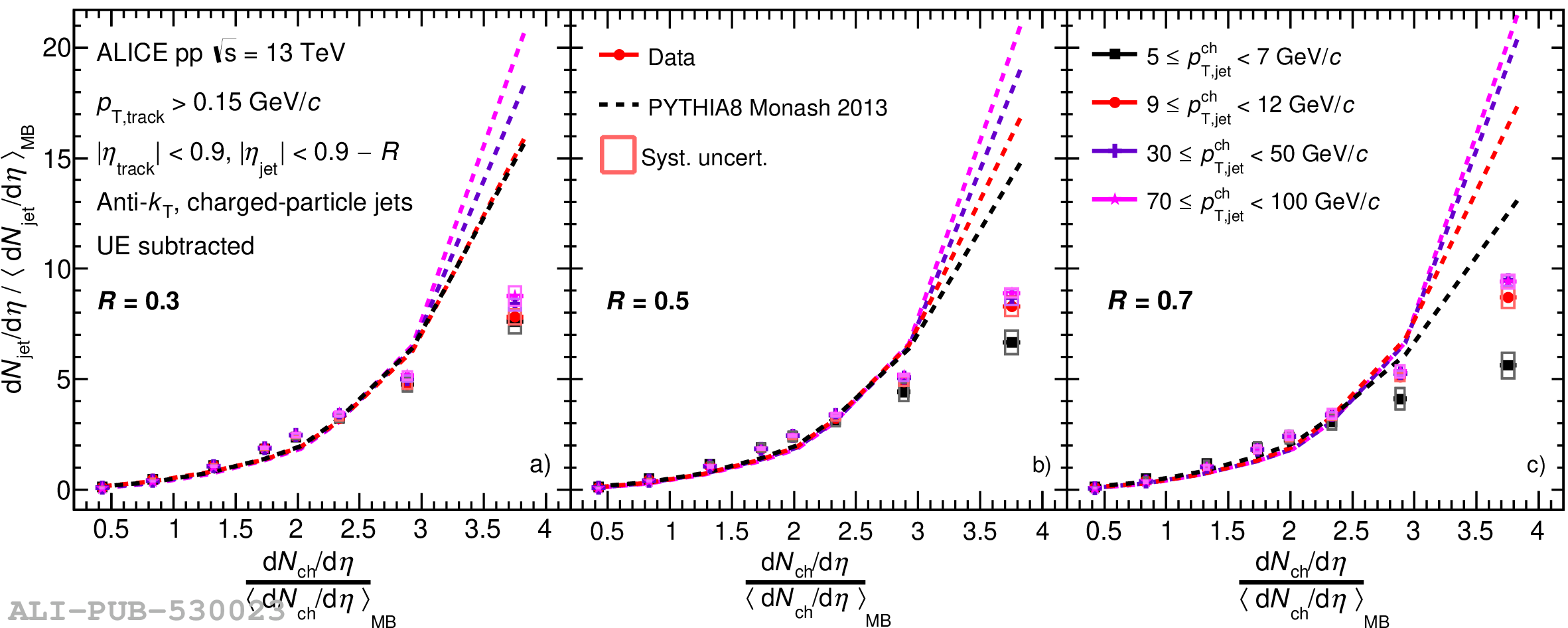}
\caption{(Color online) Comparison of self-normalised jet yields as a function of the self-normalised charged-particle multiplicity in four selected jet $\pt$ intervals ($5\leq p_{\rm T,jet}^{\rm ch}< 7\ {\rm GeV}/c$, $9\leq p_{\rm T,jet}^{\rm ch}< 12\ {\rm GeV}/c$, $30\leq p_{\rm T,jet}^{\rm ch}< 50\ {\rm GeV}/c$, and $70\leq p_{\rm T,jet}^{\rm ch}< 100\ {\rm GeV}/c$) for a given jet radii: a) $R = 0.3$, b) $R = 0.5$, c) $R = 0.7$ between data and PYTHIA8 predictions~\cite{jet5}. }
\label{HMjetRatio}
\end{figure*}

\begin{figure*}[!htb]
\includegraphics[width=.8\hsize]{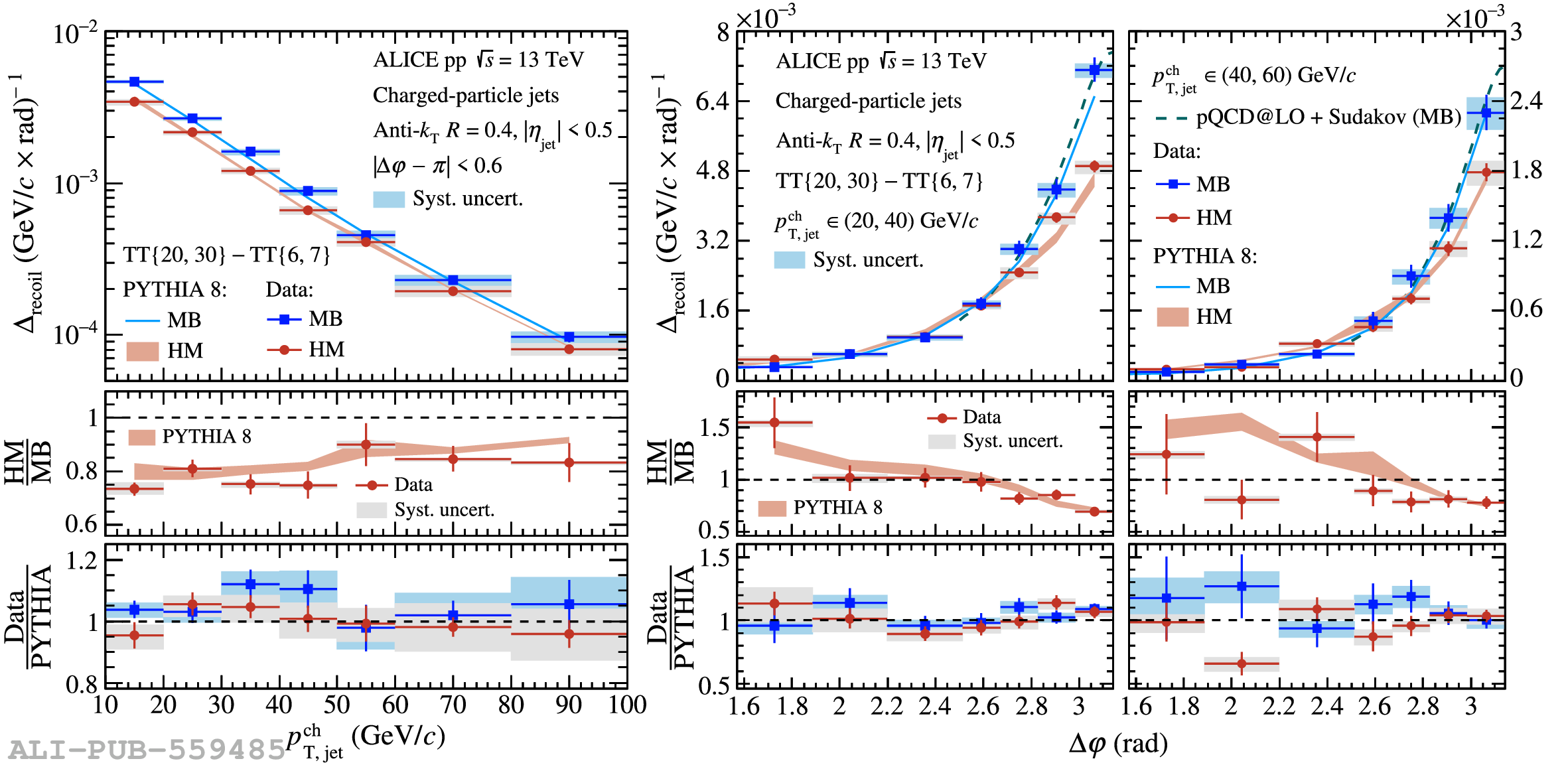}
\caption{
(Color online) Fully-corrected \Drecoil\ distributions measured in MB and HM-selected events in \pp\ collisions at $\s=13$~TeV. Left panel: \Drecoil(\pTjetch) in $|\dphi-\pi|<0.6$; middle and right panels: \Drecoil(\dphi) for $20<\pTjetch<40$~\gev\ and $40<\pTjetch<60$~\gev~\cite{jet6}. 
}
\label{HMjetDphi}
\end{figure*}

Jet quenching, a phenomenon linked to the formation of QGP in large nuclear systems, is anticipated to occur in smaller systems, though with diminished effects. At present, there is no definitive evidence, beyond experimental uncertainties, of jet quenching in small systems. This raises the question of whether the observed collective effects in these systems genuinely arise from QGP formation or from other mechanisms. Further investigations are essential to search for jet quenching effects in small systems and address this unresolved issue.

\begin{figure}[!htb]
\includegraphics[width=\linewidth]{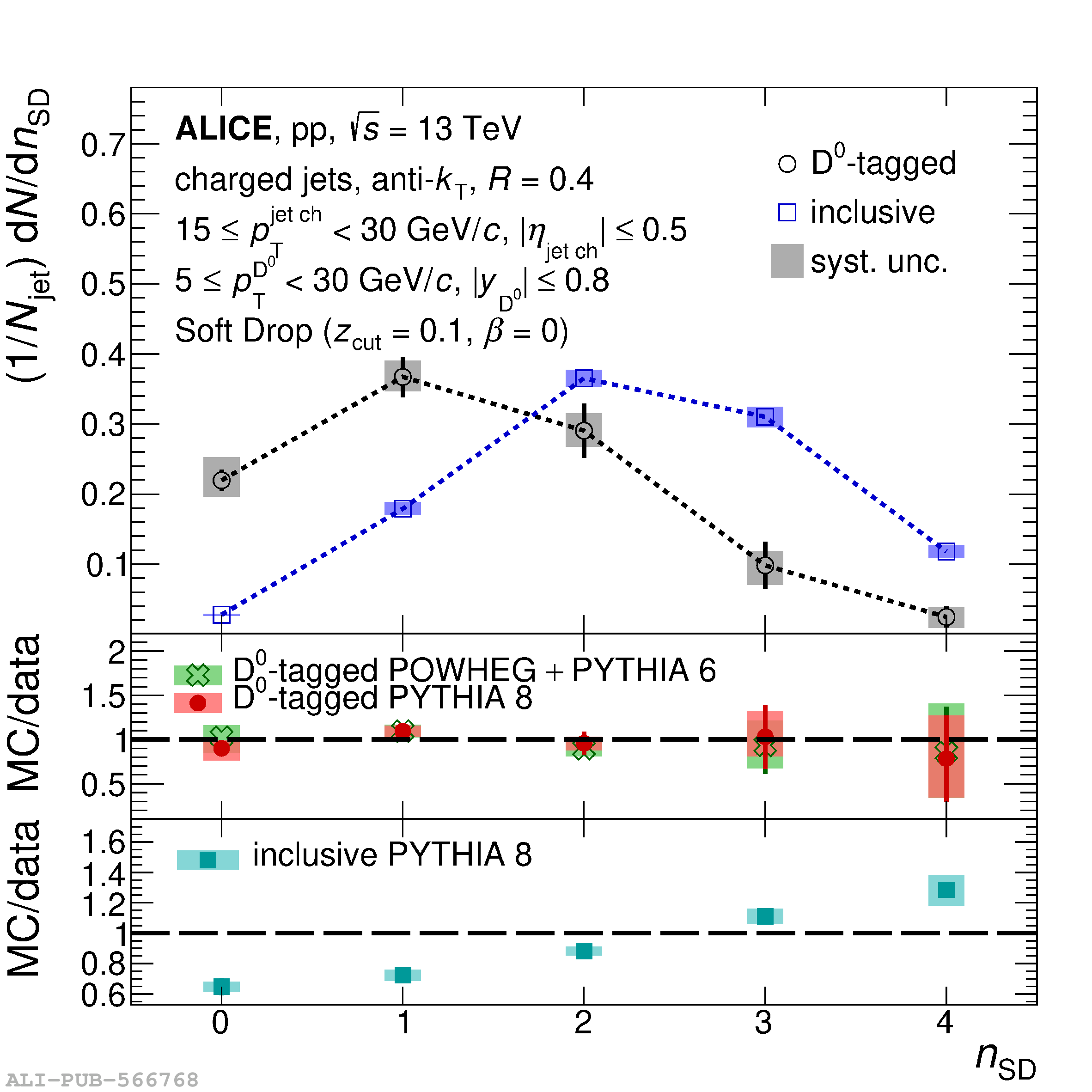}
\caption{(Color online) The $n_{SD}$ distributions of prompt \dzero-tagged jets compared to those of inclusive jets for $15 \leq \pTjet < 30$ \GeVc in $pp$ collisions at \s = 13 \TeV~\cite{jet7}.}
\label{HFJet}
\end{figure}

\begin{figure*}[!htb]
\includegraphics[width=.8\hsize]{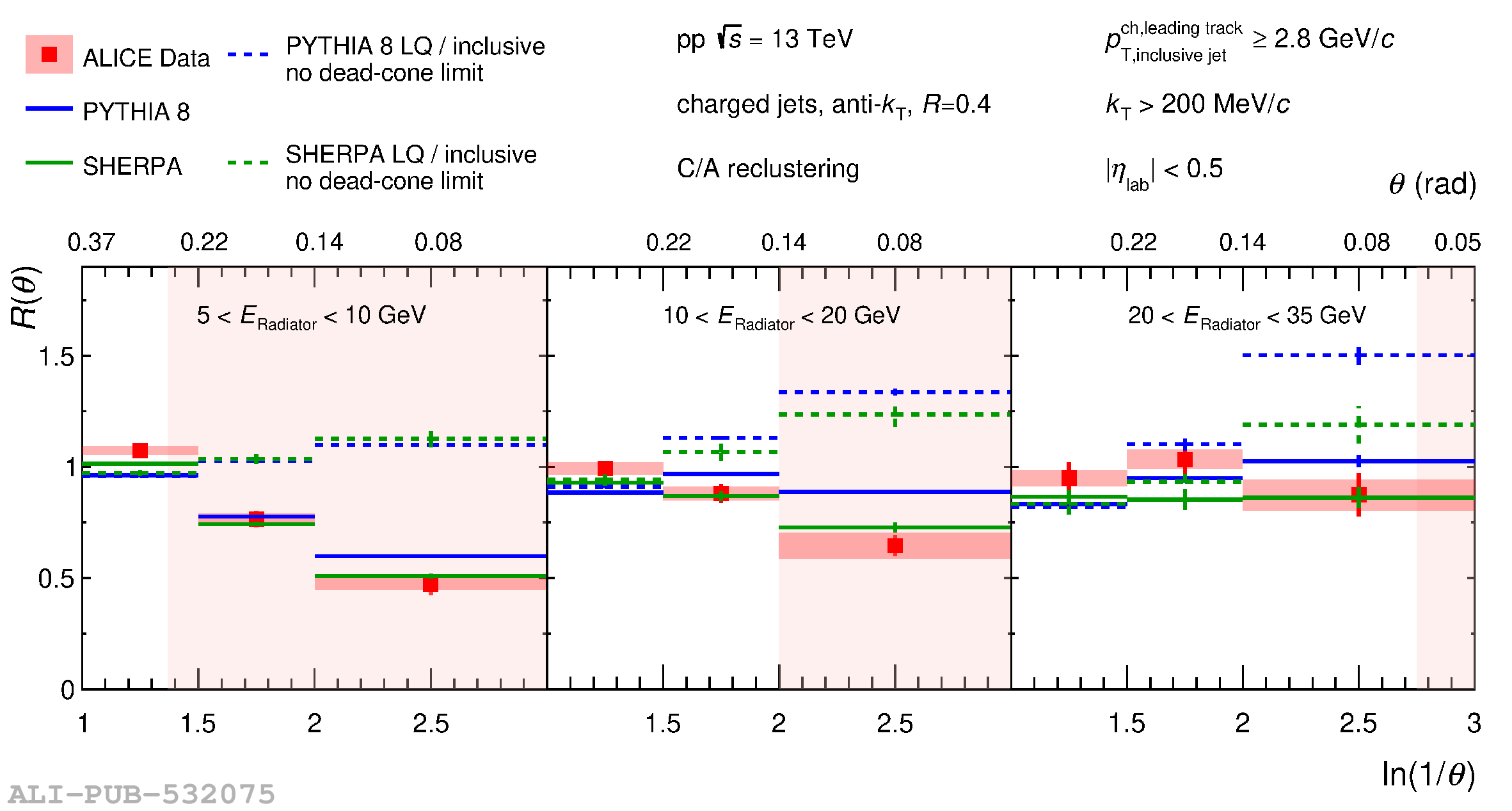}
\caption{(Color online) The nuclear modification factors for $R = 0.6 $ (left) and Double ratio of jet nuclear modification factors between $R = 0.6 $ and $R = 0.2 $ (right), shown for 0--10\%  central Pb-Pb collisions compared to theoretical calculations incorporating jet quenching~\cite{jet8}. }
\label{DeadCone}
\end{figure*}

To investigate the \(\pt\) dependence of normalized jet production as a function of self-normalized charged-particle multiplicity, Fig.~\ref{HMjetRatio} displays the self-normalized jet yields across four selected jet \(\pt\) intervals for resolution parameters \(R = 0.3\), \(0.5\), and \(0.7\). The data are also compared with PYTHIA8 predictions~\cite{jet5}. The measured jet production ratios in central rapidity exhibit an increase with multiplicity, mirroring earlier findings for identified particles using the forward multiplicity V0 estimator. The increase is less pronounced for the lowest jet \(\pt\) in the highest multiplicity interval. Current MC event generators can predict the rising trend but fail to accurately describe the absolute yields, particularly in the highest multiplicity class.

Figure~\ref{HMjetDphi} presents fully-corrected distributions of \Drecoil(\pTjetch) and \Drecoil(\dphi) measured in minimum bias (MB) and high multiplicity (HM)-selected $pp$ collisions at \(\sqrt{s} = 13\) TeV, alongside calculations from various models~\cite{jet6}. The comparison of MB and HM \Drecoil(\pTjetch) distributions shows a yield suppression in HM collisions that remains largely independent of \pTjet, though there is a slight indication of a harder recoil jet spectrum for HM events.
The \Drecoil(\dphi) distributions reveal that jet-yield suppression in HM events primarily occurs in the back-to-back configuration. The total yield is reduced, while the azimuthal distribution is broadened. This broadening might result from jet quenching, where medium-induced jet scattering is more prevalent in HM events. However, PYTHIA 8 particle-level distributions also show jet yield suppression and azimuthal broadening in HM-selected events, accurately matching the measured distributions. Since PYTHIA 8 does not account for jet quenching, this suggests that jet quenching is not the dominant factor causing the observed broadening in the data.

Moreover, the count of emissions emitted by the charm quark that meet the soft drop criterion, represented as $n_{SD}$, is determined by examining all branch splittings involving the \dzero meson. This analysis reveals a significant correlation with the number of perturbative emissions originating directly from the charm quark.

The initial measurement, correlating with the charm-quark splitting function, is depicted through the $n_{SD}$ distribution illustrated in Fig.~\ref{HFJet}. This distribution pertains to charm jets identified by a prompt \dzero{} meson within the transverse momentum range 15 $\leq \pTjet <$ 30 $\GeVc$~\cite{jet7}. 
Compared to inclusive jets, the $n_{SD}$ distribution for charm-tagged jets exhibits a noticeable shift towards smaller values. This observation suggests that, on average, charm quarks emit fewer gluons with sufficiently high $\pt$ to satisfy the Soft Drop condition during the showering process, in contrast to light and massless partons. This behavior aligns with expectations arising from the presence of a "dead cone" effect specific to charm quarks, which leads to a steeper fragmentation pattern for charm quarks relative to light quarks and gluons.

The observable utilized to detect the dead cone involves forming the ratio of the splitting angle ($\theta$) distributions between jets tagged with D$^{0}$-mesons and inclusive jets, grouped into bins of $E_{\rm{Radiator}}$. Expressing this ratio in terms of the logarithm of the inverse of the angle is appropriate, as at leading order, the QCD likelihood for a parton to split is proportional to \(\ln(1/\theta)\ln(k_{\rm{T}})\).

Figure~\ref{DeadCone} presents measurements of $R(\theta)$ in three intervals of radiator energy associated with charm quarks: $5 < E_{\rm{Radiator}} < 10$ GeV, $10 < E_{\rm{Radiator}} < 20$ GeV, and $20 < E_{\rm{Radiator}} < 35$ GeV~\cite{jet8}. A pronounced reduction in the occurrence of small-angle splittings is evident in jets tagged with D$^{0}$ mesons compared to the general jet sample.
Each plot includes a baseline corresponding to scenarios without a dead cone for every Monte Carlo generator (dashed lines). The discrepancy between the measured data points and the no dead-cone limit directly illustrates the presence of a dead cone, wherein emissions from charm quarks are suppressed. This suppression becomes more pronounced with lower radiator energies, consistent with the anticipated inverse relationship between the dead-cone angle and radiator energy.

\section{ALICE detector upgrade and future plan}\label{sec.III}

\begin{figure}[!htb]
\includegraphics[width=\linewidth]{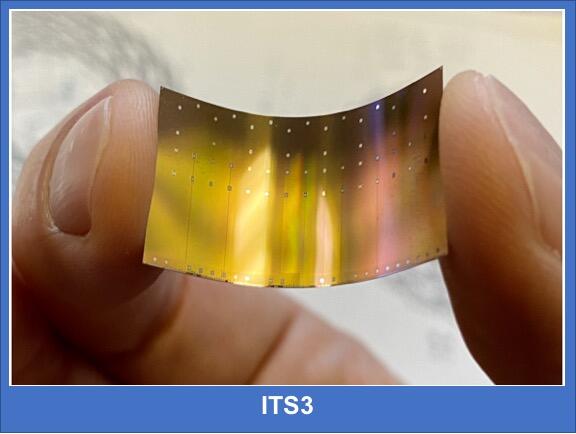}
\caption{A new bent wafer-scale ultra-thin monolithic pixel detector ITS3 is planned to be installed during the LHC LS3.}
\label{fig:its3}
\end{figure}

\begin{figure}[!htb]
\includegraphics[width=\linewidth]{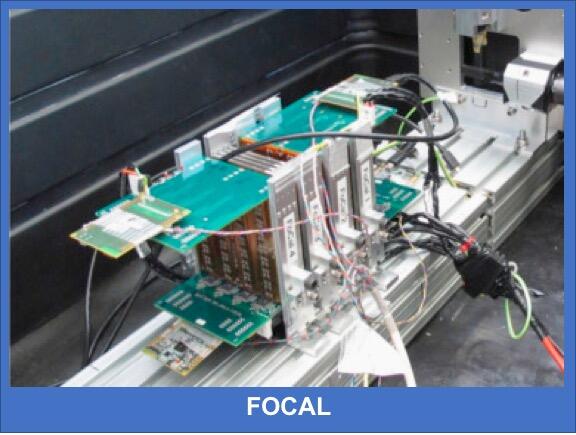}
\caption{The FoCal is a highly granular electromagnetic calorimeter combined with a conventional sampling hadronic calorimeter covering forward pseudorapidities.}
\label{fig:focal}
\end{figure}

Two detector upgrades of ALICE have recently been approved for Run 4 to be installed during the Long Shutdown 3 of the LHC, to further enhance the physics reach of the experiment. The first one is an upgrade of the innermost three layers of the Inner Tracking System (ITS3)~\cite{upgrade1}, and the second is to build a new forward calorimeter (FoCal)~\cite{upgrade2} optimized for direct photon detection in the forward direction of the ALICE detector.

The current upgraded Inner Tracking System (ITS2) is the largest pixel detector to date, with 10 m$^2$ of active silicon area and about 12.5 billion pixels. In the new Inner Tracking System (ITS3), the three inner-most layers will be replaced with ultra-thin wafer-scale silicon sensors of up to 10 cm $\times$ 26 cm, built using a novel stitching technology. The sensors will be thinned to about 50 $\mu$m and bent into truly cylindrical detector layers, which are held in place with carbon foam and cooled by a forced airflow. No further mechanical support or electrical connection (circuit board) are necessary within the active area, resulting in a material budget of 0.07\%X$_0$ per layer. These true cylinders will also allow the innermost layer to be placed closer to the interaction point at 19 mm, which is only 2 mm away from the beam pipe.

Due to the reduction of the material budget and its proximity to the interaction point, the ITS3 detector will provide unprecedented spatial resolution, improved by a factor 2 compared with that of the ITS2, and a higher reconstruction efficiency of low-momentum particle tracks. These features will significantly improve the physics performance of the ALICE detector for measurements of heavy flavor hadrons via the reconstruction of their decay topologies and those of dileptons.

The gain in performance~\cite{upgrade3} will allow for the first measurements of B$^0_{\rm s}$ and $\Lambda^0_{\rm b}$ at low transverse momenta, as well as non-prompt D$^+_{\rm s}$ and $\Xi^+_{\rm c}$ decays in heavy-ion collisions at LHC energies. The first study on the existence of the exotic c-deuteron might be in reach, and the precision of hyper-nuclei studies will also be improved. In addition, it allows to drastically reduce the background of electrons from photon conversions in the material and semi-leptonic charm-hadron decays in the study of low-mass di-electrons. Precise measurements of these observables will provide crucial information about the properties of the quark–gluon plasma formed in heavy-ion collisions.

The FoCal detector consists of an electromagnetic calorimeter (FoCal-E) and a hadronic calorimeter (FoCal-H), covering pseudo-rapidity of 3.2 $< \eta <$ 5.8, as shown in Fig.~\ref{fig:focal}. FoCal-E is a highly granular Si+W calorimeter composed of 18 layers of silicon pad sensors, each as small as 1×1 cm$^2$, and two additional silicon pixel layers with pixel size of 30×30 $\mu$m$^2$. FoCal-H is made of copper capillary tubes and scintillating fibres.

FoCal has unique capabilities~\cite{upgrade4} to measure direct photon production at forward rapidity to probe the gluon distribution in protons and nuclei at small-$x$. Furthermore, FoCal will enable to carry out inclusive and correlation measurements of photons, neutral mesons, and jets in hadronic $pp$ and $p$--Pb collisions, as well as J/$\psi$ production in ultra-peripheral p–Pb and Pb–Pb collisions, and hence significantly enhances the scope of the ALICE physics program to explore the dynamics of hadronic matter and the nature of QCD evolution at small $x$, down to $x\sim 10^{-6}$.

The ITS3 and FoCal projects have reached the important milestone of completing the Technical Design Reports~\cite{upgrade5, upgrade6}, which were endorsed by the CERN review committees in March 2024. The construction phase of ITS3 and FoCal starts now, with the detectors to be installed in early 2028 and ready for data taking in 2029.

A completely new detector, named ‘ALICE 3’, is proposed~\cite{upgrade7} for the LHC Runs 5 and 6, to enable new measurements in the heavy-flavor sector, including studies of multi-charm baryon production and the interaction potentials between heavy mesons via femtoscopy, as well as precise multi-differential measurements of di-electron emission to probe the chiral-symmetry restoration mechanism and to study the time-evolution of the QGP temperature.

ALICE 3 detector consists of a tracking system with unique pointing resolution over a large pseudo-rapidity range ($-4 < \eta < 4$), complemented by multiple sub-detector systems for particle identification, including silicon time-of-flight layers with about 20 ps resolution, a ring-imaging Cherenkov detector with high-resolution readout, a muon identification system and an electromagnetic calorimeter. To achieve an unprecedented pointing resolution at midrapidity in both the transverse and longitudinal directions, the innermost layers are constructed from wafer-scale ultra-thin silicon sensors, bent into cylinders with minimal supporting material, similarly to the ITS 3, and will be positioned inside the beam pipe as close as possible to the interaction point, on a retractable structure to leave sufficient aperture for the beams at injection energy. In the proposed apparatus with a solenoid magnetic field of $B$ = 2 T, the tracker with barrel and endcap silicon pixel layers provides a relative momentum resolution of 1-2\% over a large acceptance by measuring about 10 space points. Intensive R\&D programs are being pursued by the ALICE Collaboration to push current technological limits of silicon sensors for tracking, timing, and photon detection. 

The proposed ALICE 3 detector is conceived for studies of $pp$, $p$-A and A-A collisions at luminosities a factor of 20 to 50 times higher than possible with the current upgraded ALICE detector, enabling a rich physics program ranging from measurements with electromagnetic probes at ultra-low transverse momenta to precision physics in the charm and beauty sector.

\section{Summary} \label{sec.IV}
Since the commencement of LHC collisions in 2009, the ALICE detector has executed a highly successful data collection program. The experiment is specifically designed to investigate QCD at the LHC, utilizing the highest collision energies achievable in laboratory settings. The primary objective is to probe many-body QCD interactions at extreme temperatures by studying the formation of the QGP in heavy-ion collisions. ALICE measurements indicate that heavy-ion collisions at the LHC create conditions that far exceed those required for the formation of the QGP. The QGP formed at LHC energies have been demonstrated to undergo the most rapid expansion ever observed for a many-body system in the laboratory. ALICE has also provided an extensive mapping of hadro-chemistry in heavy-ion collisions at LHC energies, observed the energy loss of energetic partons in the presence of the QGP and the modification of their showers, revealed significant modifications in quarkonium binding within the QGP, and discovered QGP-like signatures in high-multiplicity small system collisions.
These observations underscore the substantial progress made during the period of LHC data collection. This progress has been achieved through collaborative efforts with other LHC experiments, alongside advancements in luminosity and center-of-mass energy coverage within the RHIC program. 
Looking forward, with the help of enhanced detector capabilities and increased luminosity at the LHC, ALICE is poised to unravel further mysteries of strong interactions and the properties of QCD matter in extreme conditions.

\section*{ACKNOWLEDGMENTS}
We are grateful to the ALICE Collaboration at LHC. This work is supported in part by the National Key Research and Development Program of China (Nos. 2018YFE0104600, 2018YFE0104700, 2018YFE0104800, 2018YFE0104900, 12147101, 12322508), the National Natural Science Foundation of China (No. 12061141008), the Strategic Priority Research Program of Chinese Academy of Sciences (No. XDB34000000) and the Science and Technology Commission of Shanghai Municipality (23590780100).

\vspace{5mm}
\textit{Postscript}:

This review is dedicated to Professor Wenqing Shen on the occasion of his 80th Birthday. Elected as an academician of the Chinese Academy of Sciences in 1999, Prof. Shen has held numerous academic leadership roles, including President of the Shanghai Branch of the Chinese Academy of Sciences, Deputy Director of the National Natural Science Foundation of China, and Chairman of the Shanghai Association of Science and Technology. Under his leadership, the Chinese nuclear physics community has expanded from traditional low-energy nuclear physics to high-energy nuclear physics, including activities in ALICE. In the two decades since 2000, the Chinese high-energy nuclear physics group has grown into one of the most vibrant international communities, attaining significant breakthroughs in the study of QCD matter, some of which are highlighted in this article. We also reflect on the importance of international collaborations in bringing together individuals from diverse cultural backgrounds and how teamwork can accomplish scientific objectives beyond the capabilities of individuals working alone, a principle that Prof. Shen has consistently advocated.

We hold great admiration for Prof. Shen’s outstanding leadership, far-sightedness, and substantial impact on next generations of Chinese nuclear physicists.

\bibliographystyle{utphys}
\bibliography{nst_sample_BiBCombined}

\providecommand{\href}[2]{#2}\begingroup\raggedright\begin{thebibliography}{100}

\bibitem{wilczek_quantum_1999}
F.~Wilczek, ``Quantum field theory'', \href{http://dx.doi.org/10.1103/RevModPhys.71.S85}{{\em Rev. Mod. Phys.} {\bfseries 71} (Mar., 1999) S85--S95}. \url{https://link.aps.org/doi/10.1103/RevModPhys.71.S85}. Publisher: American Physical Society.

\bibitem{gross_ultraviolet_1973}
D.~J. Gross and F.~Wilczek, ``Ultraviolet {Behavior} of {Non}-{Abelian} {Gauge} {Theories}'', \href{http://dx.doi.org/10.1103/PhysRevLett.30.1343}{{\em Phys. Rev. Lett.} {\bfseries 30} (June, 1973) 1343--1346}. \url{https://link.aps.org/doi/10.1103/PhysRevLett.30.1343}. Publisher: American Physical Society.

\bibitem{politzer_reliable_1973}
H.~D. Politzer, ``Reliable {Perturbative} {Results} for {Strong} {Interactions}?'', \href{http://dx.doi.org/10.1103/PhysRevLett.30.1346}{{\em Phys. Rev. Lett.} {\bfseries 30} (June, 1973) 1346--1349}. \url{https://link.aps.org/doi/10.1103/PhysRevLett.30.1346}. Publisher: American Physical Society.

\bibitem{Liang:2004ph}
Z.-T. Liang and X.-N. Wang, ``{Globally polarized quark-gluon plasma in non-central A+A collisions}'', \href{http://dx.doi.org/10.1103/PhysRevLett.94.102301}{{\em Phys. Rev. Lett.} {\bfseries 94} (2005) 102301}, \href{http://arxiv.org/abs/nucl-th/0410079}{{\ttfamily arXiv:nucl-th/0410079}}. [Erratum: Phys.Rev.Lett. 96, 039901 (2006)].

\bibitem{STAR:2017ckg}
{\bfseries STAR} Collaboration, L.~Adamczyk {\em et~al.}, ``{Global $\Lambda$ hyperon polarization in nuclear collisions: evidence for the most vortical fluid}'', \href{http://dx.doi.org/10.1038/nature23004}{{\em Nature} {\bfseries 548} (2017) 62--65}, \href{http://arxiv.org/abs/1701.06657}{{\ttfamily arXiv:1701.06657 [nucl-ex]}}.

\bibitem{STAR:2022fan}
{\bfseries STAR} Collaboration, M.~S. Abdallah {\em et~al.}, ``{Pattern of global spin alignment of \ensuremath{\phi} and K$^{*0}$ mesons in heavy-ion collisions}'', \href{http://dx.doi.org/10.1038/s41586-022-05557-5}{{\em Nature} {\bfseries 614} (2023) 244--248}, \href{http://arxiv.org/abs/2204.02302}{{\ttfamily arXiv:2204.02302 [hep-ph]}}.

\bibitem{Chen:2023hnb}
J.~Chen, Z.-T. Liang, Y.-G. Ma, and Q.~Wang, ``{Global spin alignment of vector mesons and strong force fields in heavy-ion collisions}'', \href{http://dx.doi.org/10.1016/j.scib.2023.04.001}{{\em Sci. Bull.} {\bfseries 68} (2023) 874--877}, \href{http://arxiv.org/abs/2305.09114}{{\ttfamily arXiv:2305.09114 [nucl-th]}}.

\bibitem{Luo:2017faz}
X.~Luo and N.~Xu, ``{Search for the QCD Critical Point with Fluctuations of Conserved Quantities in Relativistic Heavy-Ion Collisions at RHIC : An Overview}'', \href{http://dx.doi.org/10.1007/s41365-017-0257-0}{{\em Nucl. Sci. Tech.} {\bfseries 28} (2017) 112}, \href{http://arxiv.org/abs/1701.02105}{{\ttfamily arXiv:1701.02105 [nucl-ex]}}.

\bibitem{Deng2024}
X.~G. Deng, D.~Q. Fang, and Y.~G. Ma, ``Shear viscosity of nucleonic matter'', \href{http://dx.doi.org/10.1016/j.ppnp.2023.104095}{{\em Prog. Part. Nucl. Phys.} {\bfseries 136} (2024) 104095}.

\bibitem{He-NST}
W.~He, Y.~Ma, L.~Pang, H.~Song, and K.~Zhou, ``High‐energy nuclear physics meets machine learning'', \href{http://dx.doi.org/10.1007/s41365-023-01233-z}{{\em Nucl. Sci. Tech.} {\bfseries 34} (2023) 88}.

\bibitem{Ma-CPL}
Y.-G. Ma, L.-G. Pang, R.~Wang, and K.~Zhou, ``Phase transition study meets machine learning'', \href{http://dx.doi.org/10.1088/0256-307X/40/12/122101}{{\em Chin. Phys. Lett.} {\bfseries 40} (2023) 122101}.

\bibitem{SunKJ_NC}
K.-J. Sun, R.~Wang, C.~M. Ko, Y.-G. Ma, and C.~Shen, ``Unveiling the dynamics of little-bang nucleosynthesis'', \href{http://dx.doi.org/10.1038/s41467-024-45474-x}{{\em Nat. Commun.} {\bfseries 15} (2024) 1074}.

\bibitem{Chen-NST}
J.~H. Chen, X.~Dong, X.~H. He, {\em et~al.}, ``Properties of the qcd matter: An experimental review of selected results from rhic bes program'', {\em Nucl. Sci. Tech.} (2024) . this issue.

\bibitem{collaboration_alice_2008}
{\bfseries ALICE} Collaboration, ``The {ALICE} experiment at the {CERN} {LHC}'', \href{http://dx.doi.org/10.1088/1748-0221/3/08/s08002}{{\em Journal of Instrumentation} {\bfseries 3} (Aug., 2008) S08002--S08002}. \url{https://doi.org/10.1088%2F1748-0221%2F3%2F08%2Fs08002}. Publisher: IOP Publishing.

\bibitem{ALICEIntro}
{\bfseries ALICE} Collaboration, ``The {ALICE} experiment -- a journey through {QCD}'', 2022.
\newblock \url{https://arxiv.org/abs/2211.04384}.

\bibitem{PhysRevLett.116.222302}
{\bfseries ALICE} Collaboration, J.~Adam {\em et~al.}, ``Centrality dependence of the charged-particle multiplicity density at midrapidity in pb-pb collisions at $\sqrt{{s}_{NN}}=5.02\text{ }\text{ }\mathrm{TeV}$'', \href{http://dx.doi.org/10.1103/PhysRevLett.116.222302}{{\em Phys. Rev. Lett.} {\bfseries 116} (Jun, 2016) 222302}. \url{https://link.aps.org/doi/10.1103/PhysRevLett.116.222302}.

\bibitem{PhysRevC.83.024913}
B.~Alver {\em et~al.}, ``Charged-particle multiplicity and pseudorapidity distributions measured with the phobos detector in $\text{Au}+\text{Au}$, $\text{Cu}+\text{Cu}$, $\mathbf{d}+\text{Au}$, and $\mathbf{p}+\mathbf{p}$ collisions at ultrarelativistic energies'', \href{http://dx.doi.org/10.1103/PhysRevC.83.024913}{{\em Phys. Rev. C} {\bfseries 83} (Feb, 2011) 024913}. \url{https://link.aps.org/doi/10.1103/PhysRevC.83.024913}.

\bibitem{Bierlich2016}
C.~Bierlich, G.~Gustafson, and L.~L{\"o}nnblad, ``Diffractive and non-diffractive wounded nucleons and final states in pa collisions'', \href{http://dx.doi.org/10.1007/JHEP10(2016)139}{{\em Journal of High Energy Physics} {\bfseries 2016} (Oct, 2016) 139}. \url{https://doi.org/10.1007/JHEP10(2016)139}.

\bibitem{NTQCD2023_01}
Y.~Zhang, D.~Zhang, and X.~Luo, ``Experimental study of the {QCD} phase diagram in relativistic heavy-ion collisions'', \href{http://dx.doi.org/10.11889/j.0253-3219.2023.hjs.46.040001}{{\em Nuclear Techniques} {\bfseries 46} (2023) 040001}.

\bibitem{NTQCD2023_04}
S.~Wu and H.~Song, ``Critical dynamical fluctuations near the {QCD} critical point'', \href{http://dx.doi.org/10.11889/j.0253-3219.2023.hjs.46.040004}{{\em Nuclear Techniques} {\bfseries 46} (2023) 040004}.

\bibitem{NTQCD2023_05}
K.~Xu and M.~Huang, ``{QCD} critical end point and baryon number fluctuation'', \href{http://dx.doi.org/10.11889/j.0253-3219.2023.hjs.46.040005}{{\em Nuclear Techniques} {\bfseries 46} (2023) 040005}.

\bibitem{NTQCD2023_06}
Y.~Wu, X.~Li, and L.~Chen, ``Several problems in determining the {QCD} phase boundary by relativistic heavy ion collisions'', \href{http://dx.doi.org/10.11889/j.0253-3219.2023.hjs.46.040006}{{\em Nuclear Techniques} {\bfseries 46} (2023) 040006}.

\bibitem{NTQCD2023_07}
Z.~Zhu, Y.~Zhao, and D.~Hou, ``{QCD} phase structure from holographic models'', \href{http://dx.doi.org/10.11889/j.0253-3219.2023.hjs.46.040007}{{\em Nucl. Tech.} {\bfseries 46} (2023) 040007}.

\bibitem{NTQCD2023_12}
K.~Sun, L.~Chen, and C.~M. Ko, ``Light nuclei production and {QCD} phase transition in heavy-ion collisions'', \href{http://dx.doi.org/10.11889/j.0253-3219.2023.hjs.46.040012}{{\em Nucl. Tech.} {\bfseries 46} (2023) 040012}.

\bibitem{NTQCD2023_13}
Q.~Chen, G.~Ma, and J.~Chen, ``Transport model study of conserved charge fluctuations and {QCD} phase transition in heavy-ion collisions'', \href{http://dx.doi.org/10.11889/j.0253-3219.2023.hjs.46.040013}{{\em Nucl. Tech.} {\bfseries 46} (2023) 040013}.

\bibitem{ko_searching_2023}
C.~M. Ko, ``Searching for {QCD} critical point with light nuclei'', \href{http://dx.doi.org/10.1007/s41365-023-01231-1}{{\em Nucl. Sci. Tech.} {\bfseries 34} (May, 2023) 80, s41365--023--01231--1}. \url{https://link.springer.com/10.1007/s41365-023-01231-1}.

\bibitem{PhysRevD.27.140}
J.~D. Bjorken, ``Highly relativistic nucleus-nucleus collisions: The central rapidity region'', \href{http://dx.doi.org/10.1103/PhysRevD.27.140}{{\em Phys. Rev. D} {\bfseries 27} (Jan, 1983) 140--151}. \url{https://link.aps.org/doi/10.1103/PhysRevD.27.140}.

\bibitem{Adam:2016ddh}
{\bfseries ALICE} Collaboration, J.~Adam {\em et~al.}, ``{Centrality dependence of the pseudorapidity density distribution for charged particles in Pb-Pb collisions at $\sqrt{s_{\rm NN}}=5.02$ TeV}'', \href{http://dx.doi.org/10.1016/j.physletb.2017.07.017}{{\em Phys. Lett. B} {\bfseries 772} (2017) 567--577}, \href{http://arxiv.org/abs/1612.08966}{{\ttfamily arXiv:1612.08966 [nucl-ex]}}.

\bibitem{Abbas:2013bpa}
{\bfseries ALICE} Collaboration, E.~Abbas {\em et~al.}, ``{Centrality dependence of the pseudorapidity density distribution for charged particles in Pb-Pb collisions at $\sqrt{s_{\rm{NN}}}=2.76$ TeV}'',
{\em Phys. Lett.} {\bfseries B726} (2013) 610--622.

\bibitem{ALICE:2022imr}
{\bfseries ALICE} Collaboration, ``{System-size dependence of the charged-particle pseudorapidity density at $\sqrt{s_{\rm NN}} = 5.02$ TeV for pp, p-Pb, and Pb-Pb collisions}'', \href{http://arxiv.org/abs/2204.10210}{{\ttfamily arXiv:2204.10210 [nucl-ex]}}.

\bibitem{Loizides:2014vua}
C.~Loizides, J.~Nagle, and P.~Steinberg, ``{Improved version of the PHOBOS Glauber Monte Carlo}'', \href{http://dx.doi.org/10.1016/j.softx.2015.05.001}{{\em SoftwareX} {\bfseries 1-2} (2015) 13--18}, \href{http://arxiv.org/abs/1408.2549}{{\ttfamily arXiv:1408.2549 [nucl-ex]}}.

\bibitem{Braun_Munzinger_2004}
P.~Braun-Munzinger, K.~Redlich, and J.~Stachel, {\em PARTICLE PRODUCTION IN HEAVY ION COLLISIONS}, \href{http://dx.doi.org/10.1142/9789812795533_0008}{p.~491–599}.
\newblock WORLD SCIENTIFIC, Jan., 2004.
\newblock \url{http://dx.doi.org/10.1142/9789812795533_0008}.

\bibitem{BLWPhysRevC.48.2462}
E.~Schnedermann, J.~Sollfrank, and U.~Heinz, ``Thermal phenomenology of hadrons from 200a gev s+s collisions'', \href{http://dx.doi.org/10.1103/PhysRevC.48.2462}{{\em Phys. Rev. C} {\bfseries 48} (Nov, 1993) 2462--2475}. \url{https://link.aps.org/doi/10.1103/PhysRevC.48.2462}.

\bibitem{alicealpha2024}
{\bfseries ALICE} Collaboration, ``Measurement of (anti)alpha production in central pb-pb collisions at $\sqrt{s_{\rm NN}}$ = 5.02 tev'', 2023.
\newblock \url{https://arxiv.org/abs/2311.11758}.

\bibitem{aliceH3Lproductionpbpb}
``Measurement of ${}_{\Lambda}^{3}\mathrm{H}$ production in {Pb-Pb} collisions at $\sqrt{s_{\mathrm{NN}}}$ = 5.02 tev'', 2024.
\newblock \url{https://arxiv.org/abs/2405.19839}.

\bibitem{chen_measurements_2023}
J.~Chen, X.~Dong, Y.-G. Ma, and Z.~Xu, ``Measurements of the lightest hypernucleus: progress and perspective'', \href{http://dx.doi.org/10.1016/j.scib.2023.11.045}{{\em Science Bulletin} {\bfseries 68} (Dec., 2023) 3252--3260}. \url{https://www.sciencedirect.com/science/article/pii/S2095927323008149}.

\bibitem{ALICE:2019hno}
{\bfseries ALICE} Collaboration, S.~Acharya {\em et~al.}, ``{Production of charged pions, kaons, and (anti-)protons in Pb--Pb and inelastic pp collisions at $\sqrt {s_{\rm NN}}$ = 5.02 TeV}'', \href{http://dx.doi.org/10.1103/PhysRevC.101.044907}{{\em Phys. Rev. C} {\bfseries 101} (2020) 044907}, \href{http://arxiv.org/abs/1910.07678}{{\ttfamily arXiv:1910.07678 [nucl-ex]}}.

\bibitem{Nuclei_5TeV}
{\bfseries ALICE} Collaboration, S.~Acharya {\em et~al.}, ``{Light (anti)nuclei production in {Pb--Pb} collisions at $\sqrt{s_{\mathrm {NN}}}=5.02$~TeV}'', \href{http://dx.doi.org/10.1103/PhysRevC.107.064904}{{\em Phys. Rev. C} {\bfseries 107} (2023) 064904}, \href{http://arxiv.org/abs/2211.14015}{{\ttfamily arXiv:2211.14015 [nucl-ex]}}.

\bibitem{Matsui:1986dk}
T.~Matsui and H.~Satz, ``{$J\psi$ Suppression by Quark-Gluon Plasma Formation}'',
\href{http://dx.doi.org/10.1016/0370-2693(86)91404-8}{{\em Phys. Lett.} {\bfseries B178} (1986) 416--422}.

\bibitem{Adam:2015lda}
{\bfseries ALICE} Collaboration, J.~Adam {\em et~al.}, ``{Direct photon production in Pb-Pb collisions at $\sqrt{s_{\rm NN}} = 2.76$ TeV}'', \href{http://dx.doi.org/10.1016/j.physletb.2016.01.020}{{\em Phys. Lett. B} {\bfseries 754} (2016) 235--248}, \href{http://arxiv.org/abs/1509.07324}{{\ttfamily arXiv:1509.07324 [nucl-ex]}}.

\bibitem{Adler:2005}
S.~S. Adler {\em et~al.}, ``{Centrality Dependence of Direct Photon Production in $\sqrt{s_{NN}}=200~$GeV Au+Au Collisions}'', \href{http://dx.doi.org/10.1103/physrevlett.94.232301}{{\em Physical Review Letters} {\bfseries 94} (Jun, 2005) }. \url{http://dx.doi.org/10.1103/PhysRevLett.94.232301}.

\bibitem{Adare:2008ab}
{\bfseries PHENIX} Collaboration, A.~Adare {\em et~al.}, ``{Enhanced production of direct photons in Au+Au collisions at $\sqrt{s_{NN}}=200$ GeV and implications for the initial temperature}'', \href{http://dx.doi.org/10.1103/PhysRevLett.104.132301}{{\em Phys. Rev. Lett.} {\bfseries 104} (2010) 132301}, \href{http://arxiv.org/abs/0804.4168}{{\ttfamily arXiv:0804.4168 [nucl-ex]}}.

\bibitem{Aamodt:2011mr}
{\bfseries ALICE} Collaboration, K.~Aamodt {\em et~al.}, ``{Two-pion Bose-Einstein correlations in central Pb-Pb collisions at $\sqrt{s_{\rm NN}}$ = 2.76 TeV}'', \href{http://dx.doi.org/10.1016/j.physletb.2010.12.053}{{\em Phys. Lett.} {\bfseries B696} (2011) 328--337},
\href{http://arxiv.org/abs/1012.4035}{{\ttfamily arXiv:1012.4035 [nucl-ex]}}.

\bibitem{Adam:2015vna}
{\bfseries ALICE} Collaboration, J.~Adam {\em et~al.}, ``{Centrality dependence of pion freeze-out radii in Pb-Pb collisions at $\sqrt{s}_{NN}=$ 2.76 TeV}'', \href{http://dx.doi.org/10.1103/PhysRevC.93.024905}{{\em Phys. Rev. C} {\bfseries 93} (2016) 024905}, \href{http://arxiv.org/abs/1507.06842}{{\ttfamily arXiv:1507.06842 [nucl-ex]}}.

\bibitem{vanHees:2011vb}
H.~van Hees, C.~Gale, and R.~Rapp, ``{Thermal Photons and Collective Flow at the Relativistic Heavy-Ion Collider}'', \href{http://dx.doi.org/10.1103/PhysRevC.84.054906}{{\em Phys. Rev. C} {\bfseries 84} (2011) 054906}, \href{http://arxiv.org/abs/1108.2131}{{\ttfamily arXiv:1108.2131 [hep-ph]}}.

\bibitem{Paquet:2015lta}
J.-F. Paquet, C.~Shen, G.~S. Denicol, M.~Luzum, B.~Schenke, S.~Jeon, and C.~Gale, ``{Production of photons in relativistic heavy-ion collisions}'', \href{http://dx.doi.org/10.1103/PhysRevC.93.044906}{{\em Phys. Rev. C} {\bfseries 93} (2016) 044906}, \href{http://arxiv.org/abs/1509.06738}{{\ttfamily arXiv:1509.06738 [hep-ph]}}.

\bibitem{Aurenche_2006}
P.~Aurenche, J.~P. Guillet, E.~Pilon, M.~Werlen, and M.~Fontannaz, ``Recent critical study of photon production in hadronic collisions'', \href{http://dx.doi.org/10.1103/physrevd.73.094007}{{\em Physical Review D} {\bfseries 73} (May, 2006) }. \url{http://dx.doi.org/10.1103/PhysRevD.73.094007}.

\bibitem{HanburyBrown:1954amm}
R.~Hanbury~Brown and R.~Q. Twiss, ``{A New type of interferometer for use in radio astronomy}'', \href{http://dx.doi.org/10.1080/14786440708520475}{{\em Phil. Mag. Ser. 7} {\bfseries 45} (1954) 663--682}.

\bibitem{HanburyBrown:1956bqd}
R.~Hanbury~Brown and R.~Q. Twiss, ``{A Test of a new type of stellar interferometer on Sirius}'', \href{http://dx.doi.org/10.1038/1781046a0}{{\em Nature} {\bfseries 178} (1956) 1046--1048}.

\bibitem{Akkelin:1995gh}
S.~Akkelin and Y.~Sinyukov, ``{The HBT interferometry of expanding sources}'',
\href{http://dx.doi.org/10.1016/0370-2693(95)00765-D}{{\em Phys. Lett.} {\bfseries B356} (1995) 525--530}.

\bibitem{bibcf4}
{\bfseries ALICE} Collaboration, S.~Acharya {\em et~al.}, ``Investigations of anisotropic flow using multiparticle azimuthal correlations in $pp$, $p\text{\ensuremath{-}}\mathrm{Pb}$, xe-xe, and pb-pb collisions at the lhc'', \href{http://dx.doi.org/10.1103/PhysRevLett.123.142301}{{\em Phys. Rev. Lett.} {\bfseries 123} (Oct, 2019) 142301}. \url{https://link.aps.org/doi/10.1103/PhysRevLett.123.142301}.

\bibitem{bibcf1}
{\bfseries ALICE} Collaboration, S.~Acharya {\em et~al.}, ``Anisotropic flow and flow fluctuations of identified hadrons in pb--pb collisions at $\sqrt{s_{NN}}$= 5.02 tev'', \href{http://dx.doi.org/10.1007/JHEP05(2023)243}{{\em Journal of High Energy Physics} {\bfseries 2023} (May, 2023) 243}. \url{https://doi.org/10.1007/JHEP05(2023)243}.

\bibitem{bibcf2}
{\bfseries ALICE} Collaboration, J.~Adam {\em et~al.}, ``Correlated event-by-event fluctuations of flow harmonics in pb-pb collisions at $\sqrt{{s}_{NN}}=2.76\text{ }\text{ }\mathrm{TeV}$'', \href{http://dx.doi.org/10.1103/PhysRevLett.117.182301}{{\em Phys. Rev. Lett.} {\bfseries 117} (Oct, 2016) 182301}. \url{https://link.aps.org/doi/10.1103/PhysRevLett.117.182301}.

\bibitem{bibcf3}
{\bfseries ALICE} Collaboration, S.~Acharya {\em et~al.}, ``Multiharmonic correlations of different flow amplitudes in pb-pb collisions at $\sqrt{{s}_{\mathrm{NN}}}=2.76\text{ }\text{ }\mathrm{TeV}$'', \href{http://dx.doi.org/10.1103/PhysRevLett.127.092302}{{\em Phys. Rev. Lett.} {\bfseries 127} (Aug, 2021) 092302}. \url{https://link.aps.org/doi/10.1103/PhysRevLett.127.092302}.

\bibitem{bibcf5}
{Z. Moravcova (for the ALICE Collaboration)}, ``Observation of partonic ﬂow in small collision systems with {ALICE} at the {LHC}'', 2022.
\newblock {talk given at QM2022, https://indico.cern.ch/event/895086/contributions/4736573/}.

\bibitem{bibcf6}
{W. Wu (for the ALICE Collaboration)}, ``Probing partonic collectivity in pp and p--pb collisions with {ALICE}'', 2022.
\newblock {talk given at QM2022, https://indico.cern.ch/event/1043736/contributions/5363771/}.

\bibitem{bibcf7}
{\bfseries ALICE} Collaboration, S.~Acharya {\em et~al.}, ``Investigations of anisotropic flow using multiparticle azimuthal correlations in $pp$, $p\text{\ensuremath{-}}\mathrm{Pb}$, xe-xe, and pb-pb collisions at the lhc'', \href{http://dx.doi.org/10.1103/PhysRevLett.123.142301}{{\em Phys. Rev. Lett.} {\bfseries 123} (Oct, 2019) 142301}. \url{https://link.aps.org/doi/10.1103/PhysRevLett.123.142301}.

\bibitem{bibcf8}
{\bfseries ALICE} Collaboration, S.~Acharya {\em et~al.}, ``Emergence of long-range angular correlations in low-multiplicity proton-proton collisions'', \href{http://dx.doi.org/10.1103/PhysRevLett.132.172302}{{\em Phys. Rev. Lett.} {\bfseries 132} (Apr, 2024) 172302}. \url{https://link.aps.org/doi/10.1103/PhysRevLett.132.172302}.

\bibitem{bibcf90}
{\bfseries ALICE} Collaboration, S.~Acharya {\em et~al.}, ``Anisotropic flow in xe--xe collisions at snn=5.44 tev'', \href{http://dx.doi.org/https://doi.org/10.1016/j.physletb.2018.06.059}{{\em Physics Letters B} {\bfseries 784} (2018) 82--95}. \url{https://www.sciencedirect.com/science/article/pii/S037026931830515X}.

\bibitem{bibcf9}
{\bfseries ALICE} Collaboration, S.~Acharya {\em et~al.}, ``Characterizing the initial conditions of heavy-ion collisions at the lhc with mean transverse momentum and anisotropic flow correlations'', \href{http://dx.doi.org/https://doi.org/10.1016/j.physletb.2022.137393}{{\em Physics Letters B} {\bfseries 834} (2022) 137393}. \url{https://www.sciencedirect.com/science/article/pii/S0370269322005275}.

\bibitem{bibcf91}
{Z. Lu (for the ALICE Collaboration)}, ``Probing the nuclear structure with flow observables in {ALICE}'', 2023.
\newblock {talk given at IS2023, https://indico.cern.ch/event/1043736/contributions/5363770/}.

\bibitem{bibcf10}
{\bfseries ALICE} Collaboration, S.~Acharya {\em et~al.}, ``Probing the effects of strong electromagnetic fields with charge-dependent directed flow in pb-pb collisions at the lhc'', \href{http://dx.doi.org/10.1103/PhysRevLett.125.022301}{{\em Phys. Rev. Lett.} {\bfseries 125} (Jul, 2020) 022301}. \url{https://link.aps.org/doi/10.1103/PhysRevLett.125.022301}.

\bibitem{zhao_electromagnetic_2024}
J.~Zhao, J.-H. Chen, X.-G. Huang, and Y.-G. Ma, ``Electromagnetic fields in ultra-peripheral relativistic heavy-ion collisions'', \href{http://dx.doi.org/10.1007/s41365-024-01374-9}{{\em Nucl. Sci. Tech.} {\bfseries 35} (Feb., 2024) 20}. \url{https://doi.org/10.1007/s41365-024-01374-9}.

\bibitem{APS2023_072504}
Z.-F. Jiang, X.-Y. Wu, H.-Q. Yu, S.-S. Cao, and B.-W. Zhang, ``The direct flow of charged particles and the global polarization of hyperons in 200 {AGeV} {Au}+{Au} collisions at {RHIC}'', \href{http://dx.doi.org/10.7498/aps.72.20222391}{{\em Acta Phys. Sin.} {\bfseries 72} (Apr., 2023) 072504--9}. \url{https://wulixb.iphy.ac.cn/cn/article/doi/10.7498/aps.72.20222391}.

\bibitem{NTQCD2023_08}
H.~Ding, S.~Li, and J.~Liu, ``Progress on {QCD} properties in strong magnetic fields from lattice {QCD}'', \href{http://dx.doi.org/10.11889/j.0253-3219.2023.hjs.46.040008}{{\em Nucl. Tech.} {\bfseries 46} (2023) 040008}.

\bibitem{NTQCD2023_11}
J.~Yin and L.~Jinfeng, ``Phase transitions of strong interaction matter in vorticity fields'', \href{http://dx.doi.org/10.11889/j.0253-3219.2023.hjs.46.040011}{{\em Nucl. Tech.} {\bfseries 46} (2023) 040011}.

\bibitem{APS2023_112502}
X.-L. Zhao, G.-L. Ma, and Y.-G. Ma, ``Electromagnetic field effects and anomalous chiral phenomena in heavy-ion collisions at intermediate and high energy'', \href{http://dx.doi.org/10.7498/aps.72.20230245}{{\em Acta Phys. Sin.} {\bfseries 72} (June, 2023) 112502--21}. \url{https://wulixb.iphy.ac.cn/cn/article/doi/10.7498/aps.72.20230245}.

\bibitem{shou_progress_2023}
Q.-Y. Shou, J.~Zhao, H.-J. Xu, W.~Li, G.~Wang, A.-H. Tang, and F.-Q. Wang, ``Progress on the experimental search for the chiral magnetic effect, the chiral vortical effect, and the chiral magnetic wave'', \href{http://dx.doi.org/10.7498/aps.72.20230109}{{\em Acta Phys. Sin.} {\bfseries 72} (2023) 112504}. \url{https://wulixb.iphy.ac.cn/article/doi/10.7498/aps.72.20230109}.

\bibitem{bibcf15}
W.-Y. Wu, {\em et~al.}, ``Global constraint on the magnitude of anomalous chiral effects in heavy-ion collisions'', \href{http://dx.doi.org/10.1103/PhysRevC.107.L031902}{{\em Phys. Rev. C} {\bfseries 107} (Mar, 2023) L031902}. \url{https://link.aps.org/doi/10.1103/PhysRevC.107.L031902}.

\bibitem{bibcf16}
{C. Wang (for the ALICE Collaboration)}, ``Search for anomalous chiral effects in heavy-ion collisions with {ALICE}'', 2023.
\newblock {talk given at QM2023, https://indico.cern.ch/event/1139644/contributions/5502913/}.

\bibitem{bibcf11}
{\bfseries ALICE} Collaboration, S.~Acharya {\em et~al.}, ``Constraining the chiral magnetic effect with charge-dependent azimuthal correlations in pb-pb collisions at $\sqrt{s_{NN}}$= 2.76 and 5.02 tev'', \href{http://dx.doi.org/10.1007/JHEP09(2020)160}{{\em Journal of High Energy Physics} {\bfseries 2020} (Sep, 2020) 160}. \url{https://doi.org/10.1007/JHEP09(2020)160}.

\bibitem{bibcf12}
{\bfseries ALICE} Collaboration, S.~Acharya {\em et~al.}, ``Search for the chiral magnetic effect with charge-dependent azimuthal correlations in {Xe-Xe} collisions at $\sqrt{s_{\mathrm{NN}}} = 5.44$ tev'', \href{http://dx.doi.org/10.1016/j.physletb.2024.138862}{{\em Physics Letters B} {\bfseries 856} (Sept., 2024) 138862}. \url{https://www.sciencedirect.com/science/article/pii/S0370269324004209}.

\bibitem{bibcf14}
{\bfseries ALICE} Collaboration, S.~Acharya {\em et~al.}, ``Probing the chiral magnetic wave with charge-dependent flow measurements in pb-pb collisions at the lhc'', \href{http://dx.doi.org/10.1007/JHEP12(2023)067}{{\em Journal of High Energy Physics} {\bfseries 2023} (Dec, 2023) 67}. \url{https://doi.org/10.1007/JHEP12(2023)067}.

\bibitem{bibcf18}
{\bfseries ALICE} Collaboration, S.~Acharya {\em et~al.}, ``Unveiling the strong interaction among hadrons at the lhc'', \href{http://dx.doi.org/10.1038/s41586-020-3001-6}{{\em Nature} {\bfseries 588} (Dec, 2020) 232--238}. \url{https://doi.org/10.1038/s41586-020-3001-6}.

\bibitem{bibcf17}
L.~Fabbietti, V.~M. Sarti, and O.~V. Doce, ``Study of the strong interaction among hadrons with correlations at the lhc'', \href{http://dx.doi.org/https://doi.org/10.1146/annurev-nucl-102419-034438}{{\em Annual Review of Nuclear and Particle Science} {\bfseries 71} (2021) 377--402}. \url{https://www.annualreviews.org/content/journals/10.1146/annurev-nucl-102419-034438}.

\bibitem{ma_hypernuclei_2023}
Y.-G. Ma, ``Hypernuclei as a laboratory to test hyperon–nucleon interactions'', \href{http://dx.doi.org/10.1007/s41365-023-01248-6}{{\em Nuc. Sci. Tech.} {\bfseries 34} (June, 2023) 97}. \url{https://doi.org/10.1007/s41365-023-01248-6}.

\bibitem{bibcf20}
{\bfseries ALICE} Collaboration, S.~Acharya {\em et~al.}, ``Exploring the $n\lambda$-$n\sigma$ coupled system with high precision correlation techniques at the lhc'', \href{http://dx.doi.org/https://doi.org/10.1016/j.physletb.2022.137272}{{\em Physics Letters B} {\bfseries 833} (2022) 137272}. \url{https://www.sciencedirect.com/science/article/pii/S0370269322004063}.

\bibitem{bibcf21}
{\bfseries ALICE} Collaboration, S.~Acharya {\em et~al.}, ``Investigation of the p-$\sigma_0$ interaction via femtoscopy in pp collisions'', \href{http://dx.doi.org/https://doi.org/10.1016/j.physletb.2020.135419}{{\em Physics Letters B} {\bfseries 805} (2020) 135419}. \url{https://www.sciencedirect.com/science/article/pii/S0370269320302239}.

\bibitem{bibcf22}
{\bfseries ALICE} Collaboration, S.~Acharya {\em et~al.}, ``Experimental evidence for an attractive $p\text{\ensuremath{-}}\ensuremath{\phi}$ interaction'', \href{http://dx.doi.org/10.1103/PhysRevLett.127.172301}{{\em Phys. Rev. Lett.} {\bfseries 127} (Oct, 2021) 172301}. \url{https://link.aps.org/doi/10.1103/PhysRevLett.127.172301}.

\bibitem{bibcf24}
{\bfseries ALICE} Collaboration, S.~Acharya {\em et~al.}, ``Study of the $\lambda$–$\lambda$ interaction with femtoscopy correlations in pp and p–pb collisions at the lhc'', \href{http://dx.doi.org/https://doi.org/10.1016/j.physletb.2019.134822}{{\em Physics Letters B} {\bfseries 797} (2019) 134822}. \url{https://www.sciencedirect.com/science/article/pii/S0370269319305362}.

\bibitem{bibcf23}
{\bfseries ALICE} Collaboration, S.~Acharya {\em et~al.}, ``Scattering studies with low-energy kaon-proton femtoscopy in proton-proton collisions at the lhc'', \href{http://dx.doi.org/10.1103/PhysRevLett.124.092301}{{\em Phys. Rev. Lett.} {\bfseries 124} (Mar, 2020) 092301}. \url{https://link.aps.org/doi/10.1103/PhysRevLett.124.092301}.

\bibitem{bibcf25}
{\bfseries ALICE} Collaboration, S.~Acharya {\em et~al.}, ``First study of the two-body scattering involving charm hadrons'', \href{http://dx.doi.org/10.1103/PhysRevD.106.052010}{{\em Phys. Rev. D} {\bfseries 106} (Sep, 2022) 052010}. \url{https://link.aps.org/doi/10.1103/PhysRevD.106.052010}.

\bibitem{bibcf26}
{\bfseries ALICE} Collaboration, ``Studying the interaction between charm and light-flavor mesons'', 2024.
\newblock \url{https://arxiv.org/abs/2401.13541}.

\bibitem{bibcf27}
{\bfseries ALICE} Collaboration, S.~Acharya {\em et~al.}, ``Towards the understanding of the genuine three-body interaction for p--p--p and p--p--$\lambda$'', \href{http://dx.doi.org/10.1140/epja/s10050-023-00998-6}{{\em The European Physical Journal A} {\bfseries 59} (Jul, 2023) 145}. \url{https://doi.org/10.1140/epja/s10050-023-00998-6}.

\bibitem{bibcf28}
{\bfseries ALICE} Collaboration, ``Exploring the strong interaction of three-body systems at the lhc'', 2023.
\newblock \url{https://arxiv.org/abs/2308.16120}.

\bibitem{ALICE:2016flj}
{\bfseries ALICE} Collaboration, J.~Adam {\em et~al.}, ``{$J\psi$ suppression at forward rapidity in \PbPb collisions at $\sqrt{s_{NN}}=$~5~{TeV} TeV}'', \href{http://dx.doi.org/10.1016/j.physletb.2016.12.064}{{\em Phys. Lett. B} {\bfseries 766} (2017) 212--224}, \href{http://arxiv.org/abs/1606.08197}{{\ttfamily arXiv:1606.08197 [nucl-ex]}}.

\bibitem{ALICE:2023gco}
{\bfseries ALICE} Collaboration, S.~Acharya {\em et~al.}, ``{Measurements of inclusive J/\ensuremath{\psi} production at midrapidity and forward rapidity in Pb\textendash{}Pb collisions at sNN = 5.02 TeV}'', \href{http://dx.doi.org/10.1016/j.physletb.2024.138451}{{\em Phys. Lett. B} {\bfseries 849} (2024) 138451}, \href{http://arxiv.org/abs/2303.13361}{{\ttfamily arXiv:2303.13361 [nucl-ex]}}.

\bibitem{ALICE:2019tqa}
{\bfseries ALICE} Collaboration, S.~Acharya {\em et~al.}, ``{Coherent $J\psi$ photoproduction at forward rapidity in ultra-peripheral \PbPb collisions at $\sqrt{s_{\rm{NN}}}=5.02$ TeV}'', \href{http://dx.doi.org/10.1016/j.physletb.2019.134926}{{\em Phys. Lett. B} {\bfseries 798} (2019) 134926}, \href{http://arxiv.org/abs/1904.06272}{{\ttfamily arXiv:1904.06272 [nucl-ex]}}.

\bibitem{ALICE:2021gpt}
{\bfseries ALICE} Collaboration, S.~Acharya {\em et~al.}, ``{Coherent $\rm{J/\psi}$ and $\rm{\psi'}$ photoproduction at midrapidity in ultra-peripheral \PbPb collisions at $\sqrt{s_{\mathrm{NN}}}~=~5.02$ TeV}'', \href{http://dx.doi.org/10.1140/epjc/s10052-021-09437-6}{{\em Eur. Phys. J. C} {\bfseries 81} (2021) 712}, \href{http://arxiv.org/abs/2101.04577}{{\ttfamily arXiv:2101.04577 [nucl-ex]}}.

\bibitem{ALICE:2015mzu}
{\bfseries ALICE} Collaboration, J.~Adam {\em et~al.}, ``{Measurement of an excess in the yield of $J/\psi$ at very low $p_{\rm T}$ in \PbPb collisions at $\sqrt{s_{NN}}$ = 2.76 TeV}'', \href{http://dx.doi.org/10.1103/PhysRevLett.116.222301}{{\em Phys. Rev. Lett.} {\bfseries 116} (2016) 222301}, \href{http://arxiv.org/abs/1509.08802}{{\ttfamily arXiv:1509.08802 [nucl-ex]}}.

\bibitem{ALICE:2022zso}
{\bfseries ALICE} Collaboration, J.~Adam {\em et~al.}, ``{Photoproduction of low-$p_{\rm T}$ J/$\psi$ from peripheral to central Pb$-$Pb collisions at 5.02 TeV}'', \href{http://arxiv.org/abs/2204.10684}{{\ttfamily arXiv:2204.10684 [nucl-ex]}}.

\bibitem{Andronic:2019wva}
A.~Andronic, P.~Braun-Munzinger, M.~K. Koehler, K.~Redlich, and J.~Stachel, ``{Transverse momentum distributions of charmonium states with the statistical hadronization model}'', \href{http://dx.doi.org/10.1016/j.physletb.2019.134836}{{\em Phys. Lett.} {\bfseries B797} (2019) 134836},
\href{http://arxiv.org/abs/1901.09200}{{\ttfamily arXiv:1901.09200 [nucl-th]}}.

\bibitem{Zhou:2014kka}
K.~Zhou, N.~Xu, Z.~Xu, and P.~Zhuang, ``{Medium effects on charmonium production at ultrarelativistic energies available at the CERN Large Hadron Collider}'', \href{http://dx.doi.org/10.1103/PhysRevC.89.054911}{{\em Phys. Rev. C} {\bfseries 89} (2014) 054911}, \href{http://arxiv.org/abs/1401.5845}{{\ttfamily arXiv:1401.5845 [nucl-th]}}.

\bibitem{Wu:2020zbx}
B.~Wu, X.~Du, M.~Sibila, and R.~Rapp, ``{$X(3872)$transport in heavy-ion collisions}'', \href{http://dx.doi.org/10.1140/epja/s10050-021-00623-4}{{\em Eur. Phys. J. A} {\bfseries 57} (2021) 122}, \href{http://arxiv.org/abs/2006.09945}{{\ttfamily arXiv:2006.09945 [nucl-th]}}. [Erratum: Eur.Phys.J.A 57, 314 (2021)].

\bibitem{Zhao:2010nk}
X.~Zhao and R.~Rapp, ``{Charmonium in medium: from correlators to experiment}'', \href{http://dx.doi.org/10.1103/PhysRevC.82.064905}{{\em Phys. Rev. C} {\bfseries 82} (2010) 064905}, \href{http://arxiv.org/abs/1008.5328}{{\ttfamily arXiv:1008.5328 [hep-ph]}}.

\bibitem{ALICE:2022jeh}
{\bfseries ALICE} Collaboration, S.~Acharya {\em et~al.}, ``{\ensuremath{\psi}(2S) Suppression in Pb-Pb Collisions at the LHC}'', \href{http://dx.doi.org/10.1103/PhysRevLett.132.042301}{{\em Phys. Rev. Lett.} {\bfseries 132} (2024) 042301}, \href{http://arxiv.org/abs/2210.08893}{{\ttfamily arXiv:2210.08893 [nucl-ex]}}.

\bibitem{Skokov:2009qp}
V.~Skokov, A.~Y. Illarionov, and V.~Toneev, ``{Estimate of the magnetic field strength in heavy-ion collisions}'', \href{http://dx.doi.org/10.1142/S0217751X09047570}{{\em Int. J. Mod. Phys. A} {\bfseries 24} (2009) 5925--5932}, \href{http://arxiv.org/abs/0907.1396}{{\ttfamily arXiv:0907.1396 [nucl-th]}}.

\bibitem{Becattini:2007sr}
F.~Becattini, F.~Piccinini, and J.~Rizzo, ``{Angular momentum conservation in heavy ion collisions at very high energy}'', \href{http://dx.doi.org/10.1103/PhysRevC.77.024906}{{\em Phys. Rev. C} {\bfseries 77} (2008) 024906}, \href{http://arxiv.org/abs/0711.1253}{{\ttfamily arXiv:0711.1253 [nucl-th]}}.

\bibitem{ALICE:2022dyy}
{\bfseries ALICE} Collaboration, S.~Acharya {\em et~al.}, ``{Measurement of the J/\ensuremath{\psi} Polarization with Respect to the Event Plane in Pb-Pb Collisions at the LHC}'', \href{http://dx.doi.org/10.1103/PhysRevLett.131.042303}{{\em Phys. Rev. Lett.} {\bfseries 131} (2023) 042303}, \href{http://arxiv.org/abs/2204.10171}{{\ttfamily arXiv:2204.10171 [nucl-ex]}}.

\bibitem{ALICE:2019aid}
{\bfseries ALICE} Collaboration, S.~Acharya {\em et~al.}, ``{Evidence of Spin-Orbital Angular Momentum Interactions in Relativistic Heavy-Ion Collisions}'', \href{http://dx.doi.org/10.1103/PhysRevLett.125.012301}{{\em Phys. Rev. Lett.} {\bfseries 125} (2020) 012301}, \href{http://arxiv.org/abs/1910.14408}{{\ttfamily arXiv:1910.14408 [nucl-ex]}}.

\bibitem{Turbide:2003si}
S.~Turbide, R.~Rapp, and C.~Gale, ``{Hadronic production of thermal photons}'', \href{http://dx.doi.org/10.1103/PhysRevC.69.014903}{{\em Phys. Rev. C} {\bfseries 69} (2004) 014903}, \href{http://arxiv.org/abs/hep-ph/0308085}{{\ttfamily arXiv:hep-ph/0308085}}.

\bibitem{ALICE:2023jef}
{\bfseries ALICE} Collaboration, S.~Acharya {\em et~al.}, ``{Dielectron production in central Pb$-$Pb collisions at $\sqrt{s_\mathrm{NN}}$ = 5.02 TeV}'', \href{http://arxiv.org/abs/2308.16704}{{\ttfamily arXiv:2308.16704 [nucl-ex]}}.

\bibitem{ALICE:2022hvk}
{\bfseries ALICE} Collaboration, S.~Acharya {\em et~al.}, ``{Dielectron production at midrapidity at low transverse momentum in peripheral and semi-peripheral Pb\textendash{}Pb collisions at $ {\sqrt{s}}_{\textrm{NN}} $ = 5.02 TeV}'', \href{http://dx.doi.org/10.1007/JHEP06(2023)024}{{\em JHEP} {\bfseries 06} (2023) 024}, \href{http://arxiv.org/abs/2204.11732}{{\ttfamily arXiv:2204.11732 [nucl-ex]}}.

\bibitem{Rapp:1999us}
R.~Rapp and J.~Wambach, ``{Low mass dileptons at the CERN SPS: Evidence for chiral restoration?}'', \href{http://dx.doi.org/10.1007/s100500050364}{{\em Eur. Phys. J. A} {\bfseries 6} (1999) 415--420}, \href{http://arxiv.org/abs/hep-ph/9907502}{{\ttfamily arXiv:hep-ph/9907502}}.

\bibitem{Rapp:2013nxa}
R.~Rapp, ``{Dilepton Spectroscopy of QCD Matter at Collider Energies}'', \href{http://dx.doi.org/10.1155/2013/148253}{{\em Adv. High Energy Phys.} {\bfseries 2013} (2013) 148253}, \href{http://arxiv.org/abs/1304.2309}{{\ttfamily arXiv:1304.2309 [hep-ph]}}.

\bibitem{ParticleDataGroup:2022pth}
{\bfseries Particle Data Group} Collaboration, R.~L. Workman {\em et~al.}, ``{Review of Particle Physics}'', \href{http://dx.doi.org/10.1093/ptep/ptac097}{{\em PTEP} {\bfseries 2022} (2022) 083C01}.

\bibitem{ALICE:2015xmh}
{\bfseries ALICE} Collaboration, J.~Adam {\em et~al.}, ``{Direct photon production in Pb--Pb collisions at \twosevensixnn}'', \href{http://dx.doi.org/10.1016/j.physletb.2016.01.020}{{\em Phys.~Lett.} {\bfseries B754} (2016) 235--248}, \href{http://arxiv.org/abs/1509.07324}{{\ttfamily arXiv:1509.07324 [nucl-ex]}}.

\bibitem{Dong:2019byy}
X.~Dong, Y.-J. Lee, and R.~Rapp, ``{Open Heavy-Flavor Production in Heavy-Ion Collisions}'', \href{http://dx.doi.org/10.1146/annurev-nucl-101918-023806}{{\em Ann. Rev. Nucl. Part. Sci.} {\bfseries 69} (2019) 417--445}, \href{http://arxiv.org/abs/1903.07709}{{\ttfamily arXiv:1903.07709 [nucl-ex]}}.

\bibitem{Prino:2016cni}
F.~Prino and R.~Rapp, ``{Open Heavy Flavor in QCD Matter and in Nuclear Collisions}'', \href{http://dx.doi.org/10.1088/0954-3899/43/9/093002}{{\em J.~Phys.} {\bfseries G43} (2016) 093002}, \href{http://arxiv.org/abs/1603.00529}{{\ttfamily arXiv:1603.00529 [nucl-ex]}}.

\bibitem{ALICE:2012inj}
{\bfseries ALICE} Collaboration, B.~Abelev {\em et~al.}, ``{Measurement of charm production at central rapidity in proton--proton collisions at \twosevensix}'', \href{http://dx.doi.org/10.1007/JHEP07(2012)191}{{\em JHEP} {\bfseries 1207} (2012) 191}, \href{http://arxiv.org/abs/1205.4007}{{\ttfamily arXiv:1205.4007 [hep-ex]}}.

\bibitem{ALICE:2012ab}
{\bfseries ALICE} Collaboration, B.~Abelev {\em et~al.}, ``{Suppression of high transverse momentum D mesons in central Pb--Pb collisions at \twosevensixnn}'', \href{http://dx.doi.org/10.1007/JHEP09(2012)112}{{\em JHEP} {\bfseries 1209} (2012) 112}, \href{http://arxiv.org/abs/1203.2160}{{\ttfamily arXiv:1203.2160 [nucl-ex]}}.

\bibitem{ALICE:2012gkr}
{\bfseries ALICE} Collaboration, B.~Abelev {\em et~al.}, ``{${\rm D}_{\rm s}^{+}$ meson production at central rapidity in proton--proton collisions at \seven}'', \href{http://dx.doi.org/10.1016/j.physletb.2012.10.049}{{\em Phys. Lett.} {\bfseries B718} (2012) 279--294}, \href{http://arxiv.org/abs/1208.1948}{{\ttfamily arXiv:1208.1948 [hep-ex]}}.

\bibitem{ALICE:2014xjz}
{\bfseries ALICE} Collaboration, B.~B. Abelev {\em et~al.}, ``{Measurement of prompt D-meson production in p--Pb collisions at $\sqrt{s_{NN}}=$~5~{TeV}}'', \href{http://dx.doi.org/10.1103/PhysRevLett.113.232301}{{\em Phys. Rev. Lett.} {\bfseries 113} (2014) 232301}, \href{http://arxiv.org/abs/1405.3452}{{\ttfamily arXiv:1405.3452 [nucl-ex]}}.

\bibitem{ALICE:2017thy}
{\bfseries ALICE} Collaboration, S.~Acharya {\em et~al.}, ``{$\Lambda_{\rm c}^{+}$ production in pp collisions at \seven and in p--Pb collisions at $\sqrt{s_{NN}}=$~5~{TeV}}'', \href{http://dx.doi.org/10.1007/JHEP04(2018)108}{{\em JHEP} {\bfseries 1804} (2018) 108}, \href{http://arxiv.org/abs/1712.09581}{{\ttfamily arXiv:1712.09581 [nucl-ex]}}.

\bibitem{ALICE:2017dja}
{\bfseries ALICE} Collaboration, S.~Acharya {\em et~al.}, ``{First measurement of $\Xi_{\rm c}^{0}$ production in pp collisions at \seven}'', \href{http://dx.doi.org/10.1016/j.physletb.2018.03.061}{{\em Phys. Lett.} {\bfseries B781} (2018) 8--19}, \href{http://arxiv.org/abs/1712.04242}{{\ttfamily arXiv:1712.04242 [hep-ex]}}.

\bibitem{ALICE:2017pbx}
{\bfseries ALICE} Collaboration, S.~Acharya {\em et~al.}, ``{D-meson azimuthal anisotropy in midcentral Pb--Pb collisions at $\sqrt{s_{NN}}=$~5~{TeV}}'', \href{http://dx.doi.org/10.1103/PhysRevLett.120.102301}{{\em Phys. Rev. Lett.} {\bfseries 120} (2018) 102301}, \href{http://arxiv.org/abs/1707.01005}{{\ttfamily arXiv:1707.01005 [nucl-ex]}}.

\bibitem{ALICE:2018lyv}
{\bfseries ALICE} Collaboration, S.~Acharya {\em et~al.}, ``{Measurement of D$^{0}$, D$^{+}$, D$^{*+}$ and D$_{\rm s}^{+}$ production in Pb--Pb collisions at $\sqrt{s_{NN}}=$~5~{TeV}}'', \href{http://dx.doi.org/10.1007/JHEP10(2018)174}{{\em JHEP} {\bfseries 1810} (2018) 174}, \href{http://arxiv.org/abs/1804.09083}{{\ttfamily arXiv:1804.09083 [nucl-ex]}}.

\bibitem{ALICE:2018hbc}
{\bfseries ALICE} Collaboration, S.~Acharya {\em et~al.}, ``{$\Lambda_{\rm c}^{+}$ production in Pb--Pb collisions at $\sqrt{s_{NN}}=$~5~{TeV}}'', \href{http://dx.doi.org/10.1016/j.physletb.2019.04.046}{{\em Phys. Lett.} {\bfseries B793} (2019) 212--223}, \href{http://arxiv.org/abs/1809.10922}{{\ttfamily arXiv:1809.10922 [nucl-ex]}}.

\bibitem{ALICE:2012mzy}
{\bfseries ALICE} Collaboration, B.~Abelev {\em et~al.}, ``{Measurement of electrons from semileptonic heavy-flavour hadron decays in pp collisions at \seven}'', \href{http://dx.doi.org/10.1103/PhysRevD.86.112007}{{\em Phys. Rev.} {\bfseries D86} (2012) 112007}, \href{http://arxiv.org/abs/1205.5423}{{\ttfamily arXiv:1205.5423 [hep-ex]}}.

\bibitem{ALICE:2015zhm}
{\bfseries ALICE} Collaboration, J.~Adam {\em et~al.}, ``{Measurement of electrons from heavy-flavour hadron decays in p--Pb collisions at $\sqrt{s_{NN}}=$~5~{TeV}}'', \href{http://dx.doi.org/10.1016/j.physletb.2015.12.067}{{\em Phys. Lett.} {\bfseries B754} (2016) 81--93}, \href{http://arxiv.org/abs/1509.07491}{{\ttfamily arXiv:1509.07491 [nucl-ex]}}.

\bibitem{ALICE:2019nuy}
{\bfseries ALICE} Collaboration, S.~Acharya {\em et~al.}, ``{Measurement of electrons from semileptonic heavy-flavour hadron decays at midrapidity in pp and Pb--Pb collisions at $\sqrt{s_{NN}}=$~5~{TeV}}'', \href{http://dx.doi.org/10.1016/j.physletb.2020.135377}{{\em Phys. Lett.} {\bfseries B804} (2020) 135377}, \href{http://arxiv.org/abs/1910.09110}{{\ttfamily arXiv:1910.09110 [nucl-ex]}}.

\bibitem{ALICE:2020hdw}
{\bfseries ALICE} Collaboration, S.~Acharya {\em et~al.}, ``{Elliptic Flow of Electrons from Beauty-Hadron Decays in Pb--Pb Collisions at $\sqrt{s_{NN}}=$~5~{TeV}}'', \href{http://dx.doi.org/10.1103/PhysRevLett.126.162001}{{\em Phys. Rev. Lett.} {\bfseries 126} (2021) 162001}, \href{http://arxiv.org/abs/2005.11130}{{\ttfamily arXiv:2005.11130 [nucl-ex]}}.

\bibitem{ALICE:2012sxy}
{\bfseries ALICE} Collaboration, B.~Abelev {\em et~al.}, ``{Production of muons from heavy flavour decays at forward rapidity in pp and Pb--Pb collisions at \twosevensixnn}'', \href{http://dx.doi.org/10.1103/PhysRevLett.109.112301}{{\em Phys. Rev. Lett.} {\bfseries 109} (2012) 112301}, \href{http://arxiv.org/abs/1205.6443}{{\ttfamily arXiv:1205.6443 [hep-ex]}}.

\bibitem{ALICE:2015xyt}
{\bfseries ALICE} Collaboration, J.~Adam {\em et~al.}, ``{Elliptic flow of muons from heavy-flavour hadron decays at forward rapidity in Pb--Pb collisions at \twosevensixnn}'', \href{http://dx.doi.org/10.1016/j.physletb.2015.11.059}{{\em Phys. Lett.} {\bfseries B753} (2016) 41--56}, \href{http://arxiv.org/abs/1507.03134}{{\ttfamily arXiv:1507.03134 [nucl-ex]}}.

\bibitem{ALICE:2017fsl}
{\bfseries ALICE} Collaboration, S.~Acharya {\em et~al.}, ``{Production of muons from heavy-flavour hadron decays in p--Pb collisions at $\sqrt{s_{NN}}=$~5~{TeV}}'', \href{http://dx.doi.org/10.1016/j.physletb.2017.03.049}{{\em Phys. Lett.} {\bfseries B770} (2017) 459--472}, \href{http://arxiv.org/abs/1702.01479}{{\ttfamily arXiv:1702.01479 [nucl-ex]}}.

\bibitem{ALICE:2020sjb}
{\bfseries ALICE} Collaboration, S.~Acharya {\em et~al.}, ``{Production of muons from heavy-flavour hadron decays at high transverse momentum in Pb--Pb collisions at $\sqrt{s_{NN}} = 5.02$ and $2.76$~ {TeV}}'', \href{http://dx.doi.org/10.1016/j.physletb.2021.136558}{{\em Phys. Lett.} {\bfseries B820} (2021) 136558}, \href{http://arxiv.org/abs/2011.05718}{{\ttfamily arXiv:2011.05718 [nucl-ex]}}.

\bibitem{ALICE:2020try}
{\bfseries ALICE} Collaboration, S.~Acharya {\em et~al.}, ``{Inclusive heavy-flavour production at central and forward rapidity in Xe--Xe collisions at $\sqrt{s_{NN}}$ = 5.44~{TeV}}'', \href{http://dx.doi.org/10.1016/j.physletb.2021.136437}{{\em Phys. Lett.} {\bfseries B819} (2021) 136437}, \href{http://arxiv.org/abs/2011.06970}{{\ttfamily arXiv:2011.06970 [nucl-ex]}}.

\bibitem{ALICE:2021psx}
{\bfseries ALICE} Collaboration, S.~Acharya {\em et~al.}, ``{Measurement of the production cross section of prompt $\Xi_{\rm c}^{0}$ baryons at midrapidity in pp collisions at \five}'', \href{http://dx.doi.org/10.1007/JHEP10(2021)159}{{\em JHEP} {\bfseries 2110} (2021) 159}, \href{http://arxiv.org/abs/2105.05616}{{\ttfamily arXiv:2105.05616 [nucl-ex]}}.

\bibitem{ALICE:2021bli}
{\bfseries ALICE} Collaboration, S.~Acharya {\em et~al.}, ``{Measurement of the cross sections of $\Xi_{\rm c}^{0}$ and $\Xi_{\rm c}^{+}$ baryons and of the branching fraction ratio BR($\Xi_{\rm c}^{0}\rightarrow\Xi^{-}{e}^{+}\nu_{e}$) / BR($\Xi_{\rm c}^{0}\rightarrow\Xi^{-}\pi^{+}$) in pp collisions at 13 {TeV}}'', \href{http://dx.doi.org/10.1103/PhysRevLett.127.272001}{{\em Phys. Rev. Lett.} {\bfseries 127} (2021) 272001}, \href{http://arxiv.org/abs/2105.05187}{{\ttfamily arXiv:2105.05187 [nucl-ex]}}.

\bibitem{ALICE:2012vpz}
{\bfseries ALICE} Collaboration, B.~Abelev {\em et~al.}, ``{Measurement of prompt $J\psi$ and beauty hadron production cross sections at mid-rapidity in pp collisions at $\sqrt{s}=$~7~{TeV}}'', \href{http://dx.doi.org/10.1007/JHEP11(2012)065}{{\em JHEP} {\bfseries 1211} (2012) 065}, \href{http://arxiv.org/abs/1205.5880}{{\ttfamily arXiv:1205.5880 [hep-ex]}}.

\bibitem{ALICE:2015nvt}
{\bfseries ALICE} Collaboration, J.~Adam {\em et~al.}, ``{Inclusive, prompt and non-prompt J/$\psi$ production at mid-rapidity in Pb--Pb collisions at $\sqrt{s_{NN}}=$~2.76 {TeV} }'', \href{http://dx.doi.org/10.1007/JHEP07(2015)051}{{\em JHEP} {\bfseries 1507} (2015) 051}, \href{http://arxiv.org/abs/1504.07151}{{\ttfamily arXiv:1504.07151 [nucl-ex]}}.

\bibitem{ALICE:2018szk}
{\bfseries ALICE} Collaboration, S.~Acharya {\em et~al.}, ``{Prompt and non-prompt $J\psi$ production and nuclear modification at mid-rapidity in p--Pb collisions at $\sqrt{s_{NN}}=$~5~{TeV}}'', \href{http://dx.doi.org/10.1140/epjc/s10052-018-5881-2}{{\em Eur. Phys. J.} {\bfseries C78} (2018) 466}, \href{http://arxiv.org/abs/1802.00765}{{\ttfamily arXiv:1802.00765 [nucl-ex]}}.

\bibitem{ALICE:2021mgk}
{\bfseries ALICE} Collaboration, S.~Acharya {\em et~al.}, ``{Measurement of beauty and charm production in pp collisions at \five via non-prompt and prompt D mesons}'', \href{http://dx.doi.org/10.1007/JHEP05(2021)220}{{\em JHEP} {\bfseries 2105} (2021) 220}, \href{http://arxiv.org/abs/2102.13601}{{\ttfamily arXiv:2102.13601 [nucl-ex]}}.

\bibitem{ALICE:2022tji}
{\bfseries ALICE} Collaboration, S.~Acharya {\em et~al.}, ``{Measurement of beauty production via non-prompt D$^{0}$ mesons in Pb--Pb collisions at $\sqrt{s_{NN}}=$~5~{TeV}}'', \href{http://dx.doi.org/10.1007/JHEP12(2022)126}{{\em JHEP} {\bfseries 2212} (2022) 126}, \href{http://arxiv.org/abs/2202.00815}{{\ttfamily arXiv:2202.00815 [nucl-ex]}}.

\bibitem{ALICE:2020iug}
{\bfseries ALICE} Collaboration, S.~Acharya {\em et~al.}, ``{Transverse-momentum and event-shape dependence of D-meson flow harmonics in Pb--Pb collisions at $\sqrt{s_{NN}}=$~5~{TeV}}'', \href{http://dx.doi.org/10.1016/j.physletb.2020.136054}{{\em Phys. Lett.} {\bfseries B813} (2021) 136054}, \href{http://arxiv.org/abs/2005.11131}{{\ttfamily arXiv:2005.11131 [nucl-ex]}}.

\bibitem{ALICE:2021rxa}
{\bfseries ALICE} Collaboration, S.~Acharya {\em et~al.}, ``{Prompt D$^{0}$, D$^{+}$, and D$^{*+}$ production in Pb--Pb collisions at $\sqrt{s_{NN}}=$~5~{TeV}}'', \href{http://dx.doi.org/10.1007/JHEP01(2022)174}{{\em JHEP} {\bfseries 2201} (2022) 174}, \href{http://arxiv.org/abs/2110.09420}{{\ttfamily arXiv:2110.09420 [nucl-ex]}}.

\bibitem{Rapp:2009my}
R.~Rapp and H.~van Hees, \href{http://dx.doi.org/10.1142/9789814293297_0003}{``{Heavy Quarks in the Quark-Gluon Plasma}'',} pp.~111--206.
\newblock 2010.
\newblock \href{http://arxiv.org/abs/0903.1096}{{\ttfamily arXiv:0903.1096 [hep-ph]}}.

\bibitem{Svetitsky:1987gq}
B.~Svetitsky, ``{Diffusion of charmed quarks in the quark-gluon plasma}'', \href{http://dx.doi.org/10.1103/PhysRevD.37.2484}{{\em Phys. Rev.} {\bfseries D37} (1988) 2484--2491}.

\bibitem{HotQCD:2014kol}
{\bfseries HotQCD} Collaboration, A.~Bazavov {\em et~al.}, ``{Equation of state in $(2+1)$-flavor QCD}'', \href{http://dx.doi.org/10.1103/PhysRevD.90.094503}{{\em Phys. Rev.} {\bfseries D90} (2014) 094503}, \href{http://arxiv.org/abs/1407.6387}{{\ttfamily arXiv:1407.6387 [hep-lat]}}.

\bibitem{Borsanyi:2013bia}
S.~Borsanyi, Z.~Fodor, C.~Hoelbling, S.~D. Katz, S.~Krieg, and K.~K. Szabo, ``{Full result for the QCD equation of state with $2+1$ flavors}'', \href{http://dx.doi.org/10.1016/j.physletb.2014.01.007}{{\em Phys. Lett.} {\bfseries B730} (2014) 99--104}, \href{http://arxiv.org/abs/1309.5258}{{\ttfamily arXiv:1309.5258 [hep-lat]}}.

\bibitem{ALICE:2011dyt}
{\bfseries ALICE} Collaboration, K.~Aamodt {\em et~al.}, ``{Two-pion Bose-Einstein correlations in central Pb--Pb collisions at \twosevensixnn}'', \href{http://dx.doi.org/10.1016/j.physletb.2010.12.053}{{\em Phys. Lett.} {\bfseries B696} (2011) 328--337}, \href{http://arxiv.org/abs/1012.4035}{{\ttfamily arXiv:1012.4035 [nucl-ex]}}.

\bibitem{Prado:2019ste}
C.~A.~G. Prado, W.-J. Xing, S.~Cao, G.-Y. Qin, and X.-N. Wang, ``{Longitudinal dependence of open heavy flavor $\RAA$ in relativistic heavy-ion collisions}'', \href{http://dx.doi.org/10.1103/PhysRevC.101.064907}{{\em Phys. Rev.} {\bfseries C101} (2020) 064907}, \href{http://arxiv.org/abs/1911.06527}{{\ttfamily arXiv:1911.06527 [nucl-th]}}.

\bibitem{ALICE:2023gjj}
{\bfseries ALICE} Collaboration, S.~Acharya {\em et~al.}, ``{Measurement of non-prompt ${{\rm D}^{0}}$-meson elliptic flow in Pb--Pb collisions at $\sqrt{s_{NN}}=$~5~{TeV}}'', \href{http://dx.doi.org/10.1140/epjc/s10052-023-12259-3}{{\em Eur. Phys. J.} {\bfseries C83} (2023) 1123}, \href{http://arxiv.org/abs/2307.14084}{{\ttfamily arXiv:2307.14084 [nucl-ex]}}.

\bibitem{Djordjevic:2014tka}
M.~Djordjevic, M.~Djordjevic, and B.~Blagojevic, ``{RHIC and LHC jet suppression in non-central collisions}'', \href{http://dx.doi.org/10.1016/j.physletb.2014.08.063}{{\em Phys. Lett.} {\bfseries B737} (2014) 298--302}, \href{http://arxiv.org/abs/1405.4250}{{\ttfamily arXiv:1405.4250 [nucl-th]}}.

\bibitem{He:2012df}
M.~He, R.~J. Fries, and R.~Rapp, ``{$D_s$-Meson as Quantitative Probe of Diffusion and Hadronization in Nuclear Collisions}'', \href{http://dx.doi.org/10.1103/PhysRevLett.110.112301}{{\em Phys. Rev. Lett.} {\bfseries 110} (2013) 112301}, \href{http://arxiv.org/abs/1204.4442}{{\ttfamily arXiv:1204.4442 [nucl-th]}}.

\bibitem{Kuznetsova:2006bh}
I.~Kuznetsova and J.~Rafelski, ``{Heavy flavor hadrons in statistical hadronization of strangeness-rich QGP}'', \href{http://dx.doi.org/10.1140/epjc/s10052-007-0268-9}{{\em Eur. Phys. J.} {\bfseries C51} (2007) 113--133}, \href{http://arxiv.org/abs/hep-ph/0607203}{{\ttfamily arXiv:hep-ph/0607203}}.

\bibitem{Andronic:2007zu}
A.~Andronic, P.~Braun-Munzinger, K.~Redlich, and J.~Stachel, ``{Charmonium and open charm production in nuclear collisions at SPS/FAIR energies and the possible influence of a hot hadronic medium}'', \href{http://dx.doi.org/10.1016/j.physletb.2007.10.064}{{\em Phys. Lett.} {\bfseries B659} (2008) 149--155}, \href{http://arxiv.org/abs/0708.1488}{{\ttfamily arXiv:0708.1488 [nucl-th]}}.

\bibitem{Lee:2007wr}
S.~H. Lee, K.~Ohnishi, S.~Yasui, I.-K. Yoo, and C.-M. Ko, ``{$\Lambda_c$ enhancement from strongly coupled quark-gluon plasma}'', \href{http://dx.doi.org/10.1103/PhysRevLett.100.222301}{{\em Phys. Rev. Lett.} {\bfseries 100} (2008) 222301}, \href{http://arxiv.org/abs/0709.3637}{{\ttfamily arXiv:0709.3637 [nucl-th]}}.

\bibitem{Oh:2009zj}
Y.~Oh, C.~M. Ko, S.~H. Lee, and S.~Yasui, ``{Heavy baryon/meson ratios in relativistic heavy ion collisions}'', \href{http://dx.doi.org/10.1103/PhysRevC.79.044905}{{\em Phys. Rev.} {\bfseries C79} (2009) 044905}, \href{http://arxiv.org/abs/0901.1382}{{\ttfamily arXiv:0901.1382 [nucl-th]}}.

\bibitem{Das:2016llg}
S.~K. Das, J.~M. Torres-Rincon, L.~Tolos, V.~Minissale, F.~Scardina, and V.~Greco, ``{Propagation of heavy baryons in heavy-ion collisions}'', \href{http://dx.doi.org/10.1103/PhysRevD.94.114039}{{\em Phys. Rev.} {\bfseries D94} (2016) 114039}, \href{http://arxiv.org/abs/1604.05666}{{\ttfamily arXiv:1604.05666 [nucl-th]}}.

\bibitem{Plumari:2017ntm}
S.~Plumari, V.~Minissale, S.~K. Das, G.~Coci, and V.~Greco, ``{Charmed Hadrons from Coalescence plus Fragmentation in relativistic nucleus-nucleus collisions at RHIC and LHC}'', \href{http://dx.doi.org/10.1140/epjc/s10052-018-5828-7}{{\em Eur. Phys. J.} {\bfseries C78} (2018) 348}, \href{http://arxiv.org/abs/1712.00730}{{\ttfamily arXiv:1712.00730 [hep-ph]}}.

\bibitem{He:2019vgs}
M.~He and R.~Rapp, ``{Hadronization and Charm-Hadron Ratios in Heavy-Ion Collisions}'', \href{http://dx.doi.org/10.1103/PhysRevLett.124.042301}{{\em Phys. Rev. Lett.} {\bfseries 124} (2020) 042301}, \href{http://arxiv.org/abs/1905.09216}{{\ttfamily arXiv:1905.09216 [nucl-th]}}.

\bibitem{Beraudo:2022dpz}
A.~Beraudo, A.~De~Pace, M.~Monteno, M.~Nardi, and F.~Prino, ``{In-medium hadronization of heavy quarks and its effect on charmed meson and baryon distributions in heavy-ion collisions}'', \href{http://dx.doi.org/10.1140/epjc/s10052-022-10482-y}{{\em Eur. Phys. J.} {\bfseries C82} (2022) 607}, \href{http://arxiv.org/abs/2202.08732}{{\ttfamily arXiv:2202.08732 [hep-ph]}}.

\bibitem{Rapp:2018qla}
A.~Beraudo {\em et~al.}, ``{Extraction of Heavy-Flavor Transport Coefficients in QCD Matter}'', \href{http://dx.doi.org/10.1016/j.nuclphysa.2018.09.002}{{\em Nucl. Phys.} {\bfseries A979} (2018) 21--86}, \href{http://arxiv.org/abs/1803.03824}{{\ttfamily arXiv:1803.03824 [nucl-th]}}.

\bibitem{ALICE:2021bib}
{\bfseries ALICE} Collaboration, S.~Acharya {\em et~al.}, ``{Constraining hadronization mechanisms with $\Lambda_c$/$D_0$ production ratios in Pb--Pb collisions at $\sqrt{s_{NN}}=$~5~{TeV}}'', \href{http://dx.doi.org/10.1016/j.physletb.2023.137796}{{\em Phys. Lett.} {\bfseries B839} (2023) 137796}, \href{http://arxiv.org/abs/2112.08156}{{\ttfamily arXiv:2112.08156 [nucl-ex]}}.

\bibitem{ALICE:2020wla}
{\bfseries ALICE} Collaboration, S.~Acharya {\em et~al.}, ``{$\Lambda_c$ production in pp and in p--Pb collisions at $\sqrt{s_{NN}}=$~5~{TeV}}'', \href{http://dx.doi.org/10.1103/PhysRevC.104.054905}{{\em Phys. Rev.} {\bfseries C104} (2021) 054905}, \href{http://arxiv.org/abs/2011.06079}{{\ttfamily arXiv:2011.06079 [nucl-ex]}}.

\bibitem{ALICE:2022xrg}
{\bfseries ALICE} Collaboration, S.~Acharya {\em et~al.}, ``{Measurement of beauty-strange meson production in Pb--Pb collisions at $\sqrt{s_{NN}}=$~5~{TeV} via non-prompt $D_s$ mesons}'', \href{http://dx.doi.org/10.1016/j.physletb.2022.137561}{{\em Phys. Lett.} {\bfseries B846} (2023) 137561}, \href{http://arxiv.org/abs/2204.10386}{{\ttfamily arXiv:2204.10386 [nucl-ex]}}.

\bibitem{He:2014cla}
M.~He, R.~J. Fries, and R.~Rapp, ``{Heavy Flavor at the Large Hadron Collider in a Strong Coupling Approach}'', \href{http://dx.doi.org/10.1016/j.physletb.2014.05.050}{{\em Phys. Lett.} {\bfseries B735} (2014) 445--450}, \href{http://arxiv.org/abs/1401.3817}{{\ttfamily arXiv:1401.3817 [nucl-th]}}.

\bibitem{Ravagli:2007xx}
L.~Ravagli and R.~Rapp, ``{Quark Coalescence based on a Transport Equation}'', \href{http://dx.doi.org/10.1016/j.physletb.2007.07.043}{{\em Phys. Lett.} {\bfseries B655} (2007) 126--131}, \href{http://arxiv.org/abs/0705.0021}{{\ttfamily arXiv:0705.0021 [hep-ph]}}.

\bibitem{Mangano:1991jk}
M.~L. Mangano, P.~Nason, and G.~Ridolfi, ``{Heavy quark correlations in hadron collisions at next-to-leading order}'', \href{http://dx.doi.org/10.1016/0550-3213(92)90435-E}{{\em Nucl. Phys.} {\bfseries B373} (1992) 295--345}.

\bibitem{Cacciari:1998it}
M.~Cacciari, M.~Greco, and P.~Nason, ``{The $p_T$ spectrum in heavy-flavour hadroproduction.}'', \href{http://dx.doi.org/10.1088/1126-6708/1998/05/007}{{\em JHEP} {\bfseries 9805} (1998) 007}, \href{http://arxiv.org/abs/hep-ph/9803400}{{\ttfamily arXiv:hep-ph/9803400}}.

\bibitem{ALICE:2021dhb}
{\bfseries ALICE} Collaboration, S.~Acharya {\em et~al.}, ``{Charm-quark fragmentation fractions and production cross section at midrapidity in pp collisions at the LHC}'', \href{http://dx.doi.org/10.1103/PhysRevD.105.L011103}{{\em Phys. Rev.} {\bfseries D105} (2022) L011103}, \href{http://arxiv.org/abs/2105.06335}{{\ttfamily arXiv:2105.06335 [nucl-ex]}}.

\bibitem{ALICE:2014aev}
{\bfseries ALICE} Collaboration, B.~B. Abelev {\em et~al.}, ``{Beauty production in pp collisions at \twosevensix measured via semi-electronic decays}'', \href{http://dx.doi.org/10.1016/j.physletb.2014.09.026}{{\em Phys. Lett.} {\bfseries B738} (2014) 97--108}, \href{http://arxiv.org/abs/1405.4144}{{\ttfamily arXiv:1405.4144 [nucl-ex]}}.

\bibitem{ALICE:2012acz}
{\bfseries ALICE} Collaboration, B.~Abelev {\em et~al.}, ``{Measurement of electrons from beauty hadron decays in pp collisions at $\sqrt{s} = $ $7$~{TeV}}'', \href{http://dx.doi.org/10.1016/j.physletb.2013.01.069}{{\em Phys. Lett.} 13--23}, \href{http://arxiv.org/abs/1208.1902}{{\ttfamily arXiv:1208.1902 [hep-ex]}}. [Erratum: Phys.Lett.B 763, 507--509 (2016)].

\bibitem{ALICE:2021edd}
{\bfseries ALICE} Collaboration, S.~Acharya {\em et~al.}, ``{Prompt and non-prompt J/$\psi$ production cross sections at midrapidity in proton--proton collisions at $\sqrt{s} = 5.02$ and $13$~{TeV}}'', \href{http://dx.doi.org/10.1007/JHEP03(2022)190}{{\em JHEP} {\bfseries 2203} (2022) 190}, \href{http://arxiv.org/abs/2108.02523}{{\ttfamily arXiv:2108.02523 [nucl-ex]}}.

\bibitem{ALICE:2012msv}
{\bfseries ALICE} Collaboration, B.~Abelev {\em et~al.}, ``{Heavy flavour decay muon production at forward rapidity in proton--proton collisions at \seven}'', \href{http://dx.doi.org/10.1016/j.physletb.2012.01.063}{{\em Phys. Lett.} {\bfseries B708} (2012) 265--275}, \href{http://arxiv.org/abs/1201.3791}{{\ttfamily arXiv:1201.3791 [hep-ex]}}.

\bibitem{ALICE:2019rmo}
{\bfseries ALICE} Collaboration, S.~Acharya {\em et~al.}, ``{Production of muons from heavy-flavour hadron decays in pp collisions at \five}'', \href{http://dx.doi.org/10.1007/JHEP09(2019)008}{{\em JHEP} {\bfseries 1909} (2019) 008}, \href{http://arxiv.org/abs/1905.07207}{{\ttfamily arXiv:1905.07207 [nucl-ex]}}.

\bibitem{ALICE:2023sgl}
{\bfseries ALICE} Collaboration, S.~Acharya {\em et~al.}, ``{Charm production and fragmentation fractions at midrapidity in pp collisions at $\sqrt{s}=$~13~{TeV}}'', \href{http://dx.doi.org/10.1007/JHEP12(2023)086}{{\em JHEP} {\bfseries 2312} (2023) 086}, \href{http://arxiv.org/abs/2308.04877}{{\ttfamily arXiv:2308.04877 [hep-ex]}}.

\bibitem{Lisovyi:2015uqa}
M.~Lisovyi, A.~Verbytskyi, and O.~Zenaiev, ``{Combined analysis of charm-quark fragmentation-fraction measurements}'', \href{http://dx.doi.org/10.1140/epjc/s10052-016-4246-y}{{\em Eur. Phys. J.} {\bfseries C76} (2016) 397}, \href{http://arxiv.org/abs/1509.01061}{{\ttfamily arXiv:1509.01061 [hep-ex]}}.

\bibitem{ALICE:2022ruh}
{\bfseries ALICE} Collaboration, S.~Acharya {\em et~al.}, ``{Measurements of azimuthal anisotropies at forward and backward rapidity with muons in high-multiplicity p--Pb collisions at $\sqrt{s_{NN}}=$ 8.16~{TeV}}'', \href{http://dx.doi.org/10.1016/j.physletb.2023.137782}{{\em Phys. Lett.} {\bfseries B846} (2023) 137782}, \href{http://arxiv.org/abs/2210.08980}{{\ttfamily arXiv:2210.08980 [nucl-ex]}}.

\bibitem{Lin:2021mdn}
Z.-W. Lin and L.~Zheng, ``{Further developments of a multi-phase transport model for relativistic nuclear collisions}'', \href{http://dx.doi.org/10.1007/s41365-021-00944-5}{{\em Nucl. Sci. Tech.} {\bfseries 32} (2021) 113}, \href{http://arxiv.org/abs/2110.02989}{{\ttfamily arXiv:2110.02989 [nucl-th]}}.

\bibitem{Zhang:2020ayy}
C.~Zhang, C.~Marquet, G.-Y. Qin, Y.~Shi, L.~Wang, S.-Y. Wei, and B.-W. Xiao, ``{Collectivity of heavy mesons in proton--nucleus collisions}'', \href{http://dx.doi.org/10.1103/PhysRevD.102.034010}{{\em Phys. Rev.} {\bfseries D102} (2020) 034010}, \href{http://arxiv.org/abs/2002.09878}{{\ttfamily arXiv:2002.09878 [hep-ph]}}.

\bibitem{Collins:1989gx}
J.~C. Collins, D.~E. Soper, and G.~F. Sterman, ``{Factorization of Hard Processes in QCD}'', \href{http://dx.doi.org/10.1142/9789814503266_0001}{{\em Adv. Ser. Direct. High Energy Phys.} {\bfseries 5} (1989) 1--91}, \href{http://arxiv.org/abs/hep-ph/0409313}{{\ttfamily arXiv:hep-ph/0409313}}.

\bibitem{Catani:1990eg}
S.~Catani, M.~Ciafaloni, and F.~Hautmann, ``{High-energy factorization and small $x$ heavy flavor production}'', \href{http://dx.doi.org/10.1016/0550-3213(91)90055-3}{{\em Nucl. Phys.} {\bfseries B366} (1991) 135--188}.

\bibitem{Collins:1998rz}
J.~C. Collins, ``{Hard scattering factorization with heavy quarks: A General treatment}'', \href{http://dx.doi.org/10.1103/PhysRevD.58.094002}{{\em Phys. Rev.} {\bfseries D58} (1998) 094002}, \href{http://arxiv.org/abs/hep-ph/9806259}{{\ttfamily arXiv:hep-ph/9806259}}.

\bibitem{Ellis:1982cd}
R.~K. Ellis, W.~Furmanski, and R.~Petronzio, ``{Unraveling Higher Twists}'', \href{http://dx.doi.org/10.1016/0550-3213(83)90597-7}{{\em Nucl. Phys.} {\bfseries B212} (1983) 29}.

\bibitem{Alekhin:2017kpj}
S.~Alekhin, J.~Bl\"umlein, S.~Moch, and R.~Placakyte, ``{Parton distribution functions, $\alpha_{\rm s}$, and heavy-quark masses for LHC Run II}'', \href{http://dx.doi.org/10.1103/PhysRevD.96.014011}{{\em Phys. Rev.} {\bfseries D96} (2017) 014011}, \href{http://arxiv.org/abs/1701.05838}{{\ttfamily arXiv:1701.05838 [hep-ph]}}.

\bibitem{ALICE:2020wfu}
{\bfseries ALICE} Collaboration, S.~Acharya {\em et~al.}, ``{$\Lambda_c$ Production and Baryon-to-Meson Ratios in pp and p--Pb Collisions at $\sqrt{s_{NN}}=$~5~{TeV} at the LHC}'', \href{http://dx.doi.org/10.1103/PhysRevLett.127.202301}{{\em Phys. Rev. Lett.} {\bfseries 127} (2021) 202301}, \href{http://arxiv.org/abs/2011.06078}{{\ttfamily arXiv:2011.06078 [nucl-ex]}}.

\bibitem{ALICE:2021rzj}
{\bfseries ALICE} Collaboration, S.~Acharya {\em et~al.}, ``{Measurement of Prompt D$^{0}$, $\Lambda_{\rm c}^{+}$, and $\Sigma_{\rm c}^{0,++}$(2455) Production in Proton--Proton Collisions at $\sqrt{s}=$~13~{TeV}}'', \href{http://dx.doi.org/10.1103/PhysRevLett.128.012001}{{\em Phys. Rev. Lett.} {\bfseries 128} (2022) 012001}, \href{http://arxiv.org/abs/2106.08278}{{\ttfamily arXiv:2106.08278 [hep-ex]}}.

\bibitem{ALICE:2022cop}
{\bfseries ALICE} Collaboration, S.~Acharya {\em et~al.}, ``{First measurement of $\Omega_{\rm c}^{0}$ production in pp collisions at $\sqrt{s}=$~13~{TeV}}'', \href{http://dx.doi.org/10.1016/j.physletb.2022.137625}{{\em Phys. Lett.} {\bfseries B846} (2023) 137625}, \href{http://arxiv.org/abs/2205.13993}{{\ttfamily arXiv:2205.13993 [nucl-ex]}}.

\bibitem{ALICE:2024ocs}
{\bfseries ALICE} Collaboration, S.~Acharya {\em et~al.}, ``{Charm fragmentation fractions and ${\rm c}\overline{\rm c}$ cross section in p--Pb collisions at $\sqrt{s_{NN}}=$~5~{TeV}}'', \href{http://arxiv.org/abs/2405.14571}{{\ttfamily arXiv:2405.14571 [nucl-ex]}}.

\bibitem{Armesto:2018ljh}
N.~Armesto, ``{Small collision systems: Theory overview on cold nuclear matter effects}'', \href{http://dx.doi.org/10.1051/epjconf/201817111001}{{\em EPJ Web Conf.} {\bfseries 171} (2018) 11001}.

\bibitem{ALICE:2019fhe}
{\bfseries ALICE} Collaboration, S.~Acharya {\em et~al.}, ``{Measurement of prompt D$^{0}$, D$^{+}$, D$^{*+}$, and $D_s$ production in p--Pb collisions at $\sqrt{s_{NN}}=$~5~{TeV}}'', \href{http://dx.doi.org/10.1007/JHEP12(2019)092}{{\em JHEP} {\bfseries 1912} (2019) 092}, \href{http://arxiv.org/abs/1906.03425}{{\ttfamily arXiv:1906.03425 [nucl-ex]}}.

\bibitem{ALICE:2022exq}
{\bfseries ALICE} Collaboration, S.~Acharya {\em et~al.}, ``{First measurement of $\Lambda_c$ production down to $p_T = 0$ in pp and p--Pb collisions at $\sqrt{s_{NN}}=$~5~{TeV}}'', \href{http://dx.doi.org/10.1103/PhysRevC.107.064901}{{\em Phys. Rev.} {\bfseries C107} (2023) 064901}, \href{http://arxiv.org/abs/2211.14032}{{\ttfamily arXiv:2211.14032 [nucl-ex]}}.

\bibitem{ALICE:2024ozd}
{\bfseries ALICE} Collaboration, S.~Acharya {\em et~al.}, ``{Measurement of the production cross section of prompt $\Xi_{\rm c}^{0}$ baryons in p--Pb collisions at $\sqrt{s_{NN}}=$~5~{TeV}}'', \href{http://arxiv.org/abs/2405.14538}{{\ttfamily arXiv:2405.14538 [nucl-ex]}}.

\bibitem{ALICE:2022cxs}
{\bfseries ALICE} Collaboration, S.~Acharya {\em et~al.}, ``{W$^{\pm}$-boson production in p--Pb collisions at $\sqrt{s_{NN}}$ = 8.16~{TeV} and Pb--Pb collisions at $\sqrt{s_{NN}}=$~5~{TeV}}'', \href{http://dx.doi.org/10.1007/JHEP05(2023)036}{{\em JHEP} {\bfseries 2305} (2023) 036}, \href{http://arxiv.org/abs/2204.10640}{{\ttfamily arXiv:2204.10640 [nucl-ex]}}.

\bibitem{ALICE:2016uid}
{\bfseries ALICE} Collaboration, J.~Adam {\em et~al.}, ``{Measurement of electrons from beauty-hadron decays in p--Pb collisions at $\sqrt{s_{NN}}=$~5~{TeV} and Pb--Pb collisions at \twosevensixnn}'', \href{http://dx.doi.org/10.1007/JHEP07(2017)052}{{\em JHEP} {\bfseries 1707} (2017) 052}, \href{http://arxiv.org/abs/1609.03898}{{\ttfamily arXiv:1609.03898 [nucl-ex]}}.

\bibitem{ALICE:2021wct}
{\bfseries ALICE} Collaboration, S.~Acharya {\em et~al.}, ``{Measurement of inclusive charged-particle b-jet production in pp and p--Pb collisions at $\sqrt{s_{NN}}=$~5~{TeV}}'', \href{http://dx.doi.org/10.1007/JHEP01(2022)178}{{\em JHEP} {\bfseries 2201} (2022) 178}, \href{http://arxiv.org/abs/2110.06104}{{\ttfamily arXiv:2110.06104 [nucl-ex]}}.

\bibitem{ALICE:2018gyx}
{\bfseries ALICE} Collaboration, S.~Acharya {\em et~al.}, ``{Azimuthal Anisotropy of Heavy-Flavor Decay Electrons in p--Pb Collisions at $\sqrt{s_{NN}}=$~5~{TeV}}'', \href{http://dx.doi.org/10.1103/PhysRevLett.122.072301}{{\em Phys. Rev. Lett.} {\bfseries 122} (2019) 072301}, \href{http://arxiv.org/abs/1805.04367}{{\ttfamily arXiv:1805.04367 [nucl-ex]}}.

\bibitem{jet1}
{\bfseries ALICE} Collaboration, S.~Acharya {\em et~al.}, ``Measurement of the radius dependence of charged-particle jet suppression in pb–pb collisions at snn=5.02tev'', \href{http://dx.doi.org/https://doi.org/10.1016/j.physletb.2023.138412}{{\em Physics Letters B} {\bfseries 849} (2024) 138412}. \url{https://www.sciencedirect.com/science/article/pii/S0370269323007451}.

\bibitem{jet2}
{\bfseries ALICE} Collaboration, S.~Acharya {\em et~al.}, ``Modification of charged-particle jets in event-shape engineered pb–pb collisions at snn=5.02 tev'', \href{http://dx.doi.org/https://doi.org/10.1016/j.physletb.2024.138584}{{\em Physics Letters B} {\bfseries 851} (2024) 138584}. \url{https://www.sciencedirect.com/science/article/pii/S0370269324001424}.

\bibitem{jet4}
{\bfseries ALICE} Collaboration, ``Measurements of jet quenching using semi-inclusive hadron+jet distributions in pp and central pb$-$pb collisions at $\sqrt{s_{\rm NN}}=5.02$ tev'', 2023.
\newblock \url{https://arxiv.org/abs/2308.16128}.

\bibitem{jet3}
{\bfseries ALICE} Collaboration, ``Observation of medium-induced yield enhancement and acoplanarity broadening of low-$p_\mathrm{T}$ jets from measurements in pp and central pb$-$pb collisions at $\sqrt{s_{\rm NN}}=5.02$ tev'', 2023.
\newblock \url{https://arxiv.org/abs/2308.16131}.

\bibitem{jet5}
{\bfseries ALICE} Collaboration, S.~Acharya {\em et~al.}, ``Multiplicity dependence of charged-particle jet production in pp collisions at $\sqrt{s}$ = 13 {TeV}'', \href{http://dx.doi.org/10.1140/epjc/s10052-022-10405-x}{{\em The European Physical Journal C} {\bfseries 82} (Jun, 2022) 514}. \url{https://doi.org/10.1140/epjc/s10052-022-10405-x}.

\bibitem{jet6}
{\bfseries ALICE} Collaboration, S.~Acharya {\em et~al.}, ``Search for jet quenching effects in high-multiplicity pp collisions at $\sqrt{s}$ = 13 {TeV} via di-jet acoplanarity'', \href{http://dx.doi.org/10.1007/JHEP05(2024)229}{{\em Journal of High Energy Physics} {\bfseries 2024} (May, 2024) 229}. \url{https://doi.org/10.1007/JHEP05(2024)229}.

\bibitem{jet7}
{\bfseries ALICE} Collaboration, ``Measurement of the angle between jet axes in pb$-$pb collisions at $\sqrt{s_{\rm NN}} = 5.02$ tev'', 2023.
\newblock \url{https://arxiv.org/abs/2303.13347}.

\bibitem{jet8}
{\bfseries ALICE} Collaboration, S.~Acharya {\em et~al.}, ``Direct observation of the dead-cone effect in quantum chromodynamics'', \href{http://dx.doi.org/10.1038/s41586-022-04572-w}{{\em Nature} {\bfseries 605} (May, 2022) 440--446}. \url{https://doi.org/10.1038/s41586-022-04572-w}.

\bibitem{upgrade1}
L.~Musa, \href{http://dx.doi.org/10.17181/CERN-LHCC-2019-018}{``{Letter of Intent for an ALICE ITS Upgrade in LS3}'',} tech. rep., CERN, Geneva, 2019.
\newblock \url{https://cds.cern.ch/record/2703140}.

\bibitem{upgrade2}
{\bfseries ALICE} Collaboration, ``{Letter of Intent: A Forward Calorimeter (FoCal) in the ALICE experiment}'', tech. rep., CERN, Geneva, 2020.
\newblock \url{https://cds.cern.ch/record/2719928}.

\bibitem{upgrade3}
{\bfseries ALICE} Collaboration, ``{Upgrade of the ALICE Inner Tracking System during LS3: study of physics performance}'',. \url{https://cds.cern.ch/record/2868015}.

\bibitem{upgrade4}
{\bfseries ALICE} Collaboration, ``{Physics performance of the ALICE Forward Calorimeter upgrade}'',. \url{https://cds.cern.ch/record/2869141}.

\bibitem{upgrade5}
{\bfseries ALICE} Collaboration, ``{Technical Design report for the ALICE Inner Tracking System 3 - ITS3 ; A bent wafer-scale monolithic pixel detector}'', tech. rep., CERN, Geneva, 2024.
\newblock \url{https://cds.cern.ch/record/2890181}.

\bibitem{upgrade6}
``{Technical Design Report of the ALICE Forward Calorimeter (FoCal)}'', tech. rep., CERN, Geneva, 2024.
\newblock \url{https://cds.cern.ch/record/2890281}.

\bibitem{upgrade7}
{\bfseries ALICE} Collaboration, ``Letter of intent for alice 3: A next-generation heavy-ion experiment at the lhc'', 2022.
\newblock \url{https://arxiv.org/abs/2211.02491}.

\end{thebibliography}\endgroup

\end{document}